\documentclass[twocolumn,showpacs,preprintnumbers,amsmath,amssymb]{revtex4-2}
\usepackage{graphicx}
\usepackage{dcolumn}
\usepackage{color}
\usepackage{bm}
\usepackage{epsfig}
\usepackage{enumerate}
\usepackage{array}

\usepackage{multirow}
\usepackage{hyperref}
\hypersetup{
    colorlinks=true,
    linkcolor=blue,
    urlcolor=blue,
	citecolor=blue
}
\usepackage{upgreek}

\date{\today}
\begin{document}
\title{Coherent control of thermoelectric performance via engineered transmission functions in multi-dot Aharonov-Bohm heat engine}

\author{Sridhar,$^1$ Salil Bedkihal,$^2$  Malay Bandyopadhyay$^1$}
\affiliation{$^{1}$School of Basic Sciences, Indian Institute of Technology Bhubaneswar, Odisha, India 752050,\\
$^{2}$ Thayer School of Engineering, Dartmouth College, 15 Thayer Drive, Hanover, NH 03755, USA,
}\date{\today}

\begin{abstract}
  We theoretically investigate strategies for harnessing quantum interference to optimize the figure of merit $ZT$, power output, and thermodynamic efficiency in multi–quantum–dot Aharonov-Bohm (AB) thermoelectric heat engines. Within the non-equilibrium Green function formalism, we demonstrate that interference effects, such as Fano-type asymmetries, Dicke-like superradiant and subradiant modes, and multi-peaked transmission spectra, can be tailored through device geometry, magnetic flux, and dot–lead coupling to produce hybrid transmission profiles that combine features of Lorentzian, boxcar, and Fano lineshapes. Such engineered profiles enable configurations that balance the high efficiency of sharp Lorentzian resonances with the high power output of boxcar-like spectra, achieving near-optimal power–efficiency trade-offs. For symmetric quantum-dot arrays in square, pentagonal, and hexagonal configurations, we identify an optimal coupling regime, $t/\gamma\simeq 2$ (with interdot tunneling amplitude $t$ and dot–lead coupling strength $\gamma$), which yields the most favorable trade-off between power and efficiency. In particular, a hexagonal six–dot configuration achieves a $ZT \sim 30$ at dilution temperatures, while the four–dot geometry reaches $\sim 76\%$ of Carnot efficiency with output power $4.74fW$. We further find a direct correspondence between the high-$ZT$ regime and the maximal violation of the Wiedemann–Franz (WF) law. Introducing coupling asymmetry between source and drain enhances both efficiency and power. Scaling analysis reveals that efficiency increases systematically with the number of quantum dots, whereas power output reaches its maximum at intermediate system sizes. These results establish coherent control in multi-dot nanostructures as a viable pathway toward high-performance quantum thermoelectric heat engines, particularly relevant for ultralow-power electronics applications. 
\end{abstract}
\maketitle
\section{Introduction }
The efficient and optimal conversion of heat into work is a fundamental objective in the development of thermoelectric and waste-heat recovery technologies. At the nanoscale, the ability to engineer and control energy transport enables performance that surpasses the limitations of bulk systems \cite{bedkihal2025fundamental,benenti2017fundamental}. In bulk thermoelectrics, performance is fundamentally limited by the interplay of electrical and thermal transport properties, encapsulated in the thermoelectric figure of merit \cite{benenti2011thermodynamic,benenti2017fundamental,nath2014high}
\begin{equation}
  ZT=\frac{GS^2T}{\kappa},  
\end{equation}
where $G$ is the electrical conductance, $S$ the Seebeck coefficient, $T$ the operating temperature, and $\kappa$ the total thermal conductance. Achieving high $ZT$ generally requires a transmission spectrum that strongly favors energy-selective electron transport while suppressing parasitic heat conduction. However, in conventional materials, the WF law and phonon-mediated heat transport impose constraints that have historically kept $ZT$ below $\sim 3$ under ambient conditions \cite{pichanusakorn2010nanostructured}.\\
\indent
Quantum coherent effects in mesoscopic systems offer a way to circumvent some of these limitations. In nanoscale conductors such as quantum dots (QDs), molecular junctions, and nanostructures, quantum interference (QI) can drastically reshape the electronic transmission function $T(\omega)$, enabling improved energy filtering. Advances in nanofabrication and scanning probe techniques now allow exquisite control over QD energy levels, coupling strengths, and spatial arrangements, making it possible to engineer QI in a highly tunable fashion. The impact of quantum QI on transport phenomena has been extensively investigated in the fields of mesoscopic physics, quantum dots \cite{baranski2011fano,miroshnichenko2010fano,trocha2012large}, and electron transfer systems \cite{levstein1990tuning}.
The QI effect, which manifests itself as additional peaks or dips in transmission spectra, constitutes an intriguing subject for exploration. These characteristics can improve the functionality of molecular switches \cite{tada2002quantum}, sensors \cite{markussen2010electrochemical}, and thermoelectric devices \cite{stadler2011controlling,bergfield2011novel}. Thus, its range of applicability extends from fundamental research to advanced applications in sensor interferometers \cite{valkenier2014cross}.\\
\indent
Numerous studies have identified distinctive signatures of quantum interference. Specifically, sharp, symmetric transmission dips are characteristic of anti-resonances\cite{d1989half,emberly1999antiresonances,markussen2010electrochemical}, whereas Fano resonances produce asymmetric line shapes exhibiting both peaks and dips\cite{miroshnichenko2010fano,stadler2011controlling,papadopoulos2006control}, and perfectly symmetric sharp peaks correspond to Breit–Wigner resonances. Previous studies confirmed two limiting cases for $T(\omega)$ : (i) a narrow Lorentzian profile, which approaches the limit of the ideal energy filter and can maximize efficiency but at the cost of severely reduced output power; and (ii) a broad box-shaped profile, which enables large power output but with lower efficiency \cite{whitney2014most,whitney2015finding,whitney2019quantum}. Between these extremes lies a broad class of multi-peaked and asymmetric transmission profiles, arising from interference among multiple transport channels. Such intermediate line shapes offer the potential to enhance both efficiency and power output simultaneously, which is the central goal of this study: optimizing thermoelectric performance.
In this context, two coherent mechanisms of particular relevance are Fano resonances and Dicke-like interference. Fano resonances emerge from interference between a discrete resonant pathway and a continuum of states, producing a characteristic asymmetric line shape parameterized by the Fano asymmetry factor $q$ \cite{miroshnichenko2010fano,clerk2001fano}. This controllable asymmetry allows for fine-tuning of the balance between energy selectivity and overall transmission amplitude. Dicke-like interference, originating in quantum optics \cite{sitek2013dicke,dicke1953effect,wang2013enhancement}, occurs when multiple localized states couple to a common continuum, giving rise to superradiant modes (broad, strongly coupled) and subradiant modes (narrow, weakly coupled). In a transport setting, this mode separation can be exploited to suppress thermal conductance disproportionately to electrical conductance, leading to strong violations of WF law and improved $ZT$ \cite{fu2015enhancement,gonzalez2020thermoelectric}.\\
 \indent
In the present work, we consider quantum-dot interferometer geometries, such as square, pentagonal, and hexagonal arrangements threaded by Aharonov–Bohm (AB) flux as an ideal platform for studying and utilising the effects discussed in the preceding paragraph. Magnetic flux not only modifies the interference conditions but also enables coherent control of the transmission spectrum without altering structural parameters. Using the nonequilibrium Green’s function (NEGF) formalism, we systematically analyze how geometry, magnetic flux, coupling asymmetry, and system size influence the interplay of Fano and Dicke interference effects. Furthermore, by varying the ratio of interdot tunneling ($t$) to dot–lead coupling ($\gamma$), one can traverse distinct transport regimes: (i) Weak coupling ($t/\gamma>1$) where broadened transmission profiles maximize current and output power at the expense of energy selectivity, and we obtain less efficiency; (ii) Intermediate coupling ($t\sim\gamma$) where a mixture of sharp and broad features of transmission profile can yield a favorable power–efficiency trade-off; and (iii) Strong coupling ($t/\gamma<1$) where sharp resonances in the transmission profiles are favoured and high efficiency is obtained at the expense of low output power. Thus, our key findings are: {1. \bf Fano asymmetry} tuning boosts both $ZT$ and power up to an optimal $q$, beyond which power saturates. {\bf 2. Dicke-like} mode separation suppresses thermal conductance in specific geometries, yielding $ZT$ values much above bulk limits, with pronounced violations of the WF law. {\bf 3. System scaling}—with more coherently coupled QDs—enriches interference structures, enhancing efficiency with a trade-off in power. {\bf 4. Optimal performance} occurs for $t\sim 2\gamma$, where constructive interference suppresses thermal losses while preserving substantial electrical conductance.\\
\indent
The parameter regimes we explore here are closely aligned with those already realized in nanoscale thermoelectric experiments. For instance, InAS/InP nanowire quantum dots $ZT\sim 35$ at 30K \cite{prete2019thermoelectric}, comparable with our predicted $ZT\sim 30$ for the six-dot AB interferometer at $T\sim 4.8 mK$, enabled by ultra-narrow subradiant modes. In gate-defined QD AB interferometers, tunable Fano resonances with linewidths  $\Gamma \lesssim 50\,\mu\text{eV}$ and asymmetry parameters
$q\approx 1 \text{--} 10$ have been observed \cite{kobayashi2003mesoscopic,miroshnichenko2010fano,briones2017fano}. In our analysis for a single channel, we find a maximum power output of $1.4fW$ with efficiency $38\%$ in the range $q\approx 7-14$. At large $q$, efficiency increases, while power output decreases.
. In molecular junction thermoelectrics, interference has produced Seebeck coefficients  $S\approx200\mu V/K$ and power factors of 
$100\text{--}500\mu W m^{-1} K^{-2}$ \cite{finch2009giant,markussen2010relation,bergfield2009thermoelectric}; our modeled geometries achieve comparable $S\approx 130 \mu V/K$ and $PF \approx 180\mu W m^{-1} K^{-2}$ (see Appendix \ref{appendixc}) values and match these power factors in the tunneling regime $t\sim 2\gamma$. On the other hand, Coupled-QD experiments have reported Dicke-mode splitting with superradiant linewidths $\Gamma_{+}\sim 100\mu\text{eV}$, and subradiant linewidths as narrow as $\Gamma_{-}\sim 5\mu\text{eV}$ \cite{shahbazyan1994two,celardo2010transport,sheremet2023waveguide,zanner2022coherent}. Together, these correspondences underscore that the high $ZT$, strong power factors, and tunable spectral profiles predicted here are experimentally accessible with current nanofabrication and measurement capabilities.\\
\indent
These prior results motivate the present work, in which we engineer hybrid transmission functions using multi-QD Aharonov–Bohm interferometers. By tuning geometry, magnetic flux, and dot–lead coupling, we show how Fano-type asymmetries, Dicke-like modes, and multi-resonant structures can be combined to enhance both power and efficiency simultaneously. In particular, we identify an optimal intermediate coupling regime ($t/\gamma\sim2$) where efficiency and power reach their best compromise. Our predictions include exceptionally high $ZT$ values ($\sim30$ for a six-dot AB ring at dilution temperatures) and power outputs up to $4.74\text{fW}$ at efficiencies around $76\%$ of Carnot. These values significantly surpass the experimental benchmarks noted above, highlighting clear design principles for future experiments.\\
\indent
With this preamble in place, the paper is organized as follows. In Section \ref{sec:model}, we present the model and outline the methodology. Section \ref{sec:results} discusses our main findings, covering both the linear and nonlinear response regimes. Section \ref{conclusion} concludes with a summary and outlook. Additional technical details are provided in the appendices.
\section{ Models and methods}\label{sec:model}
In this section, we construct minimal yet versatile models of multi-quantum-dot Aharonov–Bohm (AB) interferometers that capture the essential physics while remaining analytically and numerically tractable to explore how quantum interference shapes thermoelectric performance systematically. These models allow us to tune geometric arrangement, interdot tunneling, and coupling to reservoirs, thereby controlling the resulting transmission profiles—from sharp Lorentzians to broad boxcar-like shapes and complex multi-peaked spectra. By formulating the problem in a noninteracting, spinless framework, we isolate the role of coherent transport and magnetic flux without the added complexity of electron–electron interactions, enabling a focused investigation of the mechanisms that maximize power and efficiency in nanoscale heat engines.
The total Hamiltonian $\mathcal{\hat{H}}$ of the whole system  is given by:
\begin{equation}\label{eq:2}
\mathcal{\hat{H}}=\mathcal{\hat{H}}_{S}+\mathcal{\hat{H}}_{B}+\mathcal{\hat{H}}_{S,B}
\end{equation}
\indent
Where $\mathcal{\hat{H}}_{S}$ is the Hamiltonian of the subsystem of a multidot quantum nanostructure. In our present study, we move beyond the extensively explored two- and three-quantum dots systems\cite{bedkihal2013flux,bandyopadhyay2021flux,bedkihal2013probe,bedkihal2012dynamics}—where the interplay of interference effects, Fano resonances, and Coulomb blockade physics has already been well explored. In this work, we study four-dot square (4QD), five-dot pentagonal (5QD), and six-dot hexagonal (6QD) nanostructures arranged in Aharonov–Bohm interferometer geometries \cite{bergfield2010giant}. These configurations are the smallest polygonal loops beyond the triangle, and each introduces qualitatively distinct interference features: the square supports multiple competing loop areas, the pentagon exhibits flux-frustrated pathways unique to odd-sided rings, and the hexagon provides the minimal setting for collective effects reminiscent of honeycomb lattices. Together, they extend the interference landscape beyond minimal systems while remaining fully coherent and experimentally tractable, making them promising platforms for optimizing thermoelectric performance.

 The Hamiltonian of the baths, ${\mathcal{\hat{H}_B}}$, describes the metallic leads, while ${\mathcal{\hat{H}_{SB}}}$ captures the coupling between the quantum dots and the reservoirs. The subsystem Hamiltonians for the 4QD, 5QD, and 6QD geometries are given by
\begin{equation}\label{eq:3}
\mathcal{\hat{H}}_{N\!QD}=%
\sum_{i=1}^{N} \mathcal{\epsilon}_{i}\,\hat{d}_{i}^{\dagger}\hat{d}_{i}
\;+\;\Biggl[\,
\sum_{\substack{i,j=1\\ i\neq j}}^{N}
t_{ij}\,e^{i\phi_{ij}}\,
\hat{d}_{i}^{\dagger}\hat{d}_{j}
\;+\;\text{H.c.}\,\Biggr]
\end{equation}
\indent
Here $N\in\{4,5,6\}$ denotes four (4QD), five (5QD), and six-quantum-dot (6QD) systems, respectively. Here, $\epsilon_{i}$ denotes the energy of the $i^{th}$ dot. $\hat{d_i}^\dagger$ and $\hat{d_i}$ are the electron creation and annihilation operators in the respective dots, and $t_{ij}$ is the tunneling strength between dots, and $\phi_{ij}$ is the AB phase factor. The Hamiltonian for the two metallic leads, source(S) and drain(D), consisting of non-interacting electrons, can be written as:
  \begin{equation}\label{eq:4}
     \mathcal{\hat{H}}_{B}=\sum_{k} \epsilon_{k,S}\hat{c}_{k,S}^{\dagger}\hat{c}_{k,S}+\sum_{k} \epsilon_{k,D}\hat{c}_{k,D}^{\dagger}\hat{c}_{k,D}
\end{equation}
\indent
where $\hat{c}_{k, S}^{\dagger}$ and $\hat{c}_{k, S}$ represent the electron creation and annihilation operators in the momentum state $k^{th}$, with energies $\epsilon_{k, S}$ and $\epsilon_{k, D}$ for the source and drain, respectively. The subsystem-bath interaction Hamiltonian can be written as
\begin{equation}\label{eq:5}
    \mathcal{\hat{H}}_{S,B} = 
\left[
\sum_{i \in {\alpha}} \sum_k v^S_{i,k} \, \hat{d}_i^\dagger \hat{c}_{k,S}
+ 
\sum_{i \in {\beta}} \sum_k v^D_{i,k} \, \hat{d}_i^\dagger \hat{c}_{k,D}
+ \text{h.c.}
\right]
\end{equation}
\indent
We denote our configurations as $N\: QD(\alpha,\beta)$ with $N$ as the number of dots in the loop, $\alpha$ and $\beta$ label the dots coupled to source (S) and drain (D), respectively. For example ($\alpha=\{1,2\}$), ($\beta=\{3,4\}$) for the configuration $4\:QD(2,2)$ (see Fig.\ref{ZT vs T 4QD(2,2)}(a1)). Here, $v_{i,k}^S$ and $v_{i,k}^D$ denote the coupling strength of the bath. The AB phases $\phi_{ij}$ satisfy the following relation \cite{bandyopadhyay2021flux,behera2023quantum}
 \begin{equation}
     \sum_{i}^N\phi_{i,i+1}=\phi=2\pi\frac{\Phi}{\Phi_{0}}
 \end{equation}
 \indent
Here, $i$ represents the site index of the AB ring, N is the total number of quantum dots in the interferometer loop, $\Phi$ is the total magnetic flux enclosed by the AB ring, and $ \Phi_0 =h/e$ is the quantum flux. In the steady state, physical observables are gauge invariant. Since the dots are present at the vertices of the square, pentagon, and hexagonal loop, we may choose the gauge as $\phi_{i,i+1}=\phi/N$. For our system, we maintain a symmetric voltage bias condition, that is, $\mu_{S}=-\mu_{D}$. However, by applying a gate voltage to each dot, we can place the levels of the dots away from the symmetric point at which $\mu_{S}-\epsilon_{i}=\epsilon_{i}-\mu_{D}$. We use natural unit conversion $\hbar=c=e=k_{B}=1$ for simplicity.\\
\indent
We use the nonequilibrium Green's function(NEGF) \cite{meir1992landauer,wang2014nonequilibrium} approach to solve the above model and compute the observables of interest. We follow the equation of motion method for the calculations (for details, see Appendix \ref{appendixA}) and obtained the retarded $[G^+(\omega)]$ and advanced $[G^-(\omega)]$ Green's functions for our system by using the quantum Langevin equation \cite{dhar2006nonequilibrium}. The Green's functions are given as
\begin{equation}
G_{i,\alpha(\beta)}^{\pm} = \Big[ \, \omega I - \mathcal{H}_{N} 
- \sum_{\alpha'} \Sigma_{\alpha',\alpha}^{\pm(S)}(\omega)\,\delta_{\alpha i} 
- \sum_{\beta'} \Sigma_{\beta',\beta}^{\pm(D)}(\omega)\,\delta_{\beta i} 
\Big]^{-1}
\end{equation}
\indent
where $\alpha,\alpha^{\prime}$ denote the connections to the source (S) and $\beta,\beta^{\prime}$  for the drain (D), with ($\alpha,\alpha^{\prime}=\{1,2\}$) and ($\beta,\beta^{\prime}=\{3,4\}$) for the $4\:QD(2,2)$ configuration as one such example. Further, I is an identity matrix, $ \Sigma_S^{\pm}(\omega)$ and $\Sigma_D^{\pm}(\omega)$ are the self energies (defined in Eq.(\ref{self_energy_eq}) of Appendix \ref{appendixA}) and the system Hamiltonian  $\mathcal{H}_{N}$ with $N\in\{4,5,6\}.$ Note that we impose energy degeneracy for the dots  \( \epsilon_i = \epsilon \) for all \( i \in \{1, \dots, 6\} \), and further consider symmetric inter-dot tunneling strength as \( t_{ij} = t \). In the wide-band limit(WBL), when the density of states(DOS) of the metallic lead is energy-independent, the real part of the self-energy term vanishes. Then we can define the hybridization matrix from the relation $\Sigma^{+}=-i\Gamma/2$. We derive the explicit expressions for the retarded Green's function and hybridization matrices for various configurations in Appendix \ref{appendixA}. 
\\ 
\indent
The transmission of electrons from reservoir $\nu$ to $\xi$ is given by the transmission coefficient \cite{datta1997electronic}.
\begin{equation}
    T_{\nu\xi}(\omega,\phi)=\mathrm{Tr}\big[\Gamma^{\nu}G^{+}(\omega,\phi)\Gamma^{\xi}G^{-}(\omega,\phi)\big].
\end{equation}
We also express the transmission probability from reservoir source(S) to drain(D) with symmetric dot-lead coupling $(\gamma_{S}=\gamma_{D}=\gamma)$ for different configurations in Appendix \ref{appendixA}. Investigating the subsystem properties and obtaining the transmission coefficient, particle currents, and heat currents through the system is interesting. 
Using the transmission coefficients $T_{\nu\xi}$, we can express the particle currents flowing from reservoir $\nu$ to the central system as \cite{sivan1986multichannel,butcher1990thermal} 
\begin{equation}\label{eq18}
    I_{\nu}=\int_{-\infty}^{\infty}d\omega\sum_{\xi\ne\nu}\big[
    T_{\nu\xi}(\omega,\phi)f_{\nu}(\omega)-
    T_{\xi\nu}(\omega,\phi)f_{\xi}(\omega)\big].
\end{equation}
Although the definition of the heat current is debatable  \cite {talkner2020colloquium}  in the strong coupling regime, we consider the conventional and usefully studied nonequilibrium Green's function approach to defining the heat current from the reservoir $\nu$ for arbitrary coupling as\cite{benenti2017fundamental,esposito2015quantum,esposito2015nature,seshadri2021entropy,bergmann2021green,topp2015steady}
\begin{equation}
\begin{split}
    Q_{\nu}=\int_{-\infty}^{\infty}d\omega(\omega-\mu_{\nu})
        \sum_{\xi\ne\nu}\big[T_{\nu\xi}(\omega,\phi)f_{\nu}(\omega) \\
         -T_{\xi\nu}(\omega,\phi)f_{\xi}(\omega)\big]
\end{split}
\end{equation}
where $f_{\nu(\xi)}(\omega)=[e^{(\omega-\mu_{\nu(\xi)})/T_{\nu(\xi)}}+1]^{-1}$ is Fermi distribution function of the reservoir $\nu(\xi)=S,D$ with $\mu_{\nu}$ and $T_{\nu}$ be the corresponding chemical potential and temperature, respectively.
To probe the nonlinear thermoelectric behavior of a two-terminal setup operating as a heat engine, we focus on two key performance indicators: the output power $P$ and the steady-state heat-to-work conversion efficiency $\eta$. In our framework, the system is driven out of equilibrium by a finite bias $\Delta\mu=\mu_D-\mu_S >0$ and a thermal gradient $\Delta T=T_S-T_D>0$. The output power $P$ arises from the net heat exchange between the reservoirs and the subsystem, and can be expressed as the total sum of all steady-state heat currents flowing across the junction \cite{benenti2017fundamental,mazza2014thermoelectric}
\begin{equation}\label{power}
    P=\sum_{\nu=S,D}Q_{\nu}=(\mu_D-\mu_S)I_S.
\end{equation}
Equation (\ref{power}) follows the laws of conservation of particle $\sum_{\nu}I_{\nu}=0$ and energy. We define the efficiency $\eta$ as the ratio of output power $P$ to the heat currents absorbed from the hot bath, and it is expressed as \cite{benenti2017fundamental,mazza2014thermoelectric}
\begin{equation}
    \eta=\frac{P}{Q_S}.
\end{equation}
The system works as a heat engine for positive output power $P>0$ with positive heat current flow from the source $Q_S>0$. The efficiency $\eta$ is bounded from above by the Carnot efficiency $\eta_C=1-T_c/T_h$ with $T_c$ and $T_h$ being the temperatures of the cold and hot bath, respectively.
\section{Results and discussion}\label{sec:results}
This section demonstrates how quantum interference—through Fano asymmetry, Dicke-like mode splitting, and multi-resonant transmissions—shapes the thermoelectric performance of multi-quantum-dot Aharonov–Bohm interferometers. We theoretically assess how variations in coupling strengths, magnetic flux, and system size influence quantum coherence and reshape the transmission spectrum, giving rise to optimal efficiency–power regimes and pronounced violations of the WF law. The analysis spans Fano resonances controlling asymmetry and efficiency–power trade-offs, Dicke-type interference suppressing thermal conductance, and nonlinear transport and scaling in larger quantum-dot arrays.
\subsection{Fano resonance }\label{sub:Fano}
\begin{figure*}[t]
    \centering
    \includegraphics[scale=0.21]{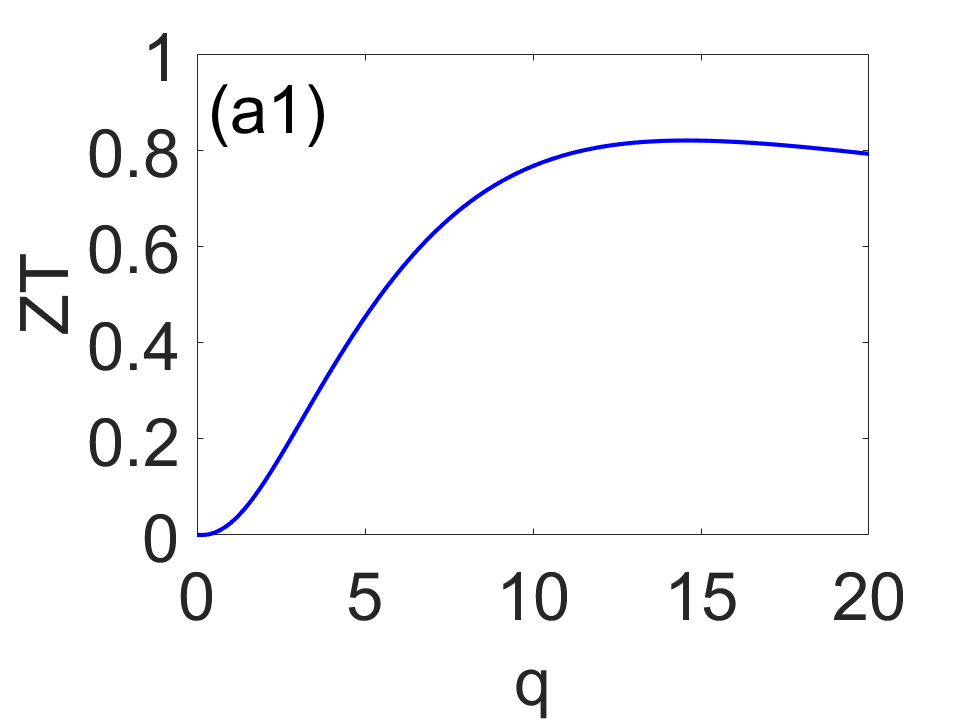}
    \includegraphics[scale=0.21]{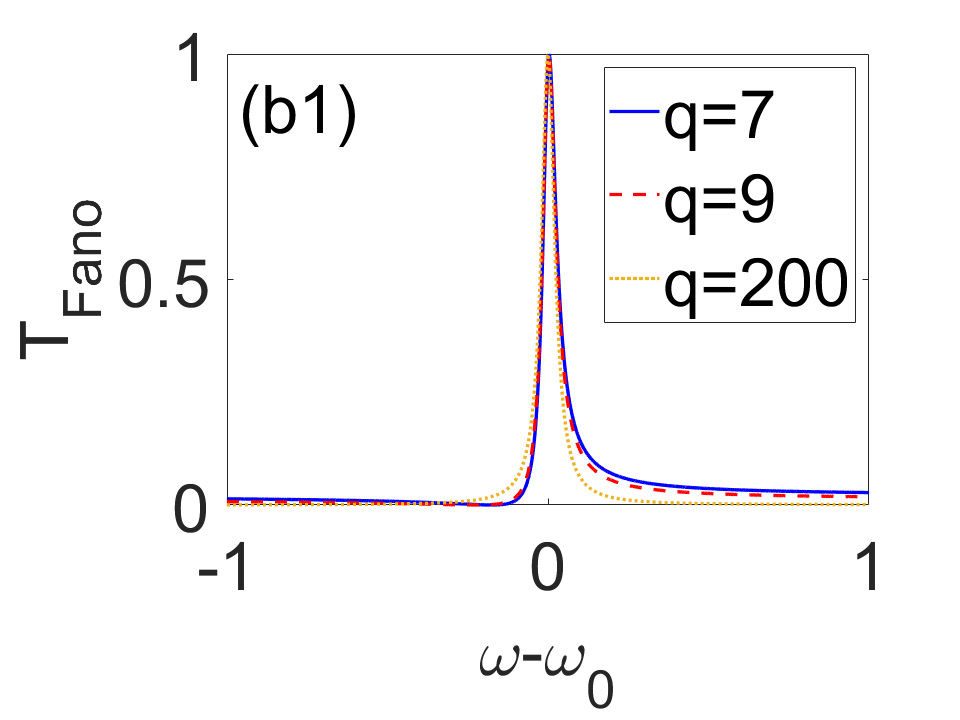}
    \includegraphics[scale=0.21]{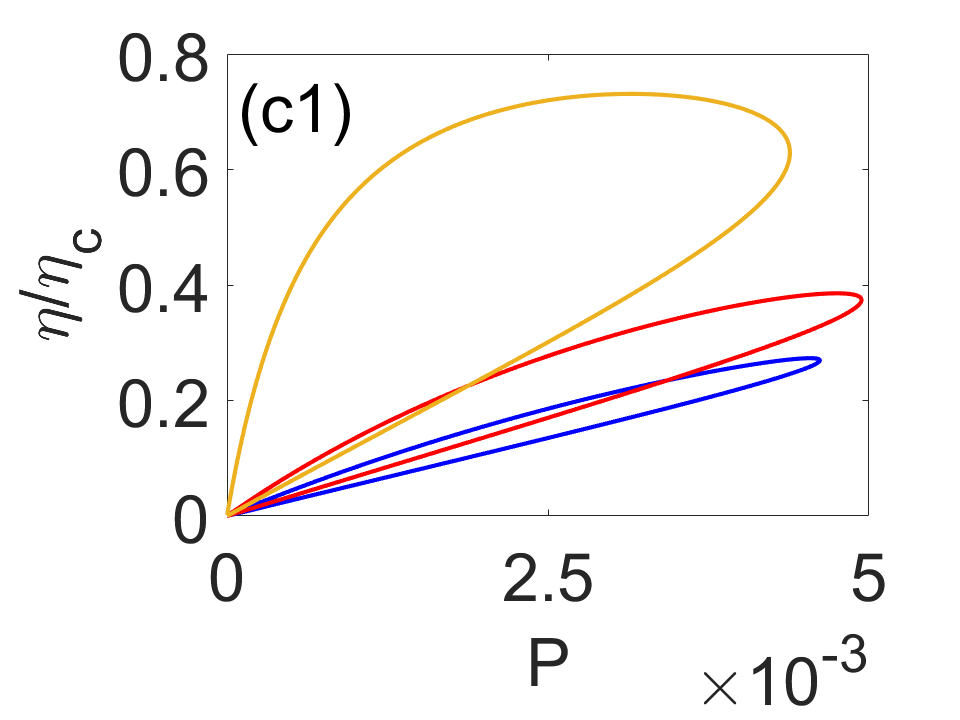}

    \caption{(a1) Figure of merit ZT as a function of q for a Fano-like lineshape with resonance. Parameters used to plot this graph are $\Gamma=0.05,\omega_0=0.3, T_S=0.4, T_D=T=0.3,\mu_D=\mu=0.2$ (b1) Normalized Fano transmission as a function of energy for different q values. (c1) Power efficiency trade-off in the non-linear regime for a single-channel Fano-resonance transmission. Parameters used $\Gamma=0.05,\omega_0=0.4, T_S=0.6, T_D=0.1,\mu_S=-\mu_D$.  }
    \label{Fig1}
\end{figure*}
Fano resonance \cite{miroshnichenko2010fano} arises from interference between resonant and non-resonant transport pathways, producing a characteristically asymmetric transmission line shape \cite{miroshnichenko2010fano,clerk2001fano}, which is defined as
\begin{equation}\label{Fano_transmission}
    T_{Fano}(\omega)=\frac{1}{1+q^2}\frac{(q+\epsilon)^2}{1+\epsilon^2}
\end{equation}
where $q$ is the asymmetry Fano parameter, a reduced energy $\epsilon$ defined by $2(\omega-\omega_0)/\Gamma$. $\omega_0$ is the resonant energy and $\Gamma$ is the resonance width.
Fano resonances in both the linear and nonlinear transport regimes are analyzed in Subsection~\ref{sub:Fano}. 
To be in the linear response regime, $\Delta$T and $\Delta\mu$ must be small. We assume that both the temperature difference
$\Delta T \equiv T_{S}-T_{D} > 0$ and the electrochemical potential difference $\Delta \mu \equiv \mu_{S}-\mu_{D} < 0 $ are small, that is, $|\Delta T | \ll T $ and
$|\Delta\mu| \ll \kappa_{B}T $ , where $\kappa_{B}$ is the Boltzmann constant. The thermodynamic forces (also known as generalized forces or affinities)
driving the electric and heat currents are given by $X_e = \Delta V$ (where $\Delta V = \Delta\mu/e $ is the applied voltage) and $X_h = \Delta T /T$
and the relationship between currents and generalized forces is linear \cite{onsager1931reciprocal} 
\begin{equation}
    \begin{aligned}
        I_{e}=\mathcal{L}_{11}X_{e}+\mathcal{L}_{12}X_{h} \\
        J_{h}=\mathcal{L}_{21}X_{e}+\mathcal{L}_{22}X_{h}
    \end{aligned}
\end{equation}
These relations are referred to as coupled phenomenological transport equations, or linear response equations, or kinetic equations, and the coefficients $\mathcal{L}_{ij}$ (i, j = 1, 2) are known as Onsager coefficients. We will define the matrix of these
coefficients as the Onsager matrix, $\mathcal{L}$, so
\begin{equation}
 \mathcal{L}=   \begin{pmatrix}
        \mathcal{L}_{11} & \mathcal{L}_{12}\\
        \mathcal{L}_{21} & \mathcal{L}_{22}
    \end{pmatrix}
\end{equation}
The $ZT$ can be expressed in terms of Onsager coefficients
\begin{equation}
    ZT=\frac{\mathcal{L}_{12}^2}{\mathcal{L}_{11}\mathcal{L}_{22}-\mathcal{L}_{12}\mathcal{L}_{21}}
\end{equation}
Where $\mathcal{L}_{ij}$ (i,j=1,2) are the Onsager coefficients for the two-terminal systems that can be written in the form of a block matrix. 
\begin{equation}
    \mathcal{L}_{i,j}=-{T}\int_{-\infty}^\infty d\omega T_{SD}\begin{pmatrix}
        1 & \omega-\mu\\
        \omega-\mu & (\omega-\mu)^2
    \end{pmatrix}f^\prime(\omega)
\end{equation}

where $T_{SD}$ is the Fano transmission function mentioned in  Eq.(\ref{Fano_transmission}) and $f^\prime(\omega)$ is the derivative of the Fermi distribution function .The ratio of thermal conductance to electrical conductance defines the Lorenz number \(L = \frac{\kappa}{GT}\). This ratio can be temperature dependent, where \(G = \mathcal{L}_{11}\) represents electrical conductance, and \(\kappa = \frac{1}{T}\left[\mathcal{L}_{22} - \frac{\mathcal{L}_{12}^2}{\mathcal{L}_{11}} \right]\) indicates thermal conductance. For a free electron gas, this relationship is described by the WF law, which establishes the universal Lorenz number \(L_0 = \frac{\pi^2}{3}\) \cite{wiley1986introduction}. However, not all systems exhibit this constant value; deviations from the WF law can occur, depending on the nature of the transmission function. Thus, the Lorenz ratio at a specific temperature is defined as \cite{bhandari2025thermoelectric}.
\begin{equation}
    \frac{L}L_0{}=\frac{3}{(\pi T)^2}\left[ \frac{\mathcal{L}_{22}}{\mathcal{L}_{11}}-\frac{\mathcal{L}_{12}^2}{\mathcal{L}_{11}^2} \right]
\end{equation}
The scaling of the $ZT$ as a function of $q$ is derived in Appendix \ref{appendix B}.
Figure~\ref{Fig1}(a1) shows the scaling of the thermoelectric $ZT$ with the Fano asymmetry parameter $q$. $ZT$ rises steadily and peaks near $q\simeq 14$, after which it declines and saturates, revealing an optimal $q$-window where energy filtering is maximized without excessive thermal leakage.\\
Figure~\ref{Fig1}(b1) presents the normalized transmission profiles. Increasing $q$ sharpens the Fano resonance, and in the large-$q$ limit ($q\sim200$), the profile reduces to a symmetric Lorentzian, marking the crossover from interference-driven to purely resonant transport.\\
Figure~\ref{Fig1}(c1) highlights the consequences for thermoelectric performance: both power and efficiency improve with $ q$ up to the optimal range, where sharp filtering aligns with sufficient transmission. Beyond this point, power saturates as a result of peak narrowing, while efficiency continues to rise from Lorentzian-like selectivity.
\subsection{Quantum coherent control : Dicke effect}\label{sub: Dicke effect}
The Dicke effect \cite{sitek2013dicke}, originally formulated in the context of quantum optics, describes the collective emission dynamics of multiple quantum emitters (such as atoms) coupled to a shared radiation field. When the emitters are indistinguishable and positioned within a wavelength of the field, their mutual coupling gives rise to correlated decay channels that split into \emph{superradiant} and \emph{subradiant} modes. The superradiant mode exhibits an enhanced decay rate due to constructive interference among the emitters, whereas the subradiant mode is strongly suppressed as destructive interference effectively protects it from radiative losses.

This cooperative mechanism finds a natural analogue in mesoscopic electronic systems, where localized quantum states—such as those in quantum dots or molecular orbitals—hybridize with a common electronic reservoir (e.g., metallic leads). In such systems, interference effects reminiscent of the Dicke effect emerge, manifesting as modified spectral line shapes, redistribution of spectral weight between broadened and narrowed resonances, and profound changes in electronic transport characteristics. These mesoscopic realizations not only extend the conceptual reach of the Dicke effect beyond photonic systems but also provide a platform to explore cooperative quantum interference in engineered nanoscale devices \cite{vorrath2003dicke, orellana2006kondo, shahbazyan1998localized, wunsch2003quasistates}.\\
To illustrate this, let us consider a single localized state (LS) with energy $\varepsilon_1$, coupled to a continuum of electronic states. The spectral function is then a Lorentzian:
\begin{equation}
    S(\omega) = \frac{1}{\pi} \frac{\Gamma}{(\omega - \varepsilon_1)^2 + \Gamma^2}, \qquad 
    \Gamma = \pi \sum_k |t_{1k}|^2 \delta(\omega - E_k),
\end{equation}
where $t_{1k}$ is the tunneling amplitude between the LS and the plane-wave state $k$ in the lead. The width $\Gamma$ encodes the finite lifetime of the LS due to tunneling-induced decay into the continuum. Now place a second LS with energy $\varepsilon_2$ at some distance from the first. The key point is that both LSs are coupled to the \emph{same} electronic continuum. The resulting spectral function becomes:
\begin{equation}
    S(\omega) = -\frac{1}{\pi} \, \mathrm{Im} \left[ \frac{1}{2} \mathrm{Tr} \left( \frac{1}{\omega - \hat{\varepsilon} + i \hat{\Gamma}} \right) \right],
\end{equation}
where $\hat{\varepsilon}$ is a $2 \times 2$ diagonal matrix with eigenvalues $\varepsilon_1$ and $\varepsilon_2$, and the non-Hermitian matrix $\hat{\Gamma}$ has elements
\begin{equation}
    \Gamma_{ij} = \pi \sum_k t_{ik} t_{jk}^* \delta(\omega - E_k).
\end{equation}
Because the tunneling amplitudes depend on the positions of the LSs through the phase $t_{ik} \sim e^{i \mathbf{k} \cdot \mathbf{r}_i}$, the off-diagonal elements of $\hat{\Gamma}$ encode quantum interference:
\begin{equation}
    \Gamma_{12} = q \sqrt{\Gamma_1 \Gamma_2}, \qquad 
    q = J_0(k_F r_{12}),
\end{equation}
where $J_0$ is the Bessel function and $r_{12} = |\mathbf{r}_1 - \mathbf{r}_2|$ is the LS separation. 

In the simplest case of identical LSs ($\varepsilon_1 = \varepsilon_2 = \varepsilon$, $\Gamma_1 = \Gamma_2 = \Gamma$), the spectral function splits into two Lorentzians:
\begin{equation}
    S(\omega) = \frac{1}{2\pi} \left[ \frac{\Gamma_+}{(\omega - \varepsilon)^2 + \Gamma_+^2} + \frac{\Gamma_-}{(\omega - \varepsilon)^2 + \Gamma_-^2} \right] 
\end{equation}
 where $\Gamma_{\pm} = (1 \pm q)\Gamma.$ These correspond to a superradiant mode with enhanced width $\Gamma_+$ and a subradiant mode with suppressed width $\Gamma_-$. The interference with the shared continuum induces collective decay properties even in the absence of direct coupling between the LSs. It has been shown that such collective effects can enhance the thermoelectric effects, increasing the figure of merit $ZT>>1$ \cite{wang2013enhancement}.\\
 \indent
In nanoscale systems such as quantum dot arrays embedded in a loop geometry (e.g., an Aharonov-Bohm ring), this Dicke-like physics becomes highly tunable via external parameters such as magnetic flux. In such settings, the superradiant mode facilitates strong charge transport while the subradiant mode, being long-lived, acts as an energy filter.
This separation of decay times offers ways to suppress parasitic heat conduction while maintaining electrical current, thus improving the $ZT$.\\
\indent
In this work, we show a strong violation of the WF law in various topologies of coupled quantum-dot systems that can be tuned using magnetic flux. We also show the enhancement of power and efficiency in the nonlinear regime.

\begin{figure*}[t]
    \centering
    \setlength{\tabcolsep}{2pt} 
    \renewcommand{\arraystretch}{1.0} 

    \begin{tabular}{cccc}
       \includegraphics[scale=0.21]{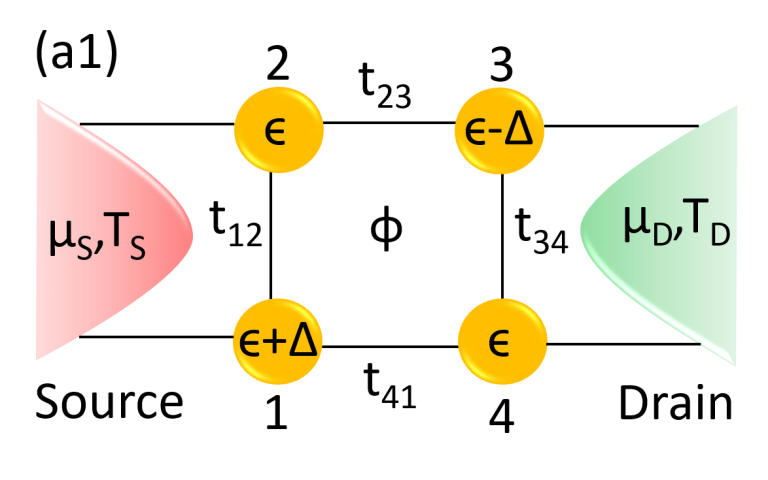}&
       \includegraphics[scale=0.21]{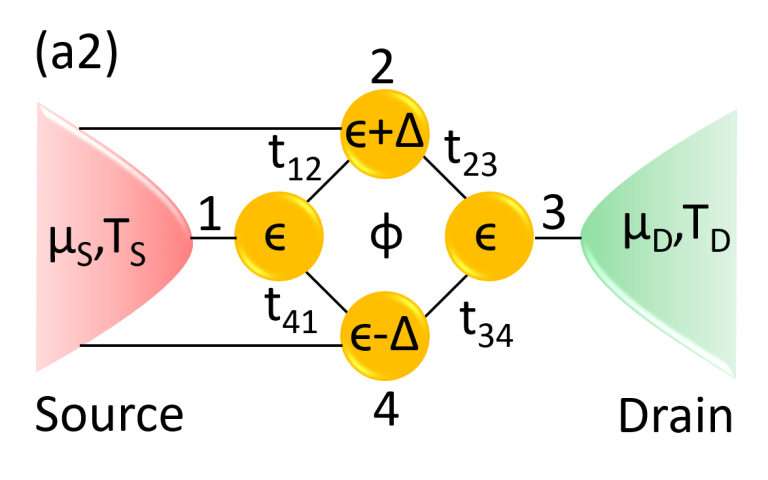}&
       \includegraphics[scale=0.21]{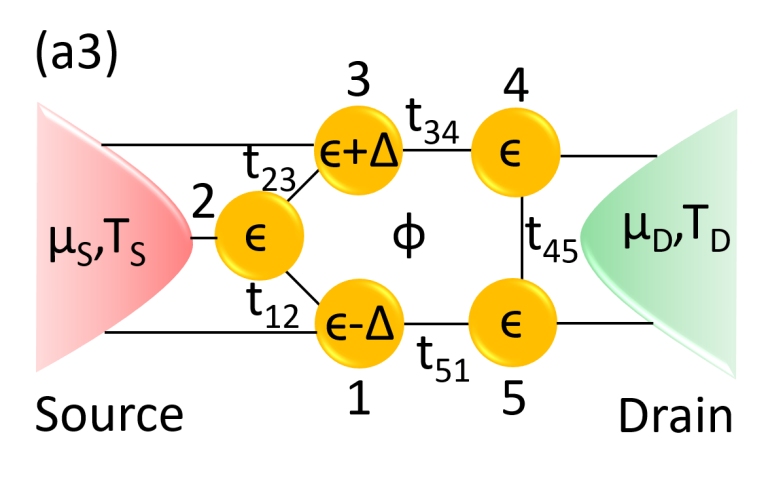}&
        \includegraphics[scale=0.21]{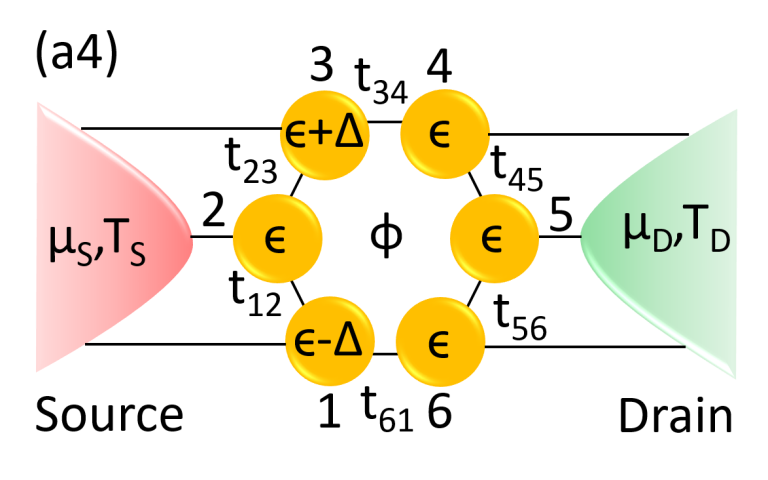}\\[-1ex]
       {\includegraphics[scale=0.21]{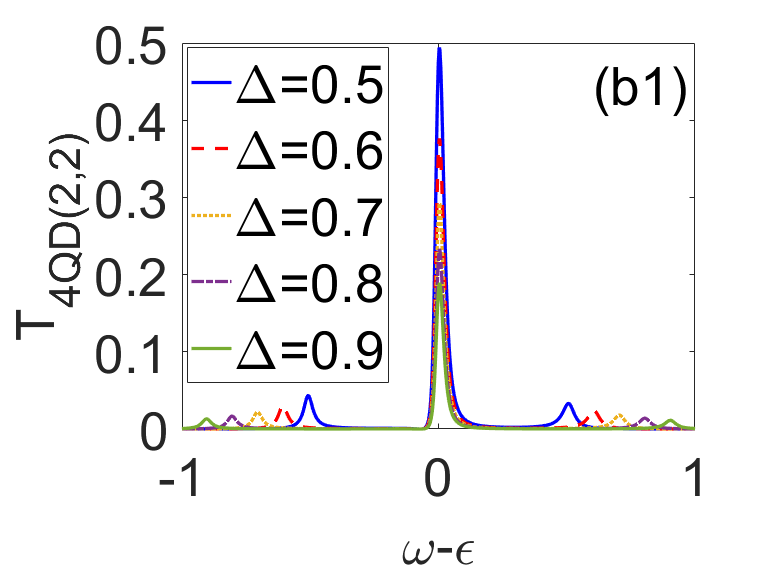}} &
        {\includegraphics[scale=0.21]{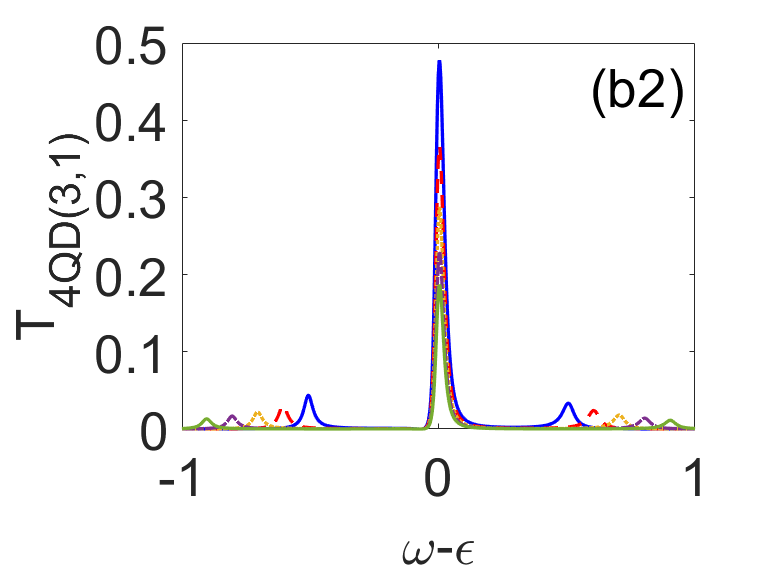}} &
        \includegraphics[scale=0.21]{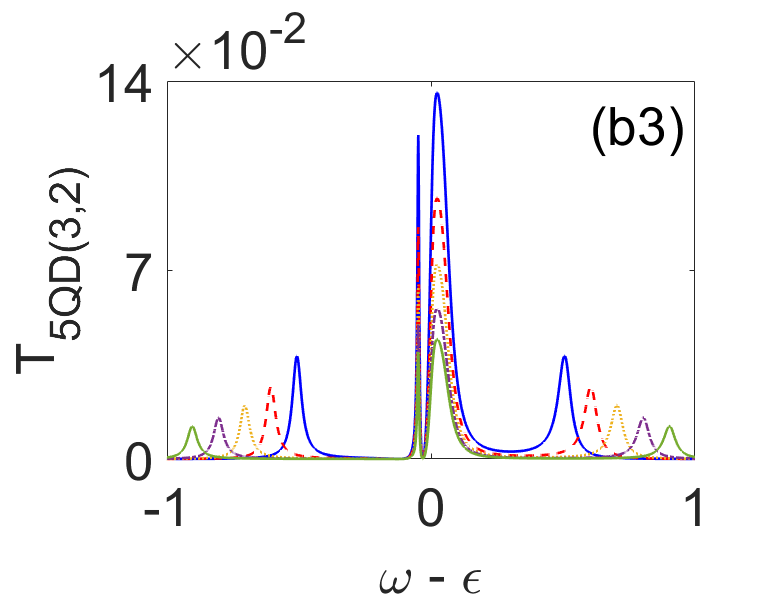}&
        {\includegraphics[scale=0.21]{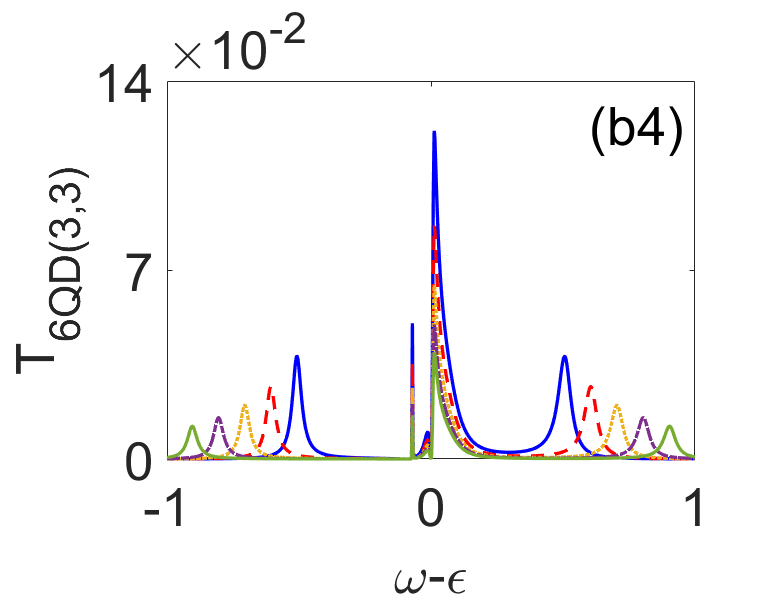}} \\[-1ex]
        {\includegraphics[scale=0.21]{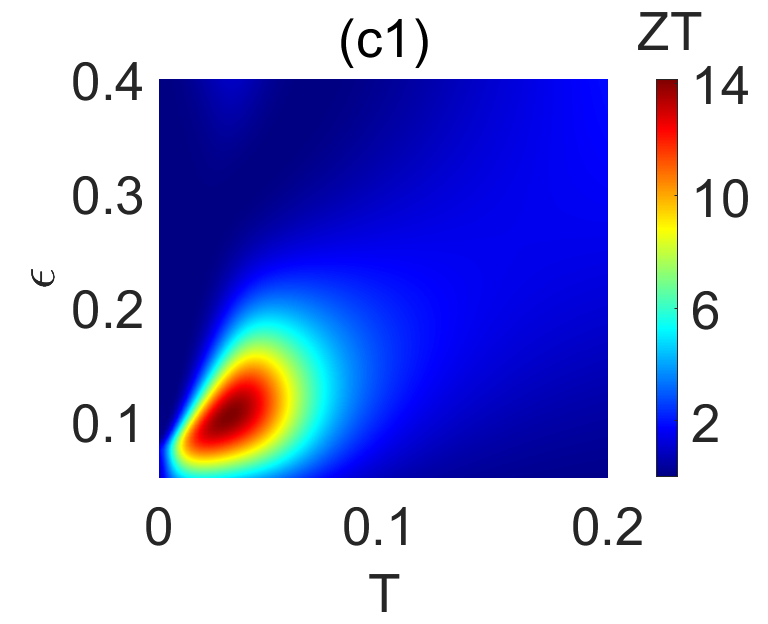}} &
        {\includegraphics[scale=0.21]{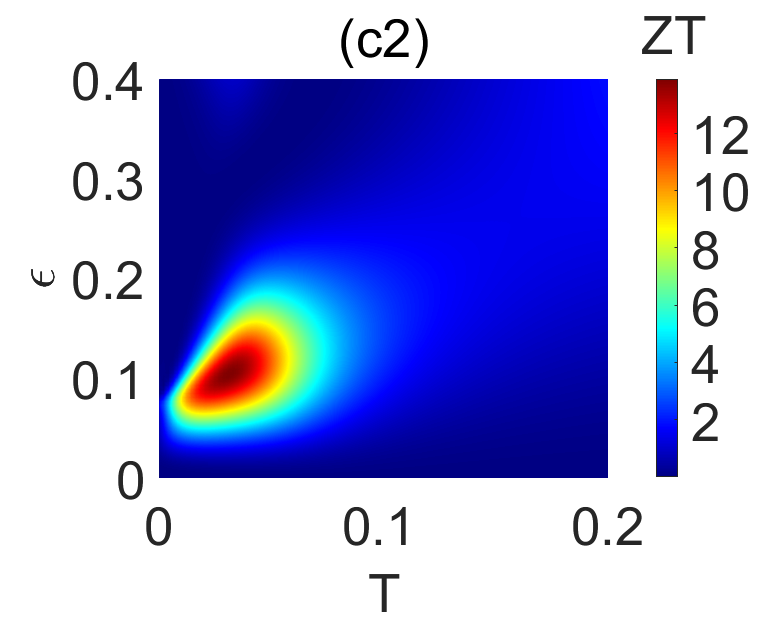}} &
        \includegraphics[scale=0.21]{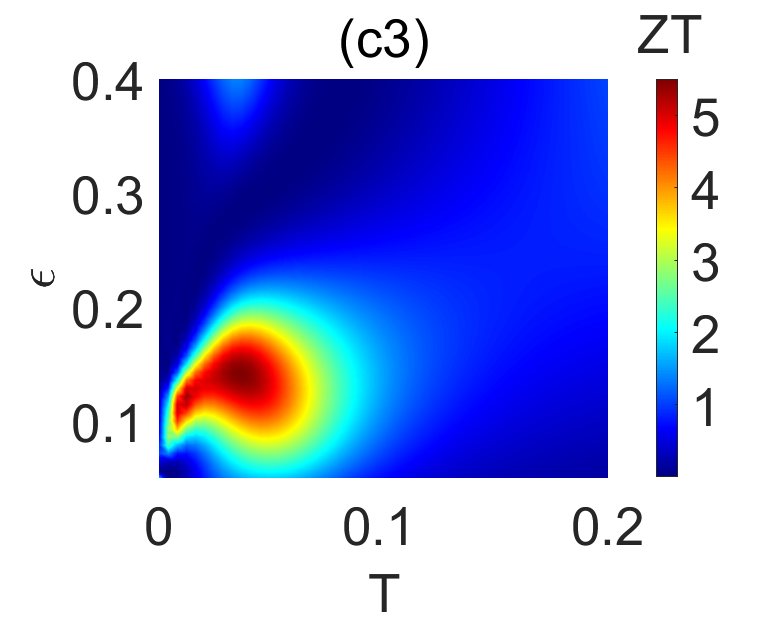}&
        {\includegraphics[scale=0.21]{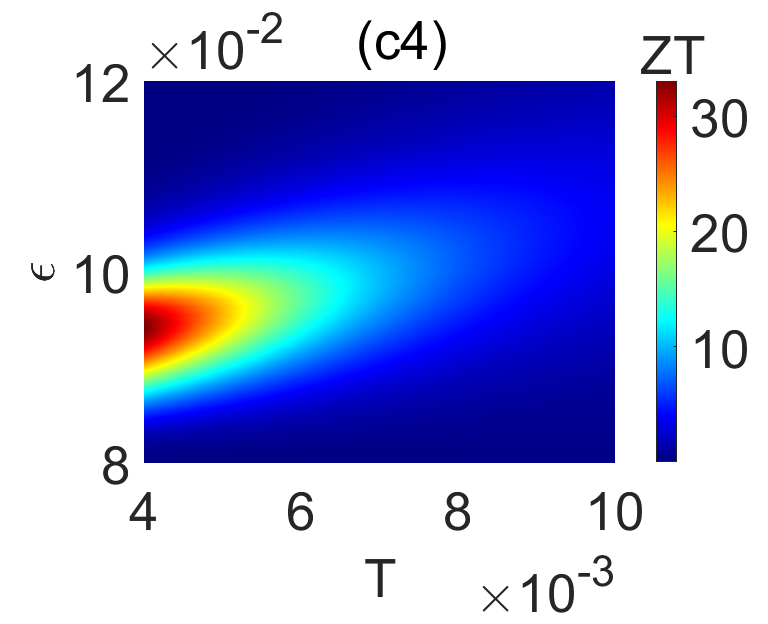}} \\[-1ex]
        {\includegraphics[scale=0.21]{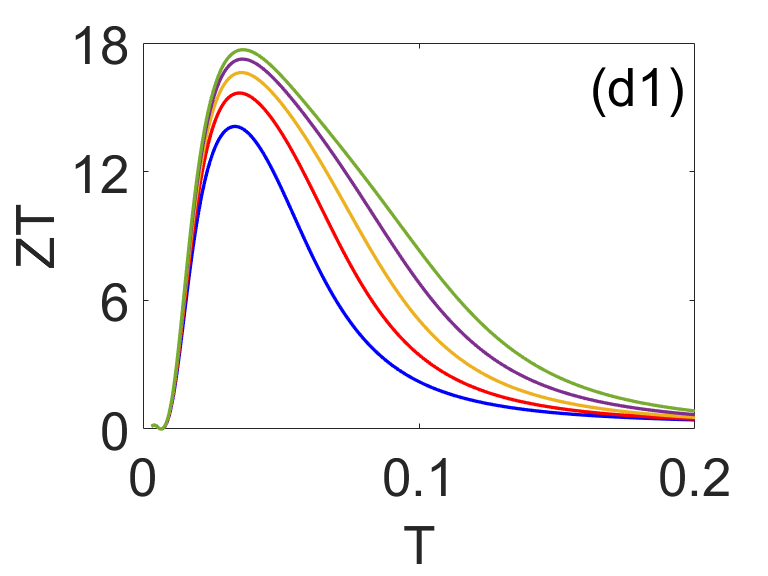}} &
        {\includegraphics[scale=0.21]{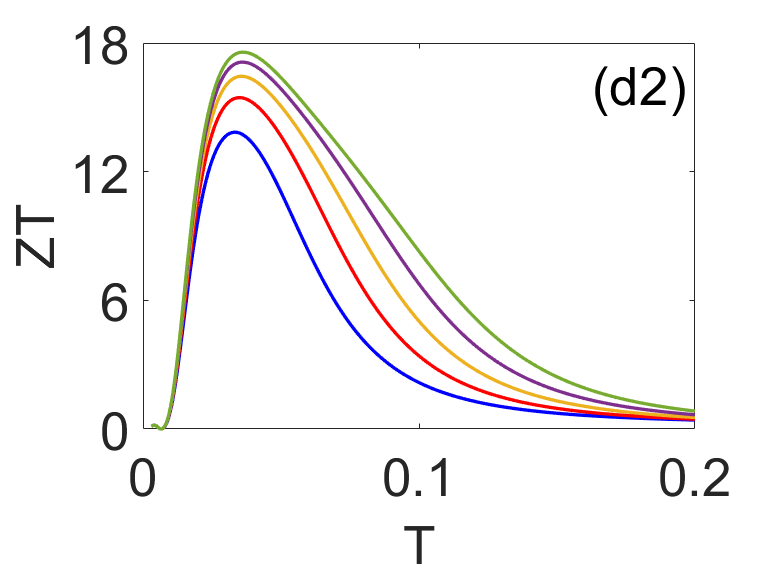}} &
        \includegraphics[scale=0.21]{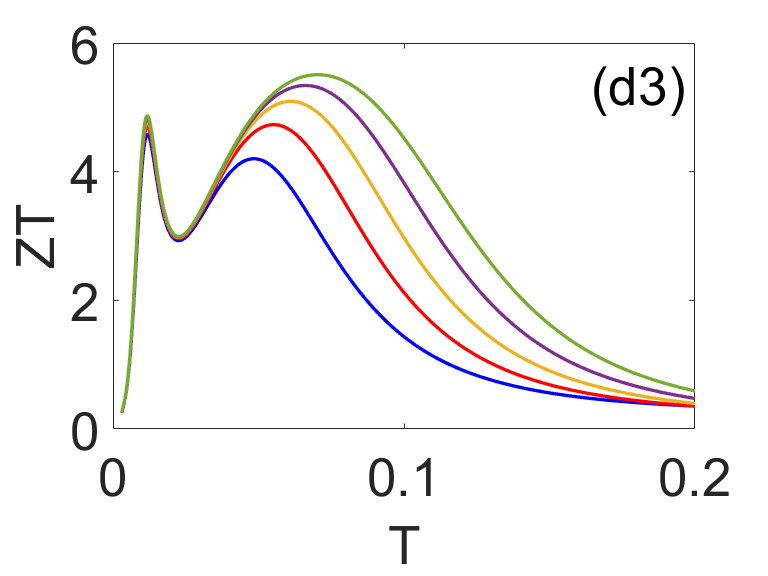}&
        {\includegraphics[scale=0.21]{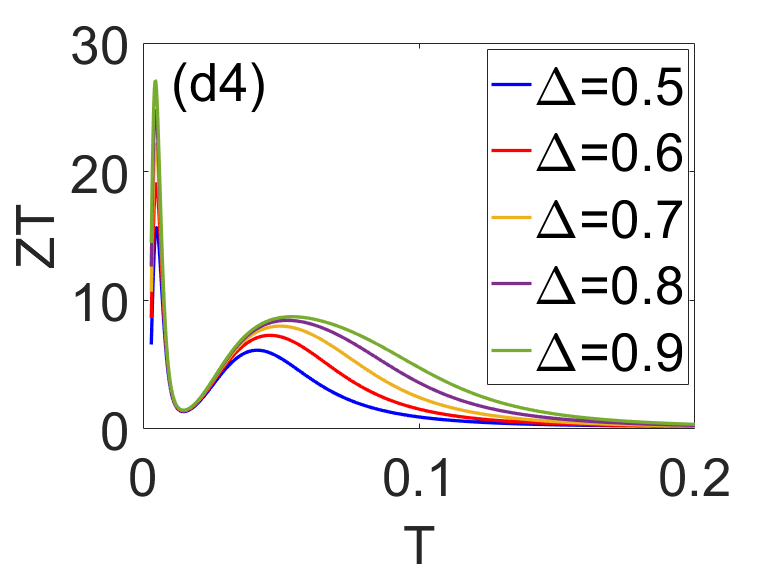}} \\
        \end{tabular}
        \caption{First row (a1--a4) shows the Aharonov-Bohm (AB) interferometer models:  (a1) 4QD(2,2), (a2) 4QD(3,1), (a3) 5QD(3,2), and (a4) 6QD(3,3). A perpendicular magnetic flux \( \phi \) threads each AB ring. The second row (b1--b4) displays the corresponding transmission spectra as functions of energy. The third row (c1--c4) presents heatmaps of the figure of merit \( ZT \) as functions of energy level \( \epsilon \) and temperature \( T \), for fixed level shift \( \Delta = 0.5 \). The fourth row (d1--d4) shows \( ZT \) as a function of temperature \( T \). Parameters used: \( \gamma = 0.05 \), \( \epsilon = 2\gamma \), \( \mu_D = \mu = 0.01 \), \( \phi = \pi \), \( \Delta = 0.5 \), and \( t/\gamma = 1 \).}
    \label{ZT vs T 4QD(2,2)}
\end{figure*}

\begin{figure*}
    \centering
    \setlength{\tabcolsep}{2pt} 
    \renewcommand{\arraystretch}{1.0} 

    \begin{tabular}{cccc}
    \includegraphics[scale=0.21]{Fig_a1.png}&
    \includegraphics[scale=0.21]{Fig_a2.png}&
    \includegraphics[scale=0.21]{Fig_a3.png}&
    \includegraphics[scale=0.21]{Fig_a4.png}\\[-1ex]
    \includegraphics[scale=0.21]{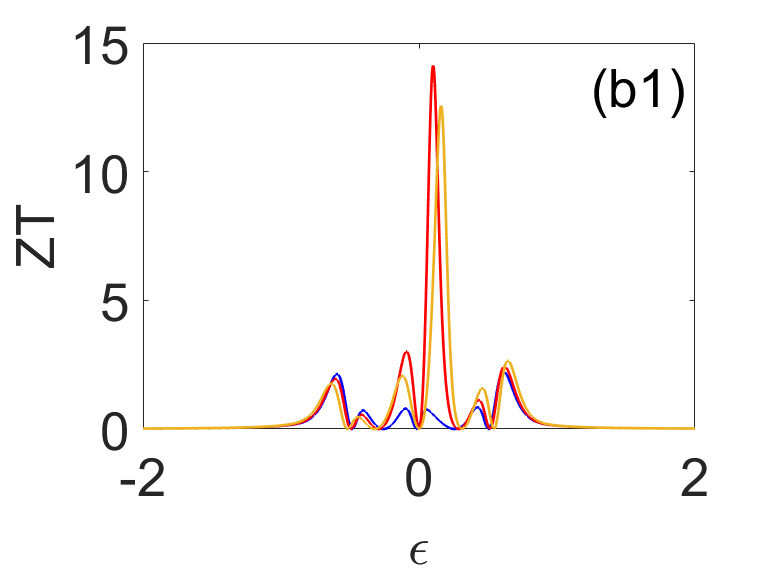}&
    \includegraphics[scale=0.21]{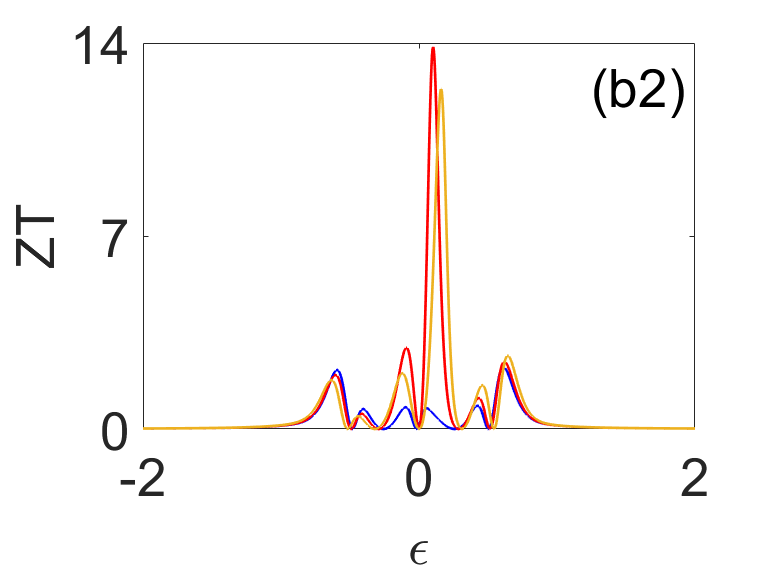}&
    \includegraphics[scale=0.21]{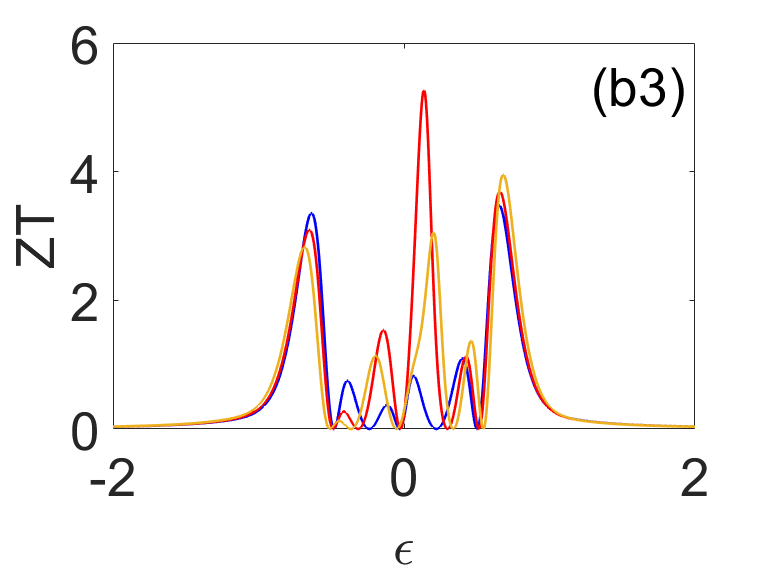}&
    \includegraphics[scale=0.21]{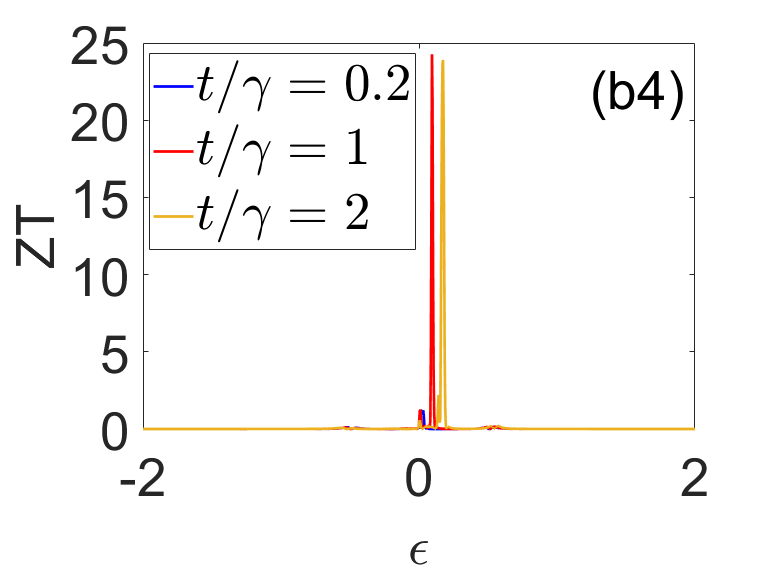}\\[-1ex]
    \includegraphics[scale=0.21]{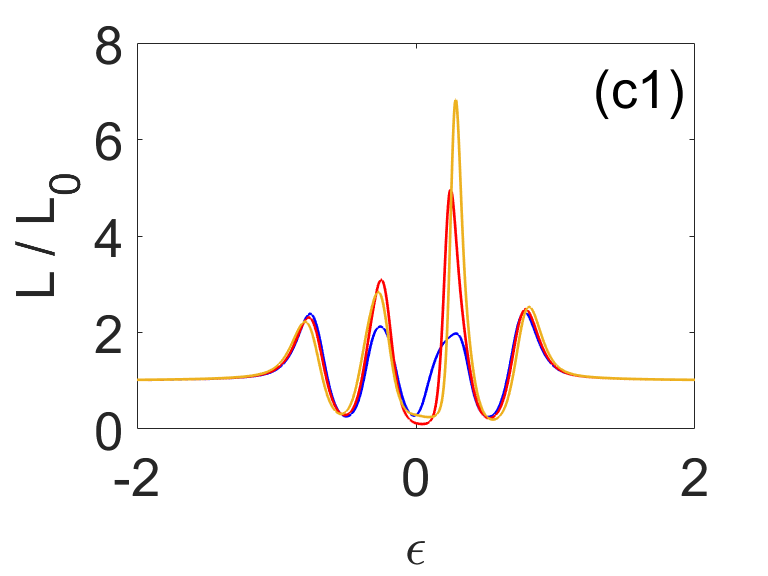}&
    \includegraphics[scale=0.21]{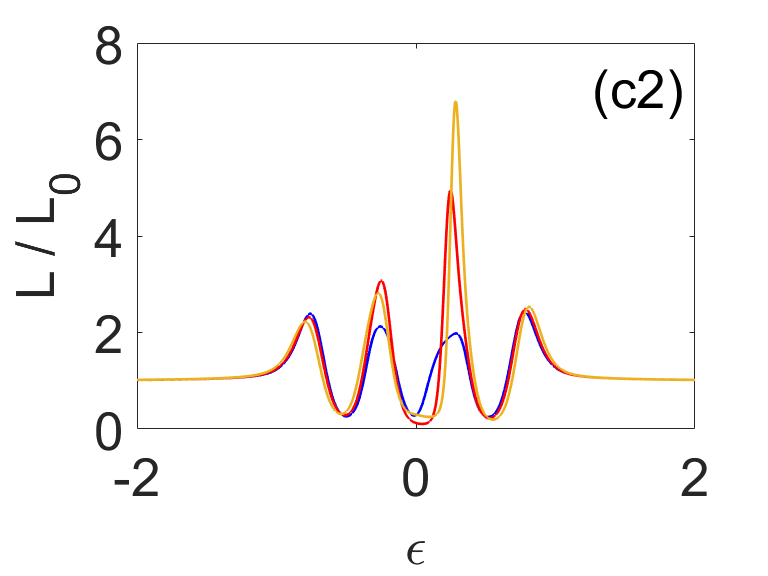}&
    \includegraphics[scale=0.21]{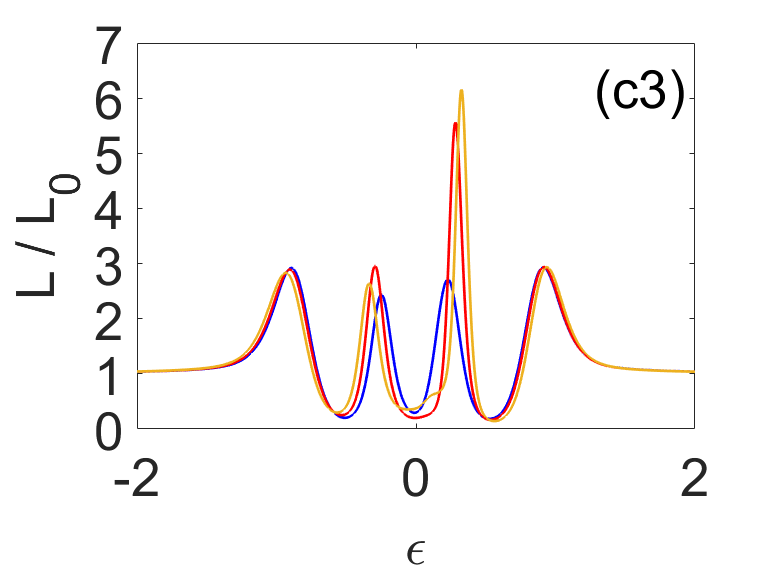}&
    \includegraphics[scale=0.21]{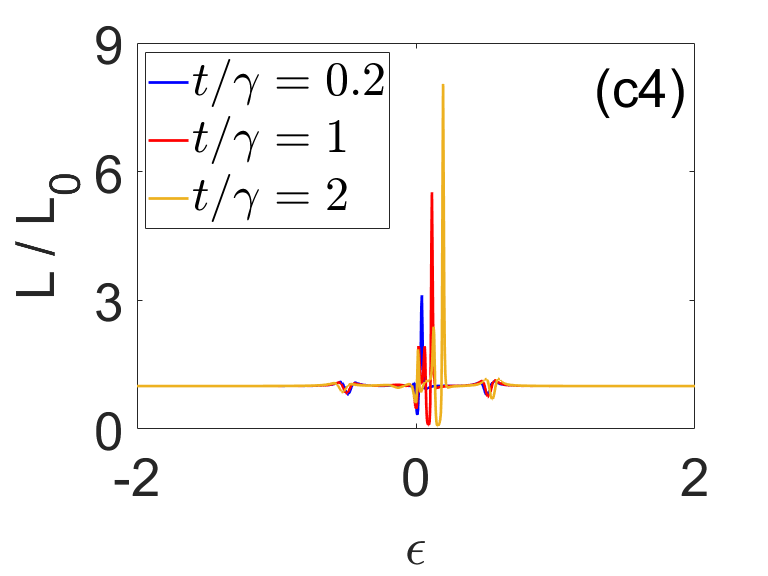}\\[-1ex]
    \includegraphics[scale=0.21]{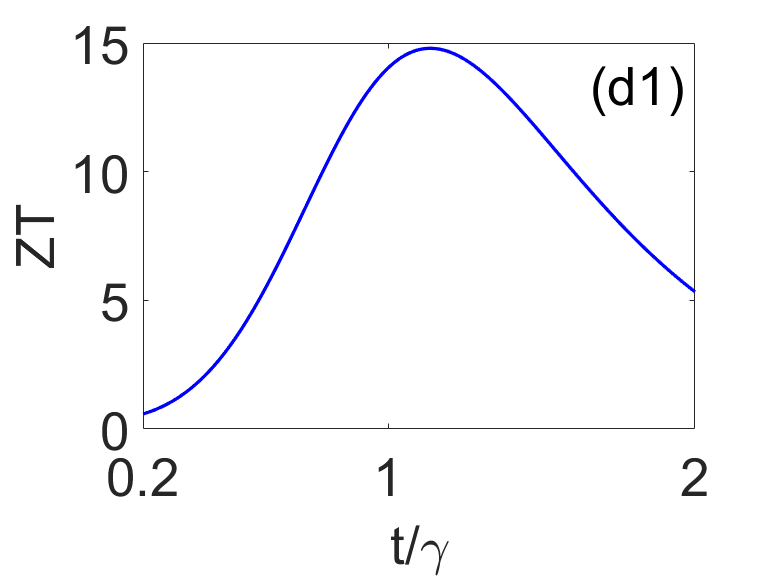}&
    \includegraphics[scale=0.21]{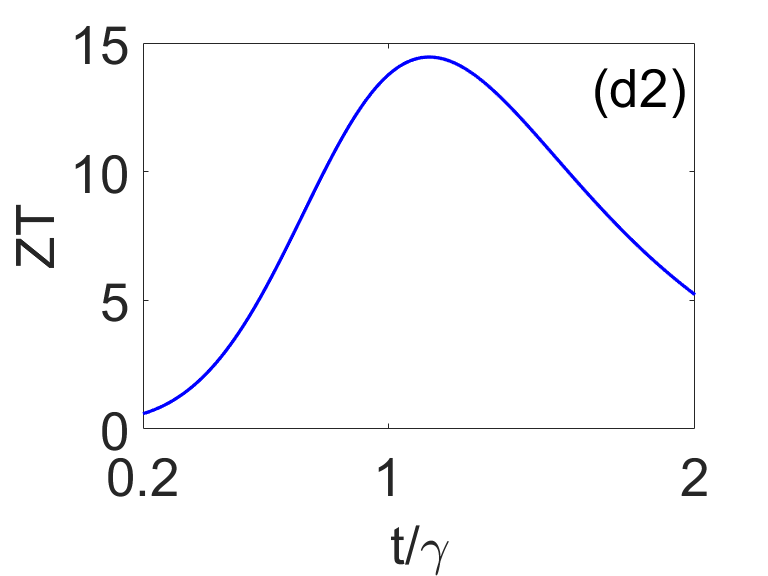}&
    \includegraphics[scale=0.21]{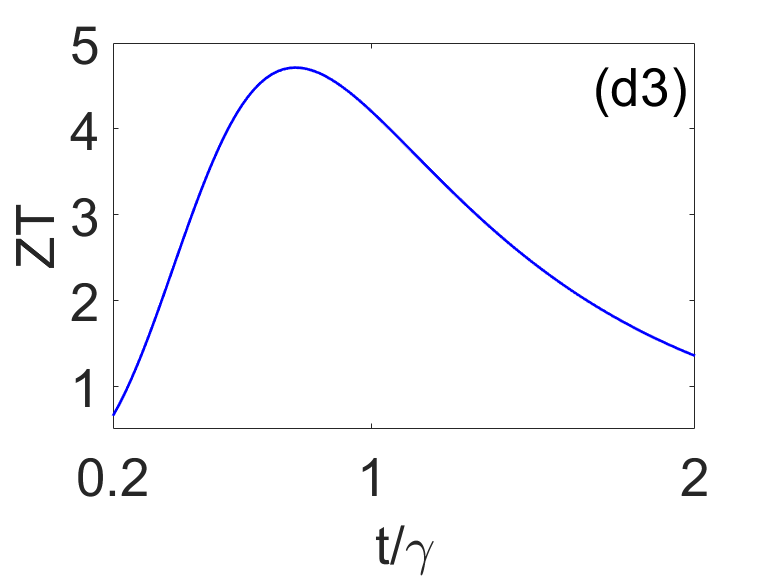}&
    \includegraphics[scale=0.21]{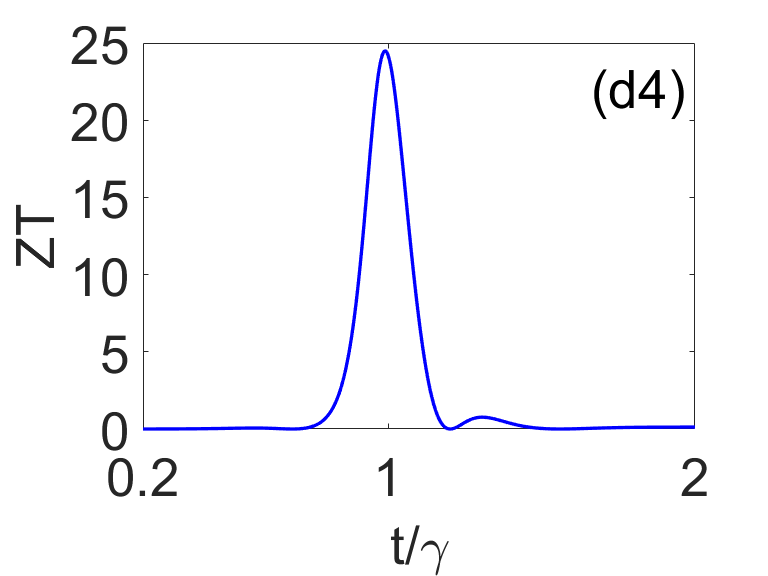}\\
   \end{tabular}
    \caption{(a1--a4) shows the quantum dot configurations, (b1)-(b4) display ZT as a function of $\epsilon$ for three different regimes. (c1)-(c4) present the Lorenz ratio as a function of $\epsilon$. (d1)-(d4) shows the ZT as a function of $t/\gamma$ for a fixed level spacing $\Delta=0.5$. Parameters used are: $\gamma=0.05,\epsilon=0.1,\phi=\pi,\Delta=0.5$, and $T=0.033$ for (a1), $T=0.0334$ for (a2), $T=0.048$ for 5QD(3,2), and $T=0.0048$ for 6QD(3,3).}
\label{Lorentz ratio results}
\end{figure*}

Figure~\ref{ZT vs T 4QD(2,2)} presents the thermoelectric performance of various quantum dot interferometer geometries: as shown in panels (a1)–(a4), respectively. The second row (panels b1–b4) illustrates the transmission probability as a function of energy offset for different values of level detuning $\Delta$. As $\Delta$ increases, the central transmission peak becomes progressively sharper while the two side peaks shift outward and decrease in amplitude. This behavior signifies a suppression of the superradiant channels and the emergence of narrow subradiant modes—states that are weakly coupled to the reservoirs due to destructive quantum interference. These subradiant modes result in narrow resonances in transmission, which enhance the energy selectivity of transport, a critical condition for achieving a high thermoelectric figure of merit. In contrast, broader peaks correspond to superradiant modes, where enhanced coupling to the leads leads to increased thermal conductance, thereby lowering \( ZT \). Notably, in the row (b2)–(b4), we also observe asymmetric Fano-like resonances, arising from interference between discrete subradiant levels and broad superradiant backgrounds, which further modulate the transmission function and enhance thermoelectric efficiency. The third row (panels c1–c4) maps the calculated $ZT$ as a function of the quantum dot level energy $\epsilon$ and temperature $T$. These plots reveal optimal regions in the $(\epsilon, T)$ plane where $ZT$ is maximized, aligning with the resonant features of the transmission in panels (b1)–(b4). The emergence of sharp features in transmission directly correlates with the localization of high-$ZT$ zones in these maps. The bottom row (panels d1–d4) shows the dependence of $ZT$ on temperature for each system and various values of $\Delta$. A non-monotonic behavior is observed: $ZT$ increases with $T$, reaches a maximum at a characteristic temperature scale ($T \sim t$), and subsequently decreases due to thermal broadening. In particular, panel (d4) for the 6QD(3,3) geometry shows a remarkably high peak of $ZT \approx 30$ at a dilution temperature of $T \approx 4.8 \text{mK}$  (or $T \approx \gamma/10$). This peak corresponds to the emergence of highly selective subradiant modes, as identified in the corresponding transmission profile (panel b4). With increasing temperature, these sharp transmission features are thermally smeared, resulting in a transition to superradiant-dominated transport and a consequent reduction in thermoelectric efficiency.
\begin{table}
    \centering
    \begin{tabular}{|c|c|c|c|}
    \hline
    Configuration and phase & Regime & $\Delta$  & $ZT_{max}$\\
    \hline
  
      \multirow{1}{*}{4QD(2,2), $\phi=2.9\pi$}
     & $t/\gamma=0.8$ & $0$ & 6.74\\
    \hline
    \multirow{6}{*}{4QD(2,2), $\phi=\pi$}
    & \multirow{6}{*}{$t/\gamma=1$ }& $0.5$ &$14.1$\\
    \cline{3-3} \cline{4-4}
     &  & $0.6$ &$15.7$\\
    \cline{3-3} \cline{4-4}
      &  & $0.7$ &$16.2$\\
    \cline{3-3} \cline{4-4}
     &  &$0.8$ & $17.25$\\
     \cline{3-3} \cline{4-4}
      &  &$0.9$ &$17.68$\\
      \hline \hline
       \multirow{1}{*}{4QD(3,1), $\phi=2.83\pi$}
     & $t/\gamma=0.8$ & $0$ & 6.07\\
    \hline
    \multirow{6}{*}{4QD(3,1), $\phi=\pi$}
    & \multirow{6}{*}{$t/\gamma=1$ }& $0.5$ &$13.84$\\
    \cline{3-3} \cline{4-4}
     &  & $0.6$ &$15.45$\\
    \cline{3-3} \cline{4-4}
      &  & $0.7$ &$16.44$\\
    \cline{3-3} \cline{4-4}
     &  &$0.8$ & $17.1$\\
     \cline{3-3} \cline{4-4}
      &  &$0.9$ &$17.57$\\
      \hline \hline
       \multirow{1}{*}{5QD(3,2), $\phi=3.67\pi$}
     & $t/\gamma=0.8$ & $0$ & 4.79\\
    \hline
    \multirow{6}{*}{5QD(3,2), $\phi=\pi$}
    & \multirow{6}{*}{$t/\gamma=1$ }& $0.5$ &$4.57$\\
    \cline{3-3} \cline{4-4}
     &  & $0.6$ &$4.73$\\
    \cline{3-3} \cline{4-4}
      &  & $0.7$ &$5.08$\\
    \cline{3-3} \cline{4-4}
     &  &$0.8$ & $5.3$\\
     \cline{3-3} \cline{4-4}
      &  &$0.9$ &$5.5$\\
      \hline \hline
       \multirow{1}{*}{6QD(3,3), $\phi=1.94\pi$}
     & $t/\gamma=0.8$ & $0$ & 5.64\\
    \hline
    \multirow{6}{*}{6QD(3,3), $\phi=\pi$}
    & \multirow{6}{*}{$t/\gamma=1$ }& $0.5$ &$15.64$\\
    \cline{3-3} \cline{4-4}
     &  & $0.6$ &$19.13$\\
    \cline{3-3} \cline{4-4}
      &  & $0.7$ &$22.17$\\
    \cline{3-3} \cline{4-4}
     &  &$0.8$ & $24.71$\\
     \cline{3-3} \cline{4-4}
      &  &$0.9$ &$27.07$\\
      \hline 
    
    \end{tabular}
   \caption{Calculated maximum thermoelectric figure of merit, $ZT_{\max}$, for different quantum dot configurations, showing the influence of coupling regime ($t/\gamma$), level shift $\Delta$, and magnetic phase $\phi$.}
    \label{ZT_table}
\end{table}

In figure~\ref{Lorentz ratio results}, panels (b1)–(b4) show the figure of merit \( ZT \) as a function of dot energy level \( \epsilon \), computed at the temperature that yields maximum \( ZT \) for each interferometer geometry (identified earlier in Fig.~\ref{ZT vs T 4QD(2,2)}(d1)–(d4)). As observed, the peak value of \( ZT \) is maximized when the inter-dot tunneling strength \( t \) is on the order of the coupling strength \( \gamma \), i.e., \( t \sim \gamma \), where quantum interference effects are optimally tuned. In particular, for the 6QD(3,3) system [panel (b4)], \( ZT \) exhibits extremely sharp peaks centered around specific values of \( \epsilon \), consistent with the narrow subradiant transmission features observed in Fig.~\ref{ZT vs T 4QD(2,2)}(b4). These sharp peaks indicate that charge carriers contributing to transport are strongly energy-filtered, which enhances thermoelectric efficiency. Similarly, the 4QD(2,2) and 4QD(3,1) systems [panels (b1) and (b2)] show central peaks that become more pronounced for intermediate tunneling strengths, further confirming the importance of subradiant modes. At higher tunneling strengths (\( t > \gamma \)), however, we observe broader peaks in the \( ZT \) profile. This broadening is associated with the emergence of superradiant modes, which enhance the thermal conductance disproportionately and degrade \( ZT \).\\
\indent
To probe the nature of thermal and electrical transport in these regimes, panels (c1)–(c4) present the Lorenz ratio \( L/L_0 \), where \( L_0 = \pi^2/3 \) is the universal Lorenz number. In all configurations, the Lorenz ratio significantly deviates from \( L_0 \) near the energy values where \( ZT \) is maximized. This strong violation of the WF law arises from quantum coherence and interference, particularly where subradiant modes dominate and thermal transport is suppressed more than charge transport. Panels (d1)–(d4) show the variation of \( ZT \) as a function of normalized tunneling strength \( t/\gamma \), evaluated at the optimal temperature for each configuration (from Fig.~\ref{ZT vs T 4QD(2,2)}(d1)–(d4)). In all cases, \( ZT \) exhibits a clear peak around \( t \sim \gamma \), reinforcing that this is the regime where destructive interference suppresses thermal conductance most effectively while preserving sharp energy-dependent charge transport.\\
\indent
Figure~\ref{Fig7}(b1)-(b4) presents the variation of the \( ZT \) as a function of \( t/\gamma \) and \( \phi/\pi \). Several sharp features and peaks are observed, with \( ZT \) values ranging between 6 and 10. In Fig.~\ref{Fig7}(b1) and (b2), a significant figure of merit is still obtained, which can be attributed to the step-like structure followed by a sharp peak in the transmission function shown in Fig.~\ref{Fig7}(c1) and (c2). In contrast, Figs.~\ref{Fig7}(c3) and (c4) display a sharp dip followed by a broad peak in the transmission profile. This combination of subradiant (sharp) and superradiant (broad) features enhances the energy filtering capability, leading to an appreciable \( ZT \) as seen in Figs.~\ref{Fig7}(b3) and (b4).

\subsection{Non linear regime}\label{sub: non linear}

\begin{figure*}[]
\centering
     \setlength{\tabcolsep}{2pt} 
    \renewcommand{\arraystretch}{1.0} 

    \begin{tabular}{cccc}
    \includegraphics[scale=0.21]{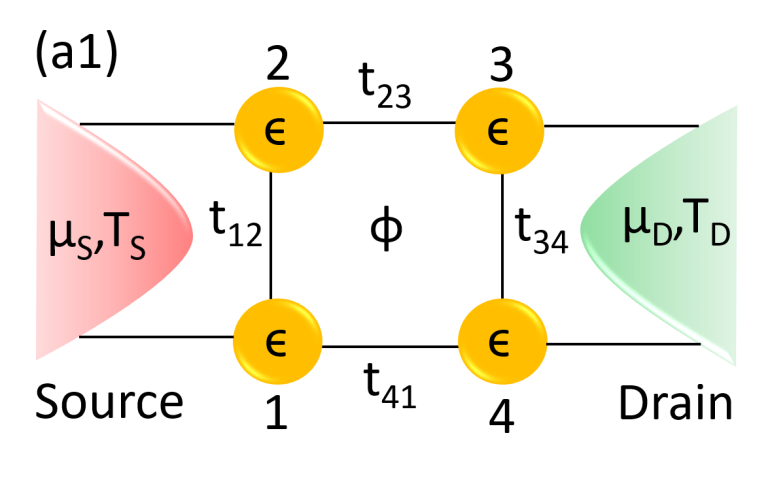}&
    \includegraphics[scale=0.21]{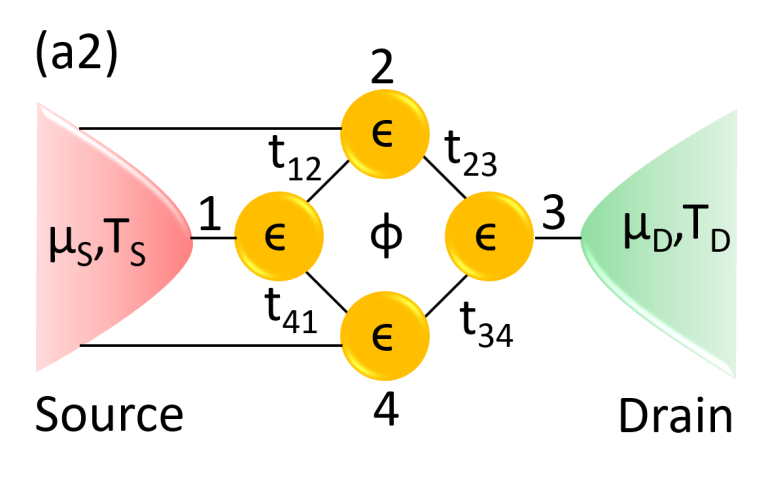}&
    \includegraphics[scale=0.21]{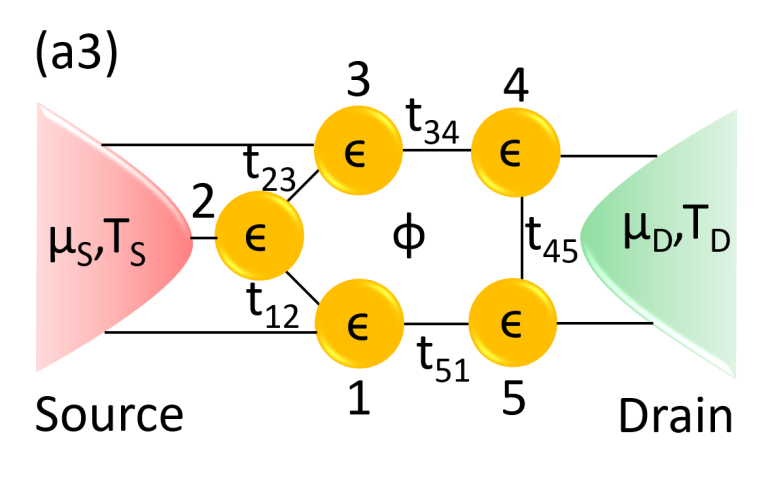}&
    \includegraphics[scale=0.21]{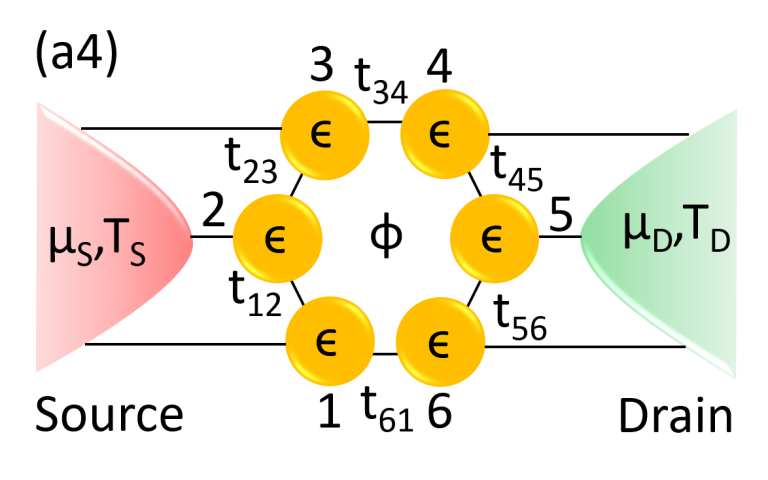}\\[-1ex]

    \includegraphics[scale=0.21]{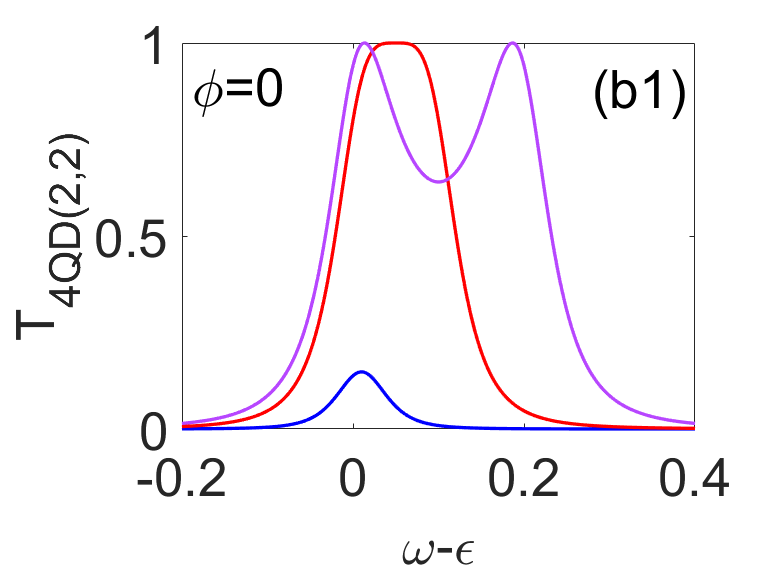}&
    \includegraphics[scale=0.21]{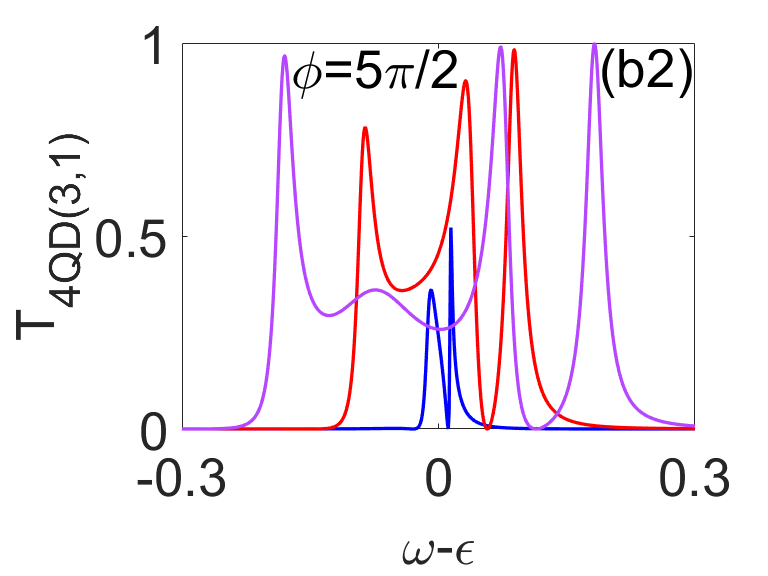}&
    \includegraphics[scale=0.21]{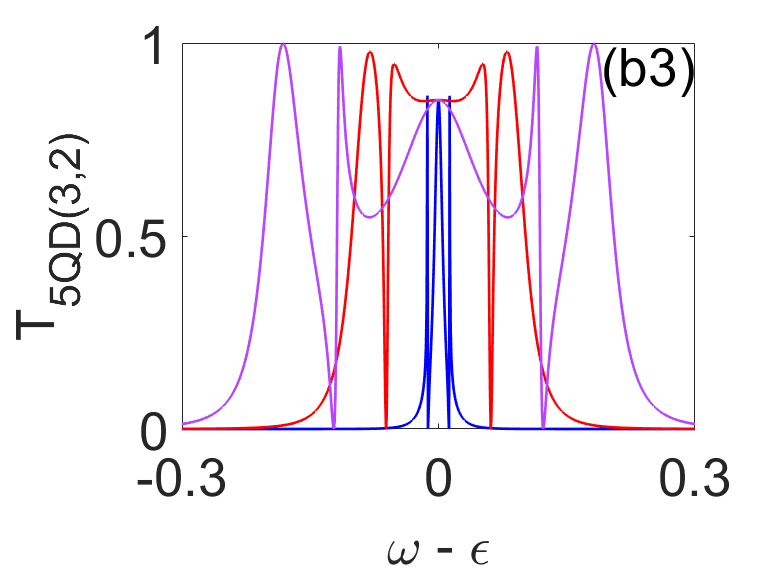}&
    \includegraphics[scale=0.21]{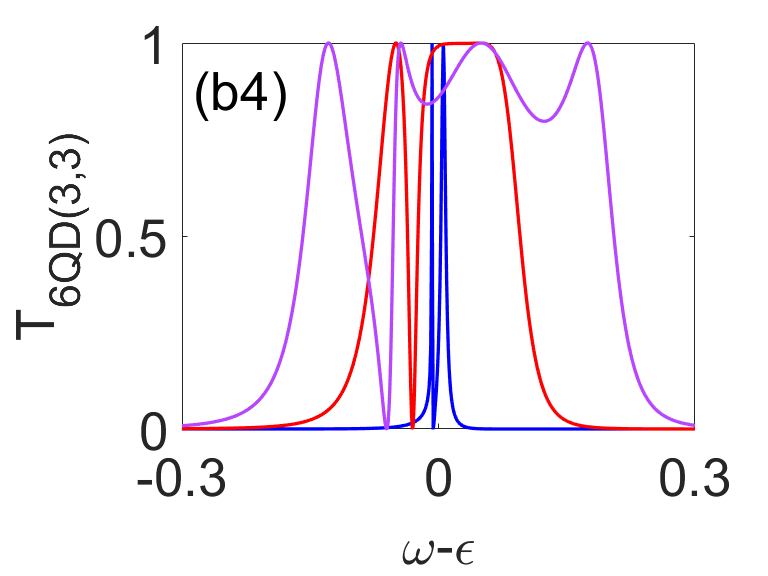}\\[-1ex]
    
    \includegraphics[scale=0.21]{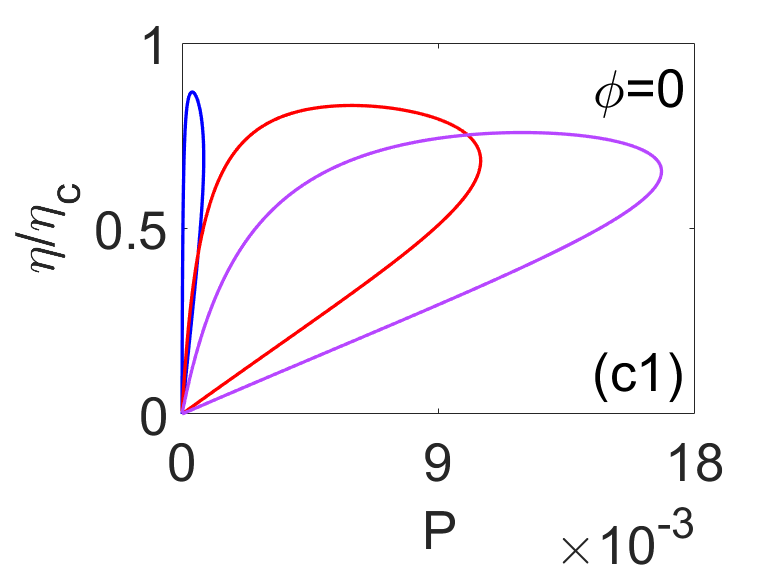}&
    \includegraphics[scale=0.21]{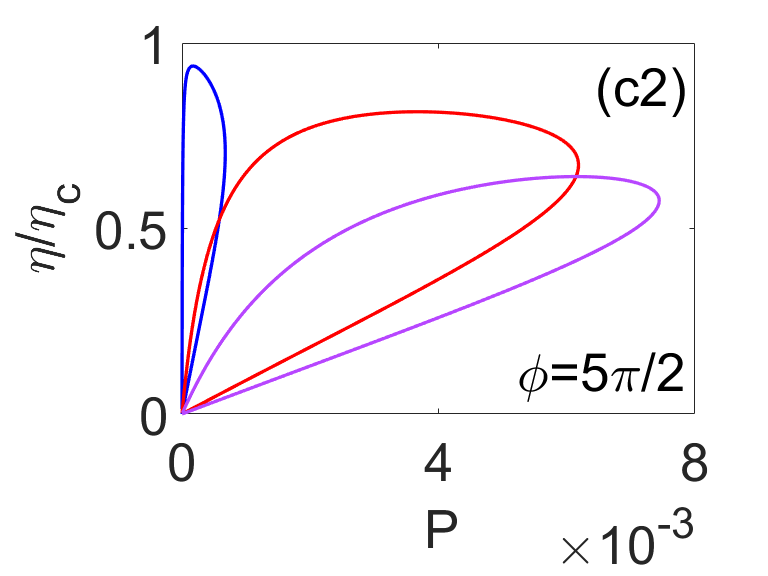}&
    \includegraphics[scale=0.21]{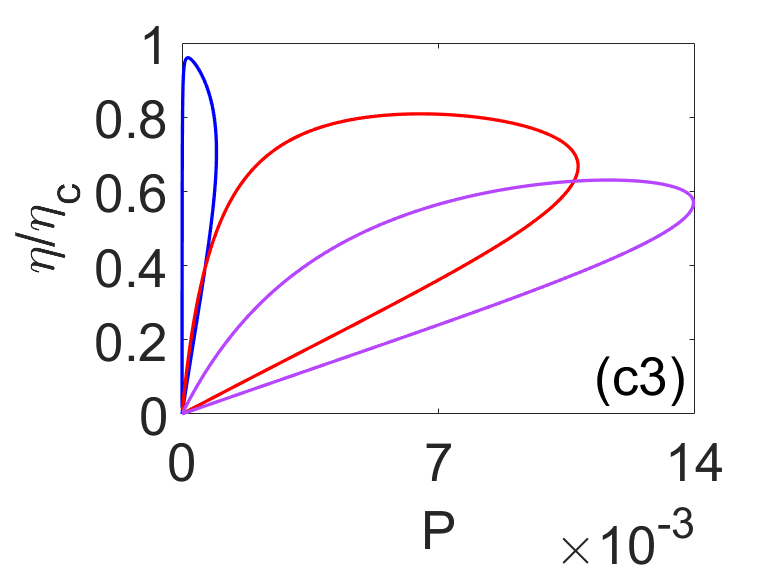}&
    \includegraphics[scale=0.21]{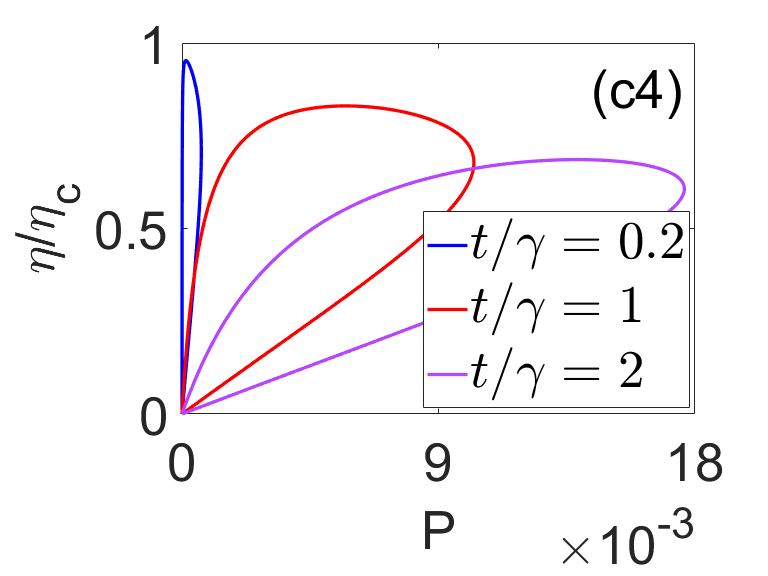}\\[-1ex]

    \includegraphics[scale=0.21]{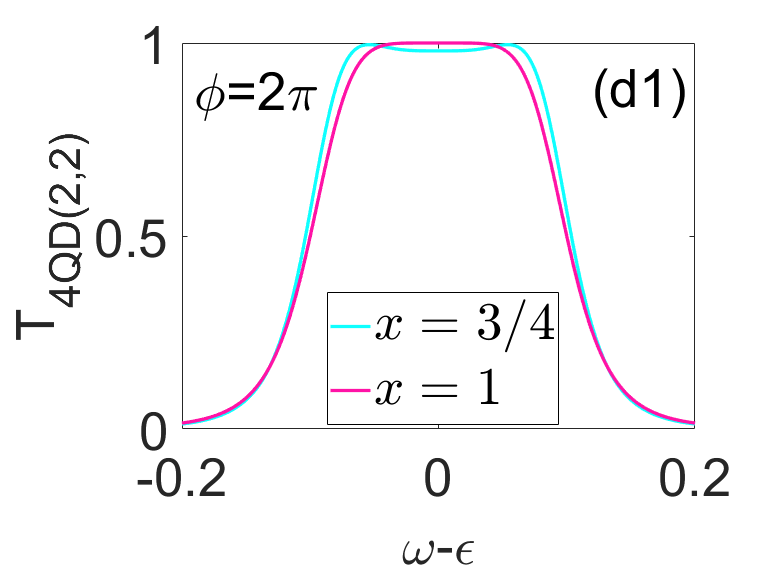}&
    \includegraphics[scale=0.21]{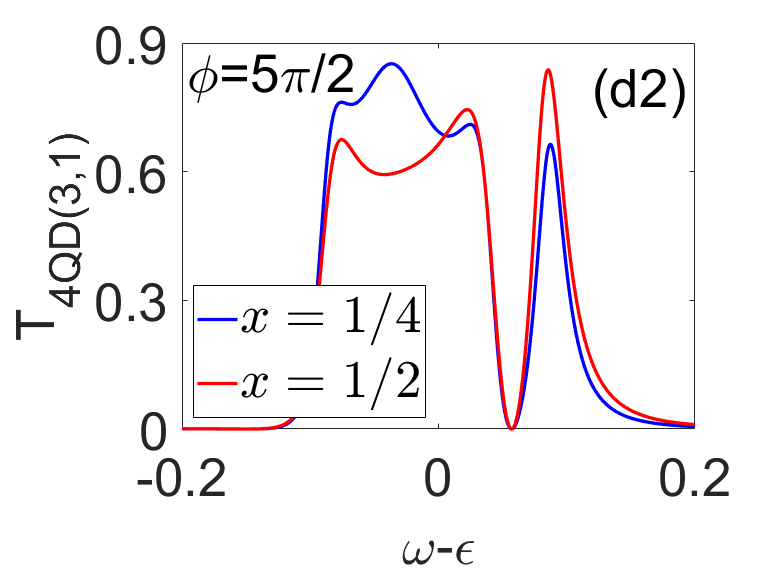}&
    \includegraphics[scale=0.21]{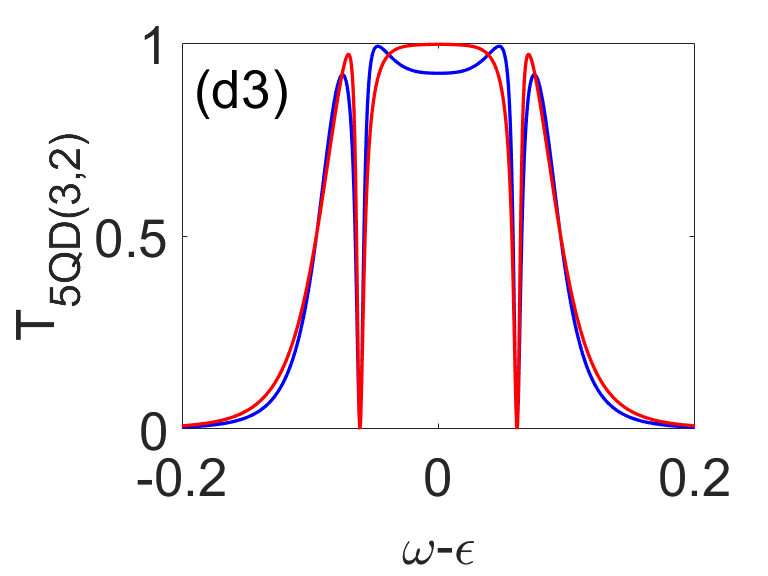}&
    \includegraphics[scale=0.21]{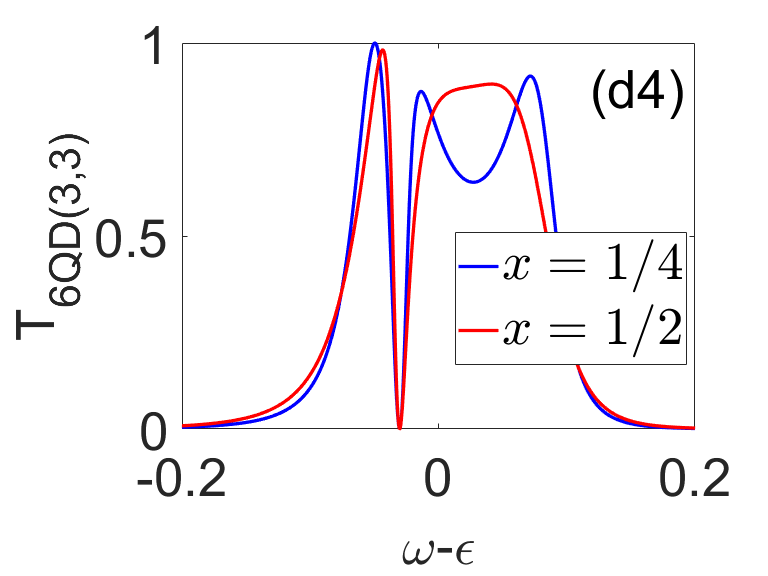}\\[-1ex]
    
    \includegraphics[scale=0.21]{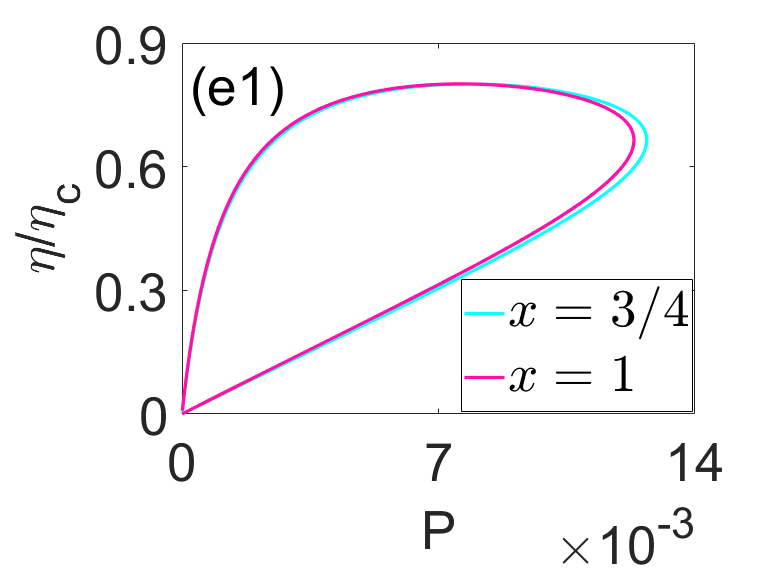}&
    \includegraphics[scale=0.21]{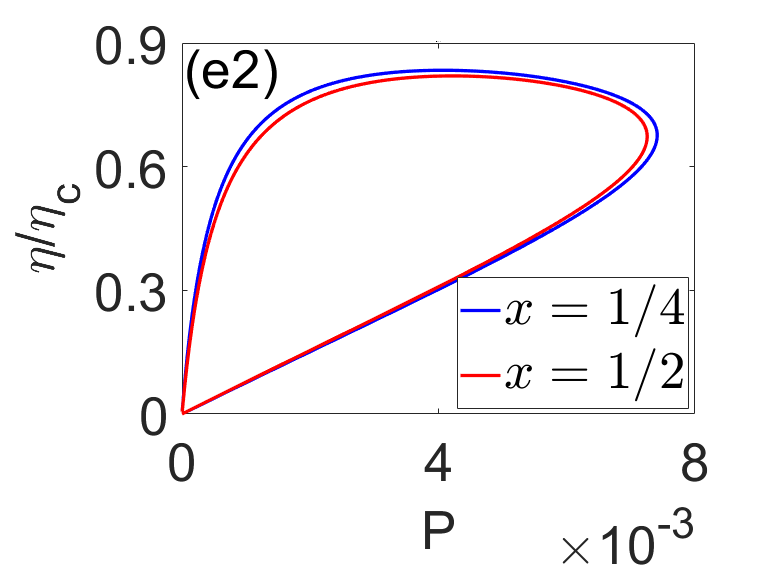}&
    \includegraphics[scale=0.21]{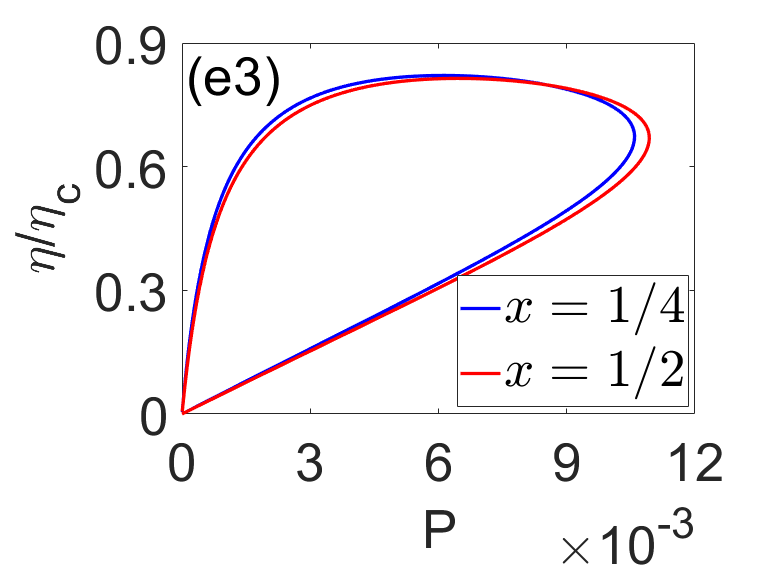}&
    \includegraphics[scale=0.21]{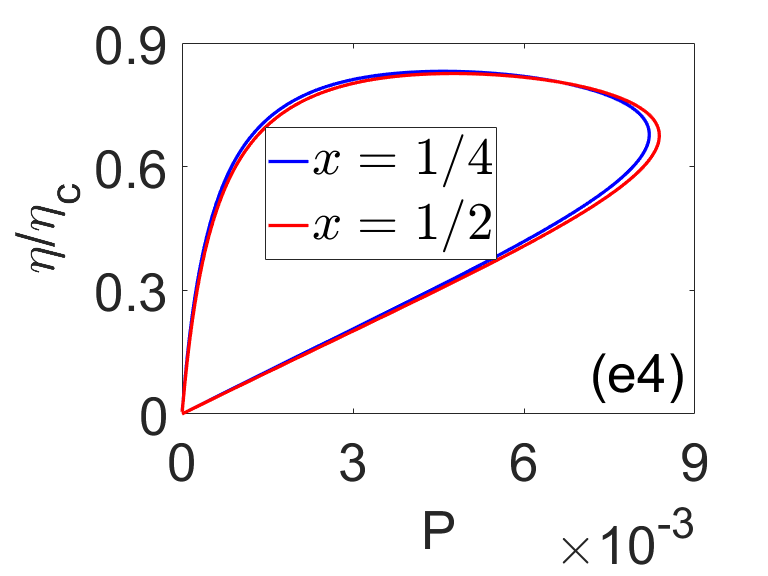}\\
     \end{tabular}
    \caption{(a1)-(a4) Configurations used in our study; (b1)-(b4) shows transmission as a function of energy at three different regimes $t/\gamma=0.2$, $t/\gamma=1$, and $t/\gamma=2$. Third panel (c1)-(c4) presents the power efficiency trade-off. fourth panel (d1)-(d4) illustrate the transmission for asymmetric coupling here $x=\gamma_S/\gamma_D$ denotes the asymmetric coupling strength with $\gamma_D=0.1$. Fifth panel (e1)-(e4) represents the corresponding power efficiency trade-off. parameters used: $\gamma=0.05,\epsilon=8\gamma,T_{S}=12\gamma,T_{D}=2\gamma,\mu_S=-\mu_D$, and phase $\phi=0$ for (b1)-(e1), $\phi=5\pi/2$ for (b2)-(e2), $\phi=5\pi/2$ for (b3)-(e4)  }
    \label{Fig4}
\end{figure*}

\begin{figure*}
    \centering
    \includegraphics[scale=0.21]{Fig_a1.png}
    \includegraphics[scale=0.21]{Fig_a2.png}
    \includegraphics[scale=0.21]{Fig_a3.png}
    \includegraphics[scale=0.21]{Fig_a4.png}
    
    \includegraphics[scale=0.21]{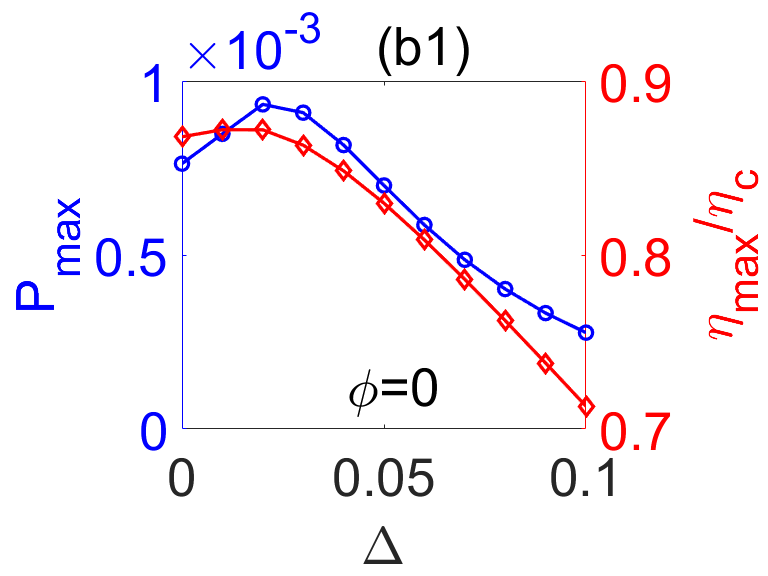}
     \includegraphics[scale=0.21]{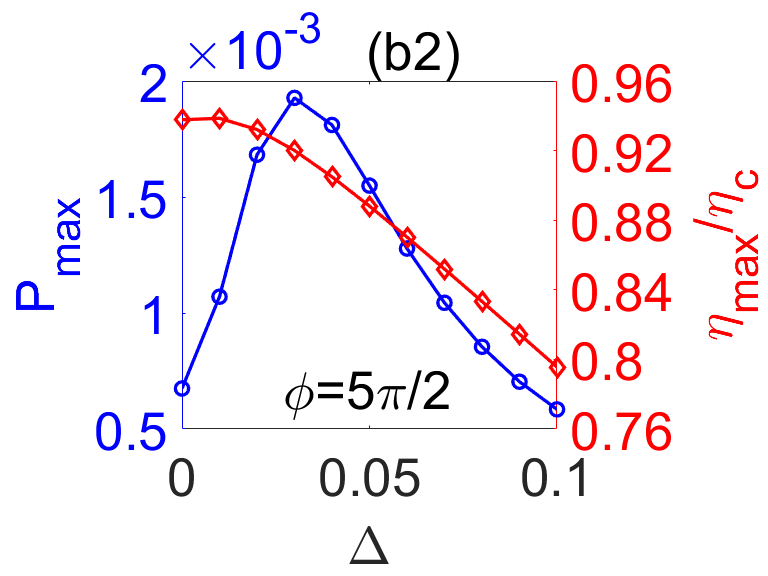}
     \includegraphics[scale=0.21]{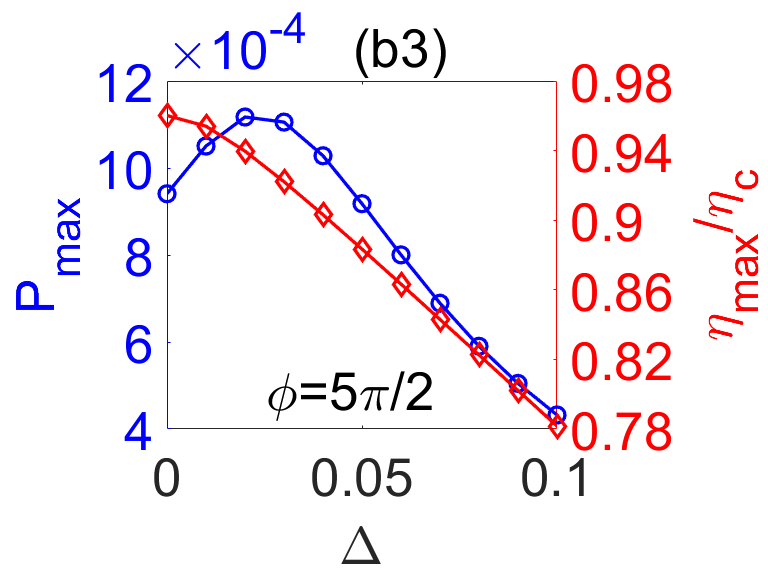}
     \includegraphics[scale=0.21]{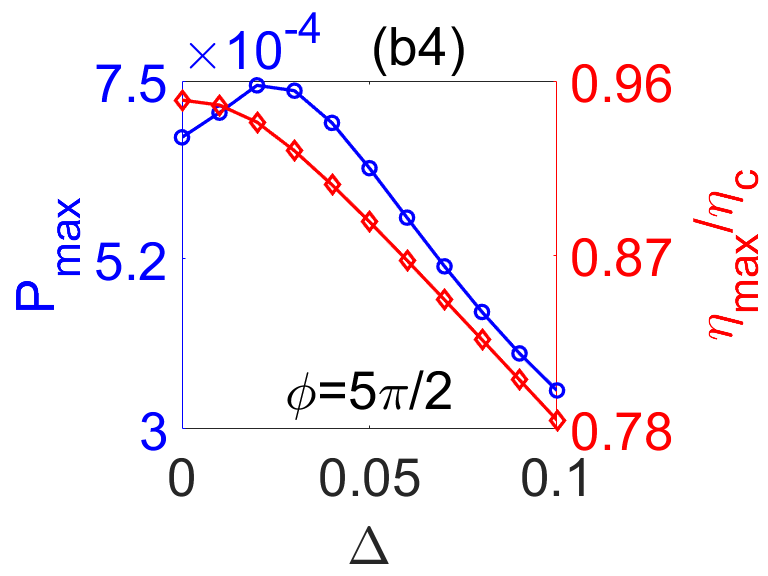}
    
     \includegraphics[scale=0.21]{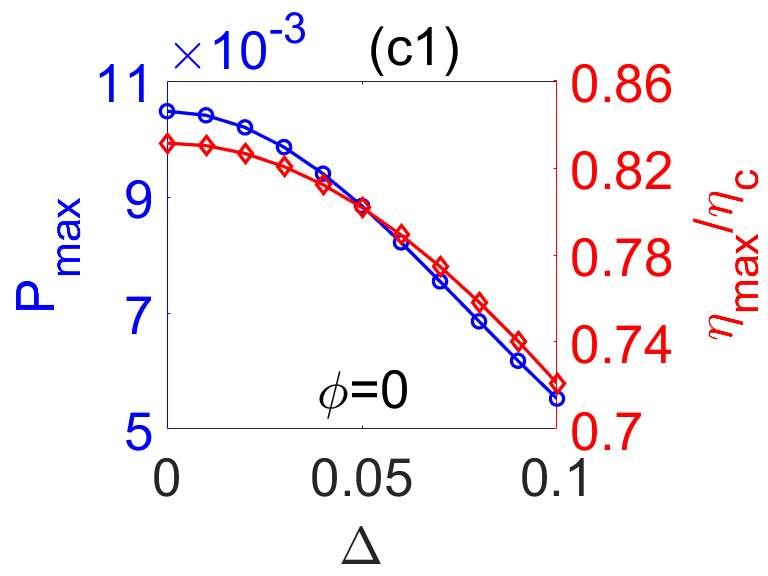}
     \includegraphics[scale=0.21]{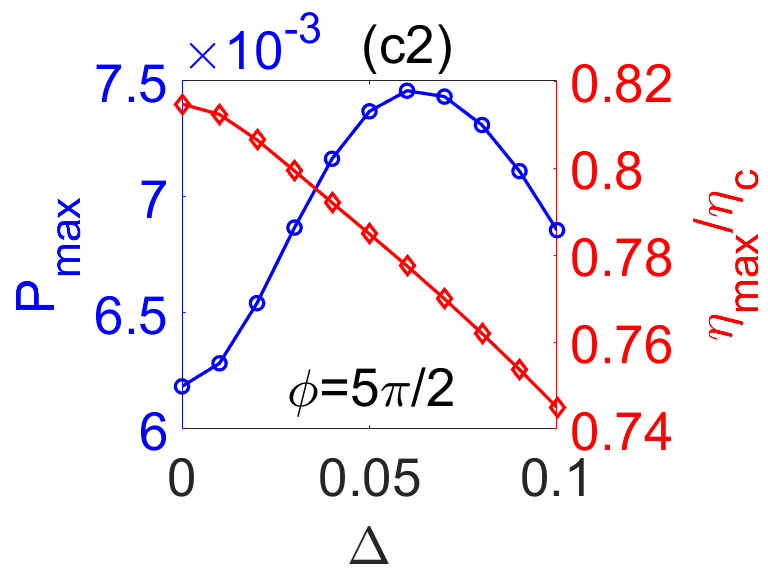}
     \includegraphics[scale=0.21]{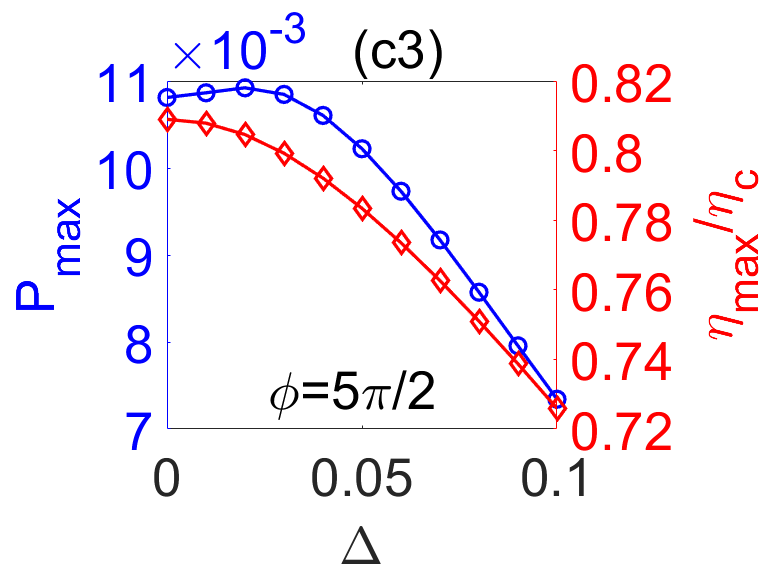}
     \includegraphics[scale=0.21]{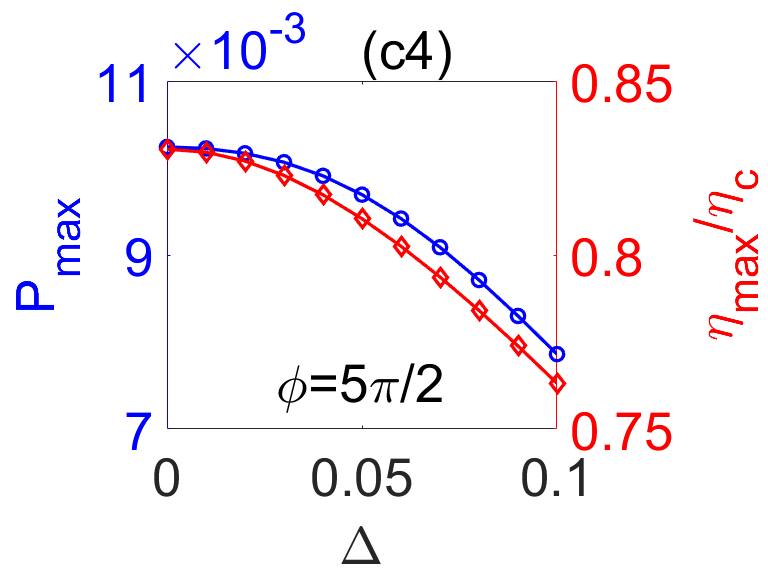}
    
      \includegraphics[scale=0.21]{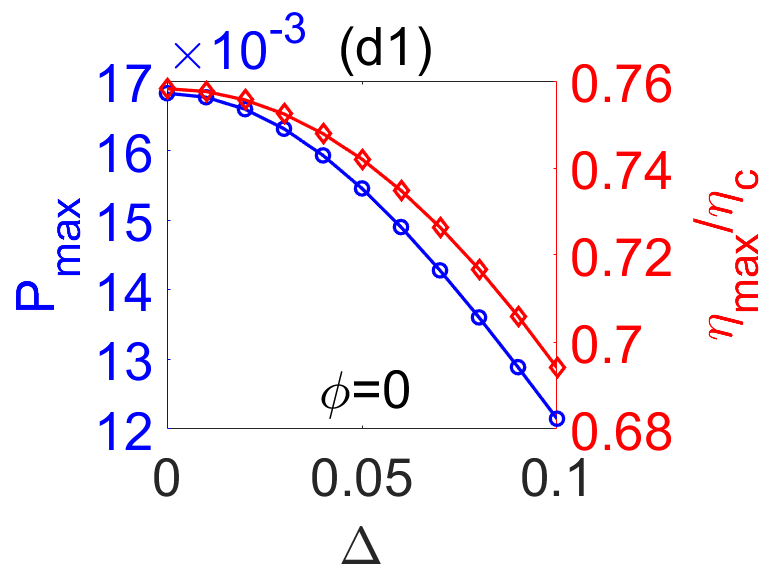}
      \includegraphics[scale=0.21]{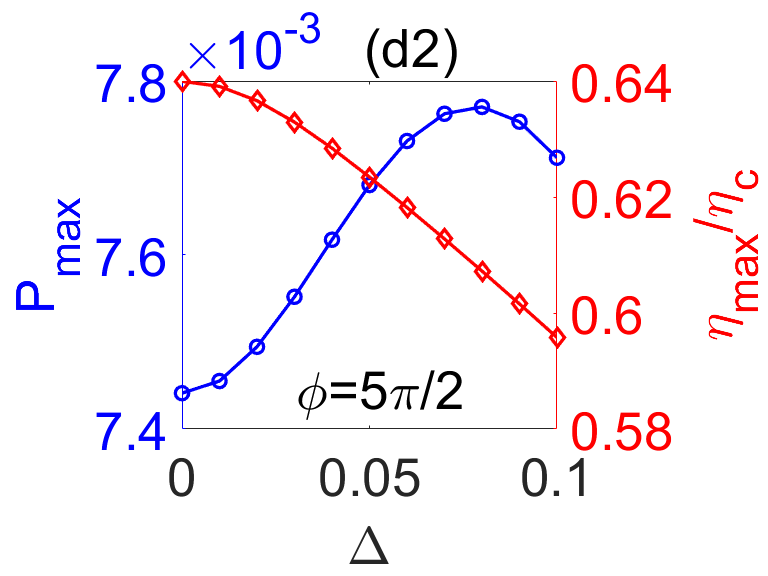}
      \includegraphics[scale=0.21]{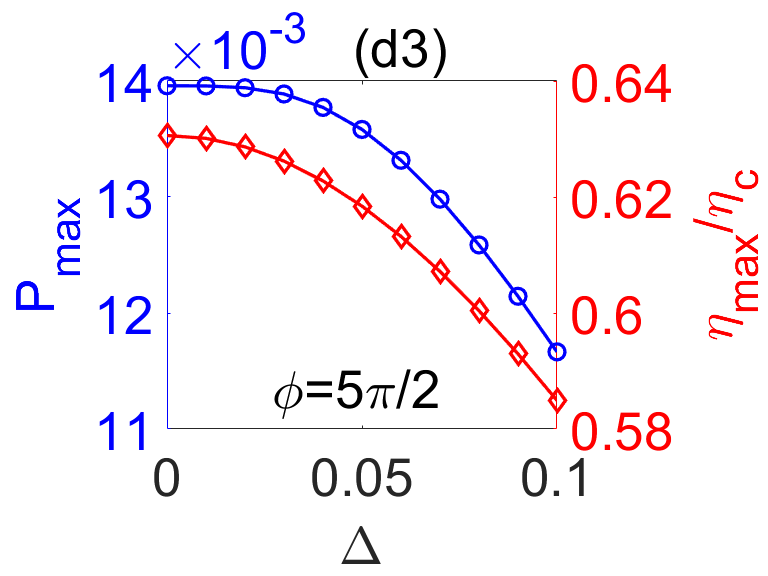}
     \includegraphics[scale=0.21]{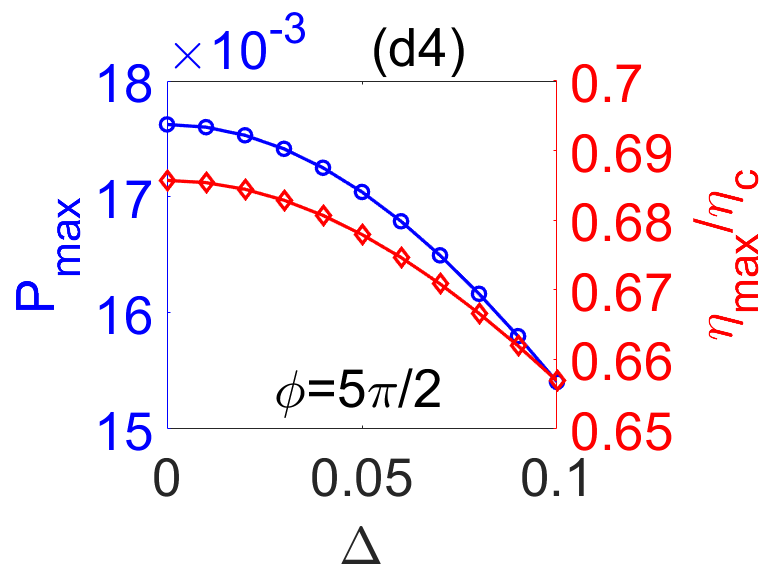}
     \caption{(a1)-(a4) manifest configurations used for the study of maximum output power \( P_{\text{max}} \) and normalized maximum efficiency \( \eta/\eta_{\text{max}} \) as functions of level shift \( \Delta \), shown for three tunneling regimes: Second panel (b1-b4) \( t/\gamma = 0.2 \), Third panel (c1-c4) \( t/\gamma = 1 \), and fourth panel (d1-d4) \( t/\gamma = 2 \). Each column corresponds to a different quantum dot configuration: 4QD(2,2) (first column), 4QD(3,1) (second column), 5QD(3,2) (third column), and 6QD(3,3) (fourth column). The parameters used are \( \gamma = 0.05 \), \( \epsilon = 8\gamma \), \( T_S = 12\gamma \), \( T_D = 2\gamma \), and \( \mu_S = -\mu_D \).}
\label{P_max vs eta_max}
\end{figure*}
To operate the device as a thermoelectric heat engine, a thermal gradient is imposed by maintaining \( T_S > T_D \), and a chemical bias is applied such that \( \mu_S < \mu_D \). These conditions drive the system into the nonlinear transport regime, where the relevant control parameters are the normalized temperature and chemical potential gradients, defined as \( \Delta T/T = (T_S - T_D)/(T_S + T_D) \) and \( \Delta \mu/\mu = (\mu_D - \mu_S)/(\mu_D + \mu_S) \), respectively.

In this regime, we analyze the power-efficiency trade-off for three distinct tunneling regimes: \( t/\gamma < 1 \), \( t \sim \gamma \), and \( t/\gamma > 1 \), where \( t \) is the inter-dot tunneling amplitude and \( \gamma \) is the dot–lead coupling strength. The quantum dot geometries under consideration are shown in Fig.~\ref{Fig4}(a1)–(a4), and the corresponding transmission functions are plotted in panels (b1)–(b4).

In the weak tunneling regime (\( t/\gamma < 1 \)), the transmission line shapes are typically narrow and Lorentzian, with features resembling asymmetric Fano resonances. These sharp peaks filter electrons very selectively, leading to high thermodynamic efficiency but limited power output due to suppressed particle currents. This behavior is seen in Fig.~\ref{Fig4}(b1)-(b4), where the blue curves yield efficiencies exceeding 87\% of the Carnot limit, albeit at low power, as reported in Table~\ref{table_3}.

In the intermediate tunneling regime (\( t \sim \gamma \)), the transmission functions become broader and boxcar-shaped, or a mixture of boxcar and Lorentzian features, depending on the configuration. These profiles are ideal for maximizing power output, as they allow a wider energy window for electron transport without compromising selectivity. This is observed for the 4QD(2,2), 5QD(3,2), and 6QD(3,3) setups, shown in Fig.~\ref{Fig4}(b1), (b2), and (b4), which yield power outputs of approximately 3 fW while maintaining high efficiencies up to 83\% of the Carnot efficiency.

Subradiant modes are more pronounced in configurations where multiple dots are weakly coupled to a common reservoir, leading to multiple narrow peaks in the transmission function, as seen in Fig.~\ref{Fig4}(b2) - (b4). These resonances enhance energy selectivity and improve efficiency, though they somewhat reduce the total power due to lower current flow.

In the strong tunneling regime (\( t/\gamma > 1 \)), the internal dynamics dominate over coupling to the reservoirs. Electrons traverse the quantum dot loop rapidly compared to their escape into the leads, allowing higher energy carriers to contribute significantly to transport. This results in broader transmission features and multiple overlapping peaks, as shown by the purple curves in Fig.~\ref{Fig4}(b1) - (b4). Such broad profiles are favorable for generating large particle currents, resulting in enhanced power output with moderately high efficiency.

\medskip

We also explore the power-efficiency characteristics under asymmetric coupling conditions, where the dot–lead coupling strengths differ: \( \gamma_S \neq \gamma_D \), with asymmetry parameter \( x = \gamma_S/\gamma_D \). Figures~\ref{Fig4}(d1)–(d4) show the transmission functions in these asymmetric configurations, and the corresponding power-efficiency trade-offs are presented in panels (e1)–(e4).

Interestingly, several configurations yield both high efficiency and substantial power output. This occurs when the transmission functions display a combination of sharp and broad peaks, effectively capturing the benefits of both subradiant and superradiant transport modes. For instance, Fig.~\ref{Fig4}(d1) demonstrates a boxcar-like transmission profile for the 4QD(2,2) configuration, resulting in a power output of 3.6 fW and efficiency reaching 80\% of the Carnot limit (see Table~\ref{table_3}). Fig.~\ref{Fig4}(d3) shows a combination of two Lorentzian and boxcar-like transmission profile for the 5QD(3,2) configuration, leads to power output of 3 fW and efficiency reaching 82\% of the Carnot limit Similar performance is observed in Figs.~\ref{Fig4}(d2) and (d4), where asymmetric line shapes enable favorable energy filtering and strong current flow.
\begin{table*}
    \centering
    \begin{tabular}{|c|c|c|c|c|}
    \hline
     \multicolumn{2}{|c|}{Fano Parameter} & $P_{max}$(fW) & $\eta_{max}/\eta_c$ & Transmission behaviour\\
     \hline
     \multicolumn{2}{|c|}{$q=7$} & 1.29& 27.33& Fano Fig.\ref{Fig1}(b1)\\
     \hline
    \multicolumn{2}{|c|}{$q=9$} & 1.375& 38.56& Fano Fig.\ref{Fig1}(b1)\\
    \hline
    \multicolumn{2}{|c|}{$q=200$} & 1.25&73.1 & Lorentzian Fig.\ref{Fig1}(b1)\\
    \hline  
    \hline 
Configuration and Phase & Regimes  &  $P_{max}$(fW) &  $\eta_{max}/\eta_c$& Transmission behaviour \\

    \hline \hline
    \multirow{4}{*}{4QD(2,2) , $\phi=0$}  &$t/\gamma<1$   & 0.21 & 87 & Single peak Fig.\ref{Fig4}(b2)  \\
    \cline{2-5}
   &$t\sim\gamma$  & 3 & 83.2 & Boxcar Fig.\ref{Fig4}(b2)\\
   \cline{2-5}
   & $t/\gamma>1$  & 4.74 & 76 & Broad two peaks Fig.\ref{Fig4}(b2)\\
   \hline\hline
   \multirow{1}{*}{4QD(2,2),$\phi=2\pi$} & $\gamma_S\neq\gamma_D$ & 3.6 & 80 & Boxcar Fig.\ref{Fig4}(d2)\\
   
    \hline \hline
     \multirow{5}{*}{4QD(3,1) , $\phi=5\pi/2$}  & $t/\gamma<1$  & 0.19 & 93.8 & Two peaks Fig.\ref{Fig4}(b3) \\
    \cline{2-5}
    & $t\sim\gamma$  & 1.76 & 81.5 & Three peaks Fig.\ref{Fig4}(b3)\\
    \cline{2-5}
   & $t/\gamma>1$  & 2.12 & 64 &  Four peaks Fig.\ref{Fig4}(b3)\\
   \cline{2-5}
   &$\gamma_S\neq\gamma_D$ &2.16 & 83.4 & Boxcar and Lorentzian Fig.\ref{Fig4}(d3)\\
    \hline \hline
     \multirow{5}{*}{5QD(3,2) , $\phi=5\pi/2$}  & $t/\gamma<1$  & 0.27 & 96.02 & Three sharp peaks Fig.\ref{Fig4}(b3) \\
    \cline{2-5}
    & $t\sim\gamma$  & 3.04 & 81 & Boxcar and Lorentzian Fig.\ref{Fig4}(b3)\\
    \cline{2-5}
   & $t/\gamma>1$  & 3.97 & 63.1 &  Five peaks Fig.\ref{Fig4}(b3)\\
   \cline{2-5}
   &$\gamma_S\neq\gamma_D$ &3 & 82.1 & Boxcar and Lorentzian Fig.\ref{Fig4}(d3)\\
    \hline \hline
   \multirow{5}{*}{6QD(3,3) , $\phi=5\pi/2$}  & $t/\gamma<1$ & 0.19 & 95.3 & Two Lorentzian peaks Fig.\ref{Fig4}(b4) \\
    \cline{2-5}
   &$t\sim\gamma$  & 2.88 & 83 & Boxcar and Lorentzian Fig.\ref{Fig4}(b4) \\
    \cline{2-5}
   &$t/\gamma>1$ & 4.96 & 68.5 & Broad Multiple peaks Fig.\ref{Fig4}(b4)\\
   \cline{2-5}
   &$\gamma_S\neq\gamma_D$ &2.35 & 82.6 & Boxcar and Lorentzian Fig.\ref{Fig4}(d4)\\
    \hline 
\end{tabular}
    \caption{ Maximum output power $P_{max}$ and maximum efficiency $\eta_{max}$ and their transmission behavior for different geometries at different regimes.}
    \label{table_3}
\end{table*}

Figure~\ref{P_max vs eta_max} illustrates the dependence of the maximum output power \( P_{\text{max}} \) and the corresponding maximum efficiency \( \eta_{\text{max}}/\eta_C \) on the energy level shift \( \Delta \), across three different tunneling regimes: \( t/\gamma = 0.2 \), \( t/\gamma = 1 \), and \( t/\gamma = 2 \). Panels (a1)–(a4) show the AB interferometer configurations used in our study:  4QD(2,2), 4QD(3,1), 5QD(3,2), and 6QD(3,3), while the remaining panels present the corresponding power and efficiency results for each configuration and regime.

For the  $4QD(3,1)$ geometry [see the column (b2)–(d2)], the output power increases with increasing \( \Delta \) up to an optimal value—around \( \Delta = 0.04\text{--}0.08 \), depending on the regime—beyond which it decreases. In all cases, this enhancement in power comes at the cost of reduced thermodynamic efficiency, which gradually decreases with increasing \( \Delta \). This trade-off reflects the broadening of the transmission function and the activation of less selective transport channels as energy levels move farther from resonance.

For the $4QD(2,2)$, $5QD(3,2)$, and $6QD(3,3)$ setups [see columns (b1)-(d1), (b3)–(d3), and (b4)–(d4),respectively], the behavior is more nuanced. In the strong coupling regime (\( t/\gamma = 0.2 \)), the output power shows a non-monotonic dependence on \( \Delta \), reaching a peak before falling off. This indicates that while a moderate level shift can enhance asymmetry and boost thermoelectric response, excessive splitting reduces the overlap of transmission peaks with the transport window, diminishing both current and power.

In the intermediate and weak coupling regimes (\( t/\gamma = 1 \) and \( t/\gamma = 2 \)), the output power for these configurations generally decreases as \( \Delta \) increases. This trend suggests that at higher tunneling strengths, the system becomes more sensitive to level detuning, where strong interdot hybridization spreads out transmission features, making energy filtering less efficient.  

\subsection{Scaling}\label{sub: Scaling}

\begin{figure*}
    \centering
    \includegraphics[scale=0.25]{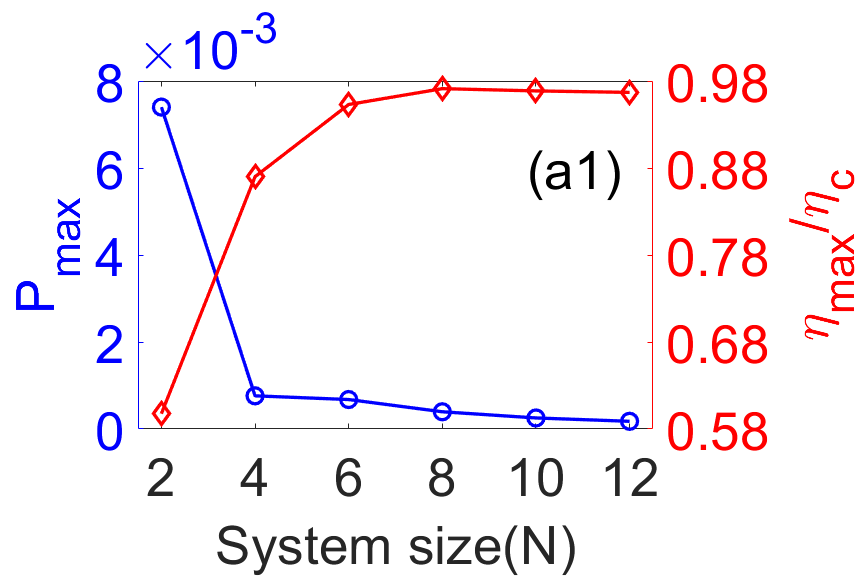}
    \includegraphics[scale=0.25]{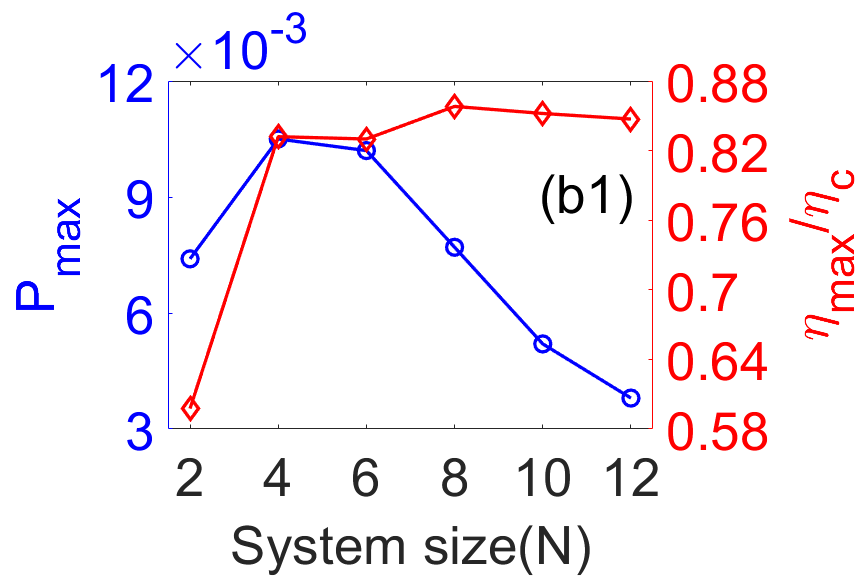}
    \includegraphics[scale=0.25]{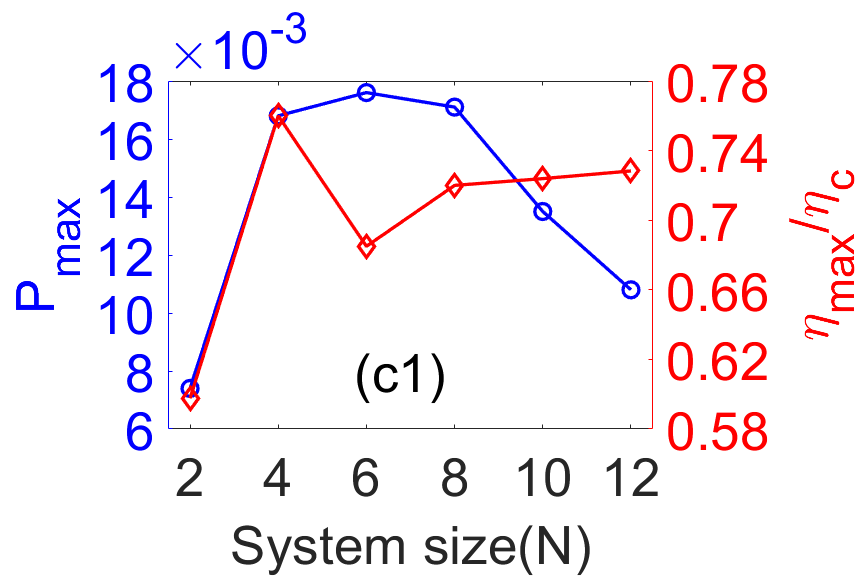}

    \caption{
Maximum output power ($P_{\max}$) and normalized maximum efficiency ($\eta_{\max}/\eta_c$) as functions of system size $N$, where $N$ denotes the total number of quantum dots symmetrically coupled to the source and drain (e.g., $N=4$ corresponds to 4QD(2,2), $N=6$ to 6QD(3,3), etc.). Results are shown for three coupling regimes: (a1) $t/\gamma = 0.2$, (b1) $t/\gamma = 1$, and (c1) $t/\gamma = 2$.}
 
    \label{Fig6}
\end{figure*}
Figure~\ref{Fig6}(a1)--(c1) shows the variation of maximum output power \( P_{\mathrm{max}} \) (left axis) and normalized maximum efficiency \( \eta_{\mathrm{max}}/\eta_c \) (right axis) as a function of system size \( N \), where \( N \) denotes the total number of quantum dots symmetrically coupled to the source and drain (e.g., 4QD(2,2), 6QD(3,3), 12QD(6,6)). Across all three regimes—\( t/\gamma = 0.2 \), \( t/\gamma = 1 \), and \( t/\gamma = 2 \) respectively. \( P_{\mathrm{max}} \) initially increases with \( N \), reaching a peak at \( N=4 \), and then decreases for larger systems. In contrast, \( \eta_{\mathrm{max}}/\eta_c \) increases with \( N \) and eventually saturates, indicating that while energy filtering improves with system size, the gain in efficiency comes at the cost of reduced power output. This trade-off highlights the existence of an optimal system size for balancing power and efficiency in quantum dot thermoelectrics.
\section{ Conclusion }\label{conclusion}
 The present work demonstrates how the performance of nanoscale thermoelectric heat engines is governed by the shape of their electronic transmission functions. While a narrow Lorentzian profile maximizes efficiency at low power and a broad boxcar profile enables high power at reduced efficiency, intermediate and hybrid profiles shaped by quantum interference can enhance both simultaneously. Utilizing nonequilibrium Green’s function calculations for multi-quantum-dot Aharonov–Bohm interferometers, we show that Fano-like asymmetries, Dicke-type superradiant and subradiant modes, and engineered multi-peaked spectra can be tuned via geometry, magnetic flux, and coupling strengths to optimize performance. Across square, pentagonal, and hexagonal quantum-dot arrays in symmetric and asymmetric configurations, we identify three coupling regimes—$t/\gamma< 1$, $t/\gamma\simeq 1$, and $t/\gamma>1$—with the $t/\gamma\simeq2$ regime yielding the best compromise between power and efficiency. For example, the 6QD (3,3) geometry achieves $ZT\simeq30$ at dilution temperatures due to highly selective subradiant modes, while the 4QD(2,2) setup exhibits boxcar-like transmission enabling both high efficiency ($\sim76\%$ of Carnot) and high power ($\sim4.74$ fW). Scaling analysis reveals that increasing system size improves efficiency via enhanced energy filtering, though power peaks at intermediate sizes.\\
 \indent
From an experimental viewpoint, the main features of our model—multi-terminal quantum dot arrays, magnetic-flux–controlled interference, and tunable dot–lead coupling—are already accessible with current nanofabrication and measurement techniques. Lateral or vertical semiconductor quantum dots (GaAs/AlGaAs, InAs nanowires) and gate-defined graphene quantum dots can be realized in the proposed triangular, square, or hexagonal AB ring geometries with precise inter-dot tunneling control. Molecular junctions and self-assembled QD arrays also offer platforms where multiple discrete levels couple coherently to common leads. The magnetic flux control can be experienced by applying perpendicular magnetic fields in the sub-Tesla range to thread AB rings of micrometer or sub-micrometer diameter with one flux quantum, enabling fine-tuning of the interference pattern.  The gate electrodes can independently be controlled to tune the inter-dot tunnel couplings $t$ and dot–lead hybridization $\gamma$, making it experimentally feasible to reach and scan across the three different regimes: $t < \gamma$, $t\sim \gamma$, and $t> \gamma$ as identified in our work. Dilution refrigerators may provide the required sub-$100 mK$ operation temperatures to access the sharp subradiant resonances with ZT peaks up to $\sim30$ predicted here. However, realizing these high-performance regimes in practice may face several challenges due to(i) decoherence from phonons and charge noise, which can broaden resonances; (ii) fabrication disorder leading to unintended asymmetries; and (iii) electron–electron interactions.\\
\indent
In conclusion, our theoretical framework provides clear design principles—control of the transmission function shape via geometry, magnetic flux, and coupling asymmetry—for building thermoelectric heat engines that approach the joint limits of efficiency and power. With current advances in nanofabrication, low-temperature transport measurements, and coherent control, experimental realization of these designs appears within reach, offering exciting prospects for quantum-coherent energy harvesting technologies. Embedding optimized QD–AB heat engines into mesoscopic circuits could enable efficient waste-heat recovery in cryogenic electronics, quantum computing environments, or nanoscale sensors.
\section*{ACKNOWLEDGMENTS}
Sridhar acknowledges the financial support received from IIT Bhubaneswar in the form of an Institute Research Fellowship. M.B. acknowledges support from the Anusandhan National Research Foundation(ANRF), India, under the MATRICES scheme, Grant No. MTR/2021/000566.
\appendix
\section{EQUATIONS of MOTION}\label{appendixA}
The model setup of the four, five, and six quantum-dot AB interferometer has been discussed in Section \ref{sec:model}. We now solve these models and calculate the observables in the non-equilibrium steady state. Since the model is noninteracting, we can use the Nonequilibrium Green's Function (NEGF) approach to calculate its steady-state characteristics \cite{meir1992landauer,wang2014nonequilibrium}. The NEGF technique has recently been widely used to investigate the transport properties in mesoscopic systems and molecular junctions \cite{fransson2010non}. We follow the equation of motion approach for the derivations \cite{dhar2006nonequilibrium}. In this method, we solve the Heisenberg equations of motion (EOM) for the bath variables and then substitute them back into the EOM for the subsystem (dots) variables. We obtain  a general quantum Langevin equation (QLE) for all the configurations   of the subsystem as follows:
\begin{align}\label{eq1}
    \frac{dd_{i}(t)}{dt}&\nonumber=-i\left[\epsilon_i d_i +\sum_{i\neq j} t_{ij} e^{i\phi_{ij}}d_j \right] -i\sum_{\alpha} \eta_{\alpha}^S(t)\delta_{i\alpha}\\&\nonumber-i\sum_{\beta}\eta_{\beta}^D(t)\delta_{i\beta}-i\int_{t_0}^t dt^\prime \sum_{\alpha,\alpha^\prime} \delta_{i\alpha}\Sigma_{\alpha,\alpha^\prime}^{+(S)}(t-t^\prime)\\&d_{\alpha^\prime}(t^\prime) -i\int_{t_0}^t dt^\prime \sum_{\beta,\beta^\prime} \delta_{i\beta}\Sigma_{\beta,\beta^\prime}^{+(D)}(t-t^\prime)d_{\beta^\prime}(t^\prime)
\end{align}
Here, we use the indices  \( i \in \{1, \dots, N\} \) with $N=\{4,5,6\}$ to identify the $ 4QD, 5QD$, and $6QD$, respectively. The summation over $\alpha,\alpha^\prime,\beta,\beta^\prime$ takes the different values for different geometry depands to the coupling between the dots and lead for instance  ($\alpha,\alpha^{\prime}=1,2$), ($\beta,\beta^{\prime}=3,4$) for the 4QD(2,2) configuration is one such example for other configuration see Fig.\ref{ZT vs T 4QD(2,2)}(a1)-(a4). The terms $\hat{\eta}_i^S$ and $\hat{\eta}_i^D$ are referred to as the noise induced on the subsystem by the source and drain, respectively, and they are expressed as
\begin{equation}
\begin{aligned}
    \eta_{i}^S=\sum_{k}\sum_{\alpha} v_{\alpha,k}^{S}\delta_{i,\alpha} g_{S,k}^{+}(t-t_{0}) c_{S,k}(t_{0}) ,\\ 
    \eta_{i}^{D}=\sum_{k}\sum_{\beta} v_{\beta,k}^{D}\delta_{i,\beta} g_{D,k}^{+}(t-t_{0}) c_{D,k}(t_{0})
    \end{aligned}
\end{equation}
The retarded Green's functions of the isolated reservoirs are given by
\begin{equation}\label{eq8}
    \begin{aligned}
        g_{Sk}^{+}(t)=-ie^{-i\epsilon_{Sk}t}\theta(t),\\
        g_{Dk}^{+}(t)=-ie^{-i\epsilon_{Dk}t}\theta(t).
    \end{aligned}
\end{equation}
For the initial condition, we take factorized states for the total density matrix $\rho_T(t_0)=\rho_S\otimes\rho_D\otimes\rho(t_0)$, with empty dots and reservoirs prepared in a grand canonical state
\begin{equation}\label{eq9}
    \hat{\rho}_{\nu}=\frac{e^{-(\hat{H}_{\nu}-\mu_{\nu}\hat{N})/T_{\nu}}}{Tr[e^{-(\hat{H}_{\nu}-\mu_{\nu}\hat{N})/T_{\nu}}]},
\end{equation}
where $T_{\nu}$ and $\mu_{\nu}$ are the temperatures and chemical potentials of the Fermi sea with $\nu=S, D$. The state of the subsystem is denoted by the reduced density matrix $\rho$. Using the initial conditions, we obtained the noise correlation as follows
\begin{equation}\label{eq10}
    \begin{aligned}
        \langle\hat{\eta}_i^{\dagger S}(t)
    \hat{\eta}_{i'}^{S}(\tau)\rangle=\sum_{k}\sum_{\alpha}V_{\alpha,k}^{S^*}\delta_{i\alpha}e^{i\omega_k(t-\tau)}\delta_{i'\alpha}V_{\alpha,k}^S\hspace{0.1cm}f_S(\omega_k),\\
    \langle\hat{\eta}_i^{\dagger D}(t)
    \hat{\eta}_{i'}^{D}(\tau)\rangle=\sum_{k}\sum_{\beta} V_{\beta,k}^{D^*}\delta_{i,\beta}e^{i\omega_k(t-\tau)}\delta_{i'\beta}V_{\beta,k}^D\hspace{0.1cm}f_D(\omega_k),
    \end{aligned}
\end{equation}
with the Fermi function $f_{\nu}=[e^{(\omega-\mu_{\nu})/T_{\nu}}+1]^{-1}$ for the reservoir $\nu=S,D$ with $\mu_{\nu}$ and $T_{\nu}$ be the corresponding chemical potential and temperature, respectively. In the Heisenberg picture, the expectation value of an observable $A$ can be obtained as
$\langle \hat{A}(t)\rangle=Tr_T[\rho_T(t_0)\hat{A}(t)]$, tracing over all degrees of freedom. The steady-state properties are obtained by taking the limits $t_0\rightarrow-\infty$ and $t\rightarrow\infty$. We can now take the Fourier transform of Eq. (\ref{eq1}) using the convolution theorem with the convention $\tilde{d}_i(\omega)=\int_{-\infty}^{\infty}dt\,d_i(t)e^{i\omega t}$
and $\tilde{\eta}_i^{\nu}(\omega)=\int_{-\infty}^{\infty}dt\,\eta_i^{\nu}(t)e^{i\omega t}$ and the result in matrix form is
 \begin{equation}
        \tilde{d_{i}}(\omega)=\sum_{\alpha}G_{i,\alpha}^{+}\tilde{\eta_{\alpha}}^{S}(\omega) +\sum_{\beta}G_{i,\beta}^{+}\tilde{\eta_{\beta}}^{D}(\omega)
    \end{equation}
    Here, the retarded Green's function is given by
    \begin{align}
     G_{i,\alpha(\beta)}^{+}&\nonumber=\left[\omega I-H_{N}- \sum_{\alpha^{'}}\Sigma_{\alpha^{'},\alpha}^{+(S)}(\omega)\delta_{\alpha i}\right.\\&\left.-\sum_{\beta^{'}}\Sigma_{\beta^{'},\beta}^{+(D)}(\omega)\delta_{\beta i} \right]^{-1}   
    \end{align}
where $N\in\{4,5,6\}$ denoting the number of sites(or QDs) in the AB ring and $I$ is  $(N\times N)$, identity matrix with $N\in\{4,5,6\}$ for 4QD, 5QD, and 6QD configurations respectively and the advanced Green's function is given by the transpose conjugate of the retarded Green's function, $G^-(\omega)=[G^+(\omega)]^{\dagger}$. The self-energies are defined as:
\begin{equation}\label{self_energy_eq}
    \begin{aligned}
        \Sigma_{\alpha,\alpha^{'}}^{\pm(S)}(t-t^{'})=\sum_{k}v_{\alpha,k}^{S} g_{S,k}^{\pm}(t-t^{'}) 
 v_{\alpha^{'},k}^{\ast S}\\
  \Sigma_{\beta,\beta^{'}}^{\pm(D)}(t-t^{'})=\sum_{k}v_{\beta,k}^{D} g_{D,k}^{\pm}(t-t^{'}) 
 v_{\beta^{'},k}^{\ast D}
    \end{aligned}
\end{equation}
Here, $g_{S}^{\pm}(\omega)$ and $g_{D}^{\pm}(\omega)$ are given by the Fourier transform of Eq. (\ref{eq8}). In the wide-band limit and when the density of states of the metallic lead is energy independent, the real part of the self-energy term vanishes. Then we can define the hybridization matrix from the relation $\Sigma^{+}=-i\Gamma/2$:
\begin{equation}\label{eqgamma}
    \begin{aligned}
\Gamma_{\alpha,\alpha'}^{S}=2\pi\sum_{k,\alpha,\alpha'}V_{\alpha',k}^{S^*}V_{\alpha,k}^{S}\, \delta(\omega-\omega_k)\\
\Gamma_{\beta,\beta'}^{D}=2\pi\sum_{k,\beta,\beta'}V_{\beta',k}^{D^*}V_{\beta,k}^{D}\, \delta(\omega-\omega_k)
  \end{aligned}  
\end{equation}
We may take $V_{\alpha,k}^{S}$  and $v_{\beta,k}^{D}$ as real constants, independent of the level index and reservoir state, resulting in $\Gamma_{\alpha,\alpha^\prime}^S=\gamma_S$  and $\Gamma_{\beta, \beta^\prime }^D=\gamma_D$, where $\gamma_{\nu}$ (energy independent) describes the coupling between the dots and metallic leads. We consider degenerate dot energies $\epsilon_i=\epsilon$ and set $t_{ij}=t$ to obtain the retarded Green's function. The matrix form of the Hamiltonian $\mathcal{H}_N$, retarded Green's function $G^{+}(\omega)$, the hybridization matrices, and the transmission function for various configurations are given in the following subsection. 

\subsection{4QD(2,2) Configuration}\label{4QD(2,2)configuration}

\begin{equation}\label{4dot_H_s_matrix}
  \mathcal{H}_{4QD}=\begin{pmatrix}
     \epsilon & t e^{\frac{i\phi}{4}} & 0 & t e^-{\frac{i\phi}{4}}\\
     t e^-{\frac{i\phi}{4}} & \epsilon & t e^{\frac{i\phi}{4}} & 0\\
     0 & t e^-{\frac{i\phi}{4}} & \epsilon & t e^{\frac{i\phi}{4}}\\
     t e^{\frac{i\phi}{4}} & 0 & t e^-{\frac{i\phi}{4}} & \epsilon
 \end{pmatrix}   
 \end{equation}
\begin{equation}\label{fqd_ret}
G^{+}(\omega)=\begin{pmatrix}
        \tilde{\omega}+\tilde{\gamma}_{S} & c+\tilde{\gamma}_{S} & 0 & c^{\ast} \\
       c^{\ast}+\tilde{\gamma}_{S} & \tilde{\omega}+\tilde{\gamma}_{S} & c & 0 \\
        0 & c^{\ast} & \tilde{\omega}+\tilde{\gamma}_{D} & c+\tilde{\gamma}_{D} \\
        c & 0 & c^{\ast}+\tilde{\gamma}_{D} & \tilde{\omega}+\tilde{\gamma}_{D}
    \end{pmatrix}^{-1}
\end{equation}
where, $\Tilde{\omega}=\omega-\epsilon$ ,$c=-t e^\frac{i\phi}{4}$ , $\Tilde{\gamma}_{S}=\frac{i\gamma_{S}}{2} $ ,$\Tilde{\gamma}_{D}=\frac{i\gamma_{D}}{2}$ and $c^{\ast}$ is the complex conjugate of c.

    \begin{equation}\label{4dot_hybdz}
    \Gamma^{S}=\gamma_{S}\begin{pmatrix}
    1 & 1 & 0 & 0\\
    1 & 1 & 0 & 0\\
     0 & 0 & 0 & 0\\
      0 & 0 & 0 & 0
\end{pmatrix}\hspace{0.5cm}
\Gamma^{D}=\gamma_{D}\begin{pmatrix}
    0 & 0 & 0 & 0\\
    0 & 0 & 0 & 0\\
     0 & 0 & 1 & 1\\
      0 & 0 & 1 & 1
      \end{pmatrix}
\end{equation}
 \begin{equation}\label{T_4QD(2,2)}
    \mathrm{T}_{4QD(2,2)}(\omega,\phi)=\frac{4 \gamma ^2 t^2 \left[2 t \sin ^2\left(\frac{\phi }{4}\right)-\tilde{\omega} \cos \left(\frac{\phi }{4}\right)\right]^2}{\Delta_{1}(\omega,\phi)}
\end{equation}
where
\begin{align}
    \Delta_{1}(\omega,\phi)\nonumber=&\left(\gamma^2+\tilde{\omega}^2-4t\tilde{\omega}\cos\frac{\phi}{4}+4t^2\cos^2\frac{\phi}{4} \right)\times\\& \left( \gamma^2\tilde{\omega}^2+\tilde{\omega}^4 -8t^2\tilde{\omega}^2\sin^2\frac{\phi}{4}+16t^4\sin^4\frac{\phi}{4}\right)
\end{align}
\subsection{4QD(3,1) Configuration}\label{4QD(3,1)configuration}
\begin{equation}\label{4QD(3,1)_ret}
    G^{+}(\omega)=\begin{pmatrix}
        \tilde{\omega}+\tilde{\gamma}_{S} & c+\tilde{\gamma}_{S} & 0 & c^{\ast}+\tilde{\gamma}_{S}\\
        c^{\ast}+\tilde{\gamma}_{S} & \tilde{\omega}+\tilde{\gamma}_{S} & c & \tilde{\gamma}_{S} \\
        0 & c^{\ast} & \tilde{\omega}+\tilde{\gamma}_{D} & c \\
        c+\tilde{\gamma}_{S} & \tilde{\gamma}_{S} & c^{\ast} & \tilde{\omega}+\tilde{\gamma}_{S}
    \end{pmatrix}^{-1}
\end{equation}
\begin{equation}\label{4QD(3,1)_hybdz}
    \Gamma^{S}=\gamma_{S}\begin{pmatrix}
    1 & 1 & 0 & 1\\
    1 & 1 & 0 & 1\\
    0 & 0 & 0 & 0\\
    1 & 1 & 0 & 1
\end{pmatrix}\hspace{0.5cm}
\Gamma^{D}=\gamma_{D}\begin{pmatrix}
    0 & 0 & 0 & 0\\
    0 & 0 & 0 & 0\\
    0 & 0 & 1 & 0\\
    0 & 0 & 0 & 0
\end{pmatrix}
\end{equation}
\begin{equation}\label{T_4QD(3,1)}
    \begin{split}
        \mathrm{T}_{4QD(3,1)}(\omega,\phi)=\frac{4t^2\gamma^2}{\Delta_{2}(\omega,\phi)}\Bigg[(t^2-\tilde{\omega}^2)\cos\frac{\phi}{4}-\\t\bigg(\tilde{\omega}\cos\frac{\phi}{2}+t\cos\frac{3\phi}{4}\bigg)\Bigg]^2
    \end{split}
\end{equation}
where
\begin{align}
    \Delta_2(\omega,\phi)\nonumber&=\left[-t\gamma^2\tilde{\omega}\cos\frac{\phi}{4}+\frac{1}{4} \left( 8t^4+2t^2(\gamma^2-8\tilde{\omega}^2)\right.\right.\\ \nonumber&\left.\left.+\tilde{\omega}^2(4\tilde{\omega}^2-3\gamma^2)-2t^2\gamma^2\cos\frac{\phi}{2}-8t^4\cos\phi   \right)   \right]^2\\ \nonumber&+\left[-4t^2\gamma\tilde{\omega}+2\gamma\tilde{\omega}^3+2\gamma t(\tilde{\omega}^2-t^2)\cos\frac{\phi}{4}\right.\\ &\left.+2\gamma\tilde{\omega}t^2\cos\frac{\phi}{2}+2t^3\gamma\cos\frac{3\phi}{4}  \right]^2
\end{align}
\subsection{5QD(3,2) Configuration}\label{5QD(3,2)configuration}
\begin{equation}\label{5dot_H_s_matrix}
  \mathcal{H}_{5QD}=\begin{pmatrix}
     \epsilon & t e^{\frac{i\phi}{5}} &0& 0 & t e^-{\frac{i\phi}{5}}\\
     t e^-{\frac{i\phi}{5}} & \epsilon & t e^{\frac{i\phi}{5}} & 0&0\\
     0 & t e^-{\frac{i\phi}{5}} & \epsilon & t e^{\frac{i\phi}{5}}&0\\
     t e^{\frac{i\phi}{5}} & 0&0 & t e^-{\frac{i\phi}{5}} & \epsilon
 \end{pmatrix}   
 \end{equation}
\begin{equation}\label{5QD(3,2)_ret}
G^{+}(\omega)=\begin{pmatrix}
        \tilde{\omega}+\tilde{\gamma}_{S} & c+\tilde{\gamma}_{S} & \tilde{\gamma}_{S} & 0&c^{\ast}\\
        c^{\ast}+\tilde{\gamma}_{S} & \tilde{\omega}+\tilde{\gamma}_{S} & c+\tilde{\gamma}_{S} & 0 & 0 \\
        \tilde{\gamma}_{S} & c^{\ast}+\tilde{\gamma}_{S} & \tilde{\omega}+\tilde{\gamma}_{S} & c &0\\0 & 0 & c^{\ast} & \tilde{\omega}+\tilde{\gamma}_{D} & c+\tilde{\gamma}_{D}\\
        c & 0 & 0 &c^{\ast}+\tilde{\gamma}_{D}& \tilde{\omega}+\tilde{\gamma}_{D}
    \end{pmatrix}^{-1}
\end{equation}
\begin{equation}\label{5QD(3,2)_hybdz}
    \Gamma^{S}=\gamma_{S}\begin{pmatrix}
    1 & 1 & 1 & 0&0\\
    1 & 1 & 1 & 0&0\\
    1 & 1 & 1 & 0&0\\
    0 & 0 & 0 & 0&0\\
    0 & 0 & 0 & 0&0
\end{pmatrix}\hspace{0.5cm}
\Gamma^{D}=\gamma_{D}\begin{pmatrix}
    0 & 0 & 0 & 0&0\\
    0 & 0 & 0 & 0&0\\
    0 & 0 & 0 & 0&0\\
    0 & 0 & 0 & 1&1\\
    0 & 0 & 0 & 1&1
\end{pmatrix}
\end{equation}
\begin{align}\label{T_5QD(3,2)}
   \mathrm{T}_{5QD(3,2)}(\omega,\phi)&\nonumber=\frac{4t^2\gamma^2}{\Delta_{3}(\omega,\phi)}\left[(\tilde{\omega}^2-2t^2)\tilde{\omega}\cos\frac{\phi}{5}-2t(\tilde{\omega}^2-\right.\\&\left.t^2)\cos\frac{2\phi}{5}-2t^2\tilde{\omega}\cos\frac{3\phi}{5}-t^3\cos\frac{4\phi}{5}\right]^2  
\end{align}
where
\begin{align}
    \Delta_3(\omega,\phi)&\nonumber=\left[\frac{1}{2} \left( 10t^4\tilde{\omega}+2t^2\tilde{\omega}(\gamma^2-5\tilde{\omega}^2)+2\tilde{\omega}^5-3\gamma^2\tilde{\omega}^3\right.\right.\\ \nonumber&\left.\left.-7t\gamma^2\tilde{\omega}^2\cos\frac{\phi}{5} +4t^3\gamma^2\cos^3\frac{\phi}{5}-3t^2\tilde{\omega}\gamma^2\cos\frac{2\phi}{5}\right.\right.\\ \nonumber&\left.\left.-4t^5\cos\frac{\phi}{5}\right)\right]^2+\left[\frac{\gamma}{2}\left(5(t^2-\tilde{\omega}^2)^2-5t^2\tilde{\omega}^2\right.\right.\\ \nonumber&\left.\left.+6t(\tilde{\omega}^3-2t^2\tilde{\omega})\cos\frac{\phi}{5}+2t^2(\tilde{\omega}^2-t^2)\cos\frac{2\phi}{5}\right.\right.\\ &\left.\left.+2t^3\tilde{\omega}\cos\frac{3\phi}{5}+6t^4\cos\frac{4\phi}{5}      \right)     \right]^2
\end{align}
\subsection{6QD(3,3) Configuration}\label{6QD(3,3)configuration}
\begin{equation}\label{6dot_H_s_marix}
     \mathcal{H}_{6QD}=\begin{pmatrix}
         \epsilon & t e^\frac{i\phi}{6} & 0 & 0 & 0 & t e^{-\frac{i\phi}{6}}\\
         t e^{-\frac{i\phi}{6}} & \epsilon & t e^\frac{i\phi}{6} & 0 & 0 & 0 \\
         0 & t e^{-\frac{i\phi}{6}} & \epsilon & t e^\frac{i\phi}{6} & 0 & 0\\
         0 & 0 & t e^{-\frac{i\phi}{6}} & \epsilon & t e^\frac{i\phi}{6} & 0\\
         0 & 0 & 0 & t e^{-\frac{i\phi}{6}} & \epsilon & t e^\frac{i\phi}{6}\\
         t e^\frac{i\phi}{6} & 0 & 0 & 0 & t e^{-\frac{i\phi}{6}} & \epsilon
     \end{pmatrix}
 \end{equation}
\begin{equation}\label{6QD_ret}
  G^{+}(\omega)=   \begin{pmatrix}
        A & B\\
        C & D
    \end{pmatrix}^{-1}
\end{equation}
where A, B, ,C and D are the block matrices of size $(3\times3)$ are given as follows:
$$A=\begin{pmatrix}
 \tilde{\omega}+\tilde{\gamma}_{S} & c+\tilde{\gamma}_{S} & \tilde{\gamma}_{S}\\
 c^{\ast}+\tilde{\gamma}_{S} & \tilde{\omega}+\tilde{\gamma}_{S} & c+\tilde{\gamma}_{S}\\
  \tilde{\gamma}_{S}  & c^{\ast}+\tilde{\gamma}_{S}  & \tilde{\omega}+\tilde{\gamma}_{S}
 \end{pmatrix}; B=\begin{pmatrix}
     0 & 0 & c^\ast\\
     0 & 0 & 0 \\
     c & 0 & 0
 \end{pmatrix}
$$
$$
C=\begin{pmatrix}
    0 & 0 & c^\ast\\
    0 & 0 & 0\\
    c & 0 & 0
\end{pmatrix} ; D= \begin{pmatrix}
    \tilde{\omega}+\tilde{\gamma}_{D} & c+\tilde{\gamma}_{D} & \tilde{\gamma}_{D}\\
    c^{\ast}+\tilde{\gamma}_{D} & \tilde{\omega}+\tilde{\gamma}_{D} & c+\tilde{\gamma}_{D}\\
    \tilde{\gamma}_{D} & c^{\ast}+\tilde{\gamma}_{D} & \tilde{\omega}+\tilde{\gamma}_{D}

\end{pmatrix}
$$
\begin{equation}\label{6dot_hybdz}
    \Gamma^S =\gamma_S\begin{pmatrix}
        1 & 1 & 1 & 0 & 0 & 0\\
        1 & 1 & 1 & 0 & 0 & 0\\
        1 & 1 & 1 & 0 & 0 & 0\\
        0 & 0 & 0 & 0 & 0 & 0\\
        0 & 0 & 0 & 0 & 0 & 0\\
        0 & 0 & 0 & 0 & 0 & 0\\
    \end{pmatrix}; \Gamma^D=\gamma_D\begin{pmatrix}
       0 & 0 & 0 & 0 & 0 & 0\\
       0 & 0 & 0 & 0 & 0 & 0\\
       0 & 0 & 0 & 0 & 0 & 0\\
       0 & 0 & 0 & 1 & 1 & 1\\
       0 & 0 & 0 & 1 & 1 & 1\\
        0 & 0 & 0 & 1 & 1 & 1\\
    \end{pmatrix}
\end{equation}
\begin{align}\label{T_6QD(3,3)}
    \mathrm{T}_{6QD(3,3)}(\omega,\phi)&\nonumber=\frac{4t^2\gamma^2}{\Delta_{4}(\omega,\phi)}\left[(\tilde{\omega}^2-2t^2)\cos\frac{\phi}{6}+t^2\cos\frac{\phi}{2}\right.\\&\left.-2t\tilde{\omega}\sin^2\frac{\phi}{6} \right]^2
\end{align}
where
\begin{align}
    \Delta_{4}(\omega,\phi)&\nonumber=2t^2\gamma^2\left[(\tilde{\omega}^2-2t^2)\cos\frac{\phi}{6}+t^2\cos\frac{\phi}{2}\right.\\& \nonumber\left.-2t\tilde{\omega}\sin^2\frac{\phi}{6}  \right]^2 +\frac{\gamma^2}{2}\left[(4t^3-2t\tilde{\omega}^2 )\cos\frac{\phi}{6}+3\tilde{\omega}^3\right.\\&\nonumber \left.-7t^2\tilde{\omega}-2t^2\tilde{\omega}\cos\frac{\phi}{3}+4t^3\cos\frac{\phi}{2}  \right]^2+\frac{9\gamma^4}{16}\left[-3(\tilde{\omega}^2\right.\\& \nonumber \left.-t^2)-2t\tilde{\omega}\cos\frac{\phi}{6}+2t^2\cos\frac{\phi}{3}  \right]^2+\left(2t\cos\frac{\phi}{6}-\tilde{\omega}\right)^2\\ & \left[3t^2\tilde{\omega}-\tilde{\omega}^3-2t^3\cos\frac{\phi}{2} \right]^2
\end{align}

\begin{figure*}
    \centering
    \includegraphics[scale=0.21]{Fig_1.png}
    \includegraphics[scale=0.21]{Fig_2.png}
    \includegraphics[scale=0.21]{Fig_3.png}
    \includegraphics[scale=0.21]{Fig_4.png}
  
    \includegraphics[scale=0.21]{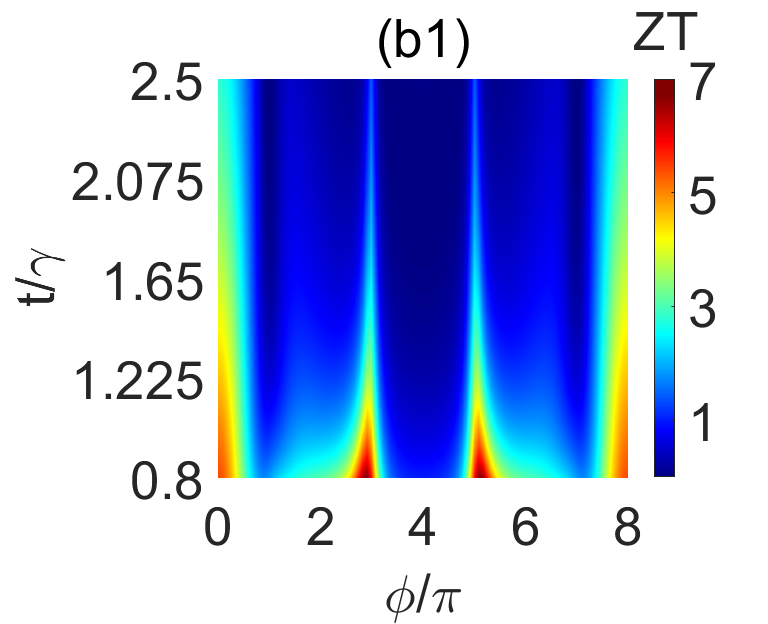}
    \includegraphics[scale=0.21]{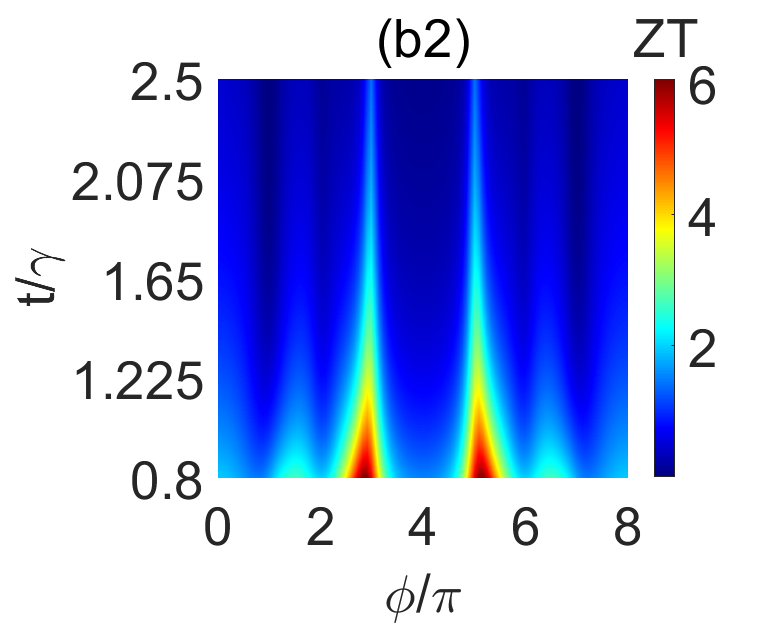}
    \includegraphics[scale=0.21]{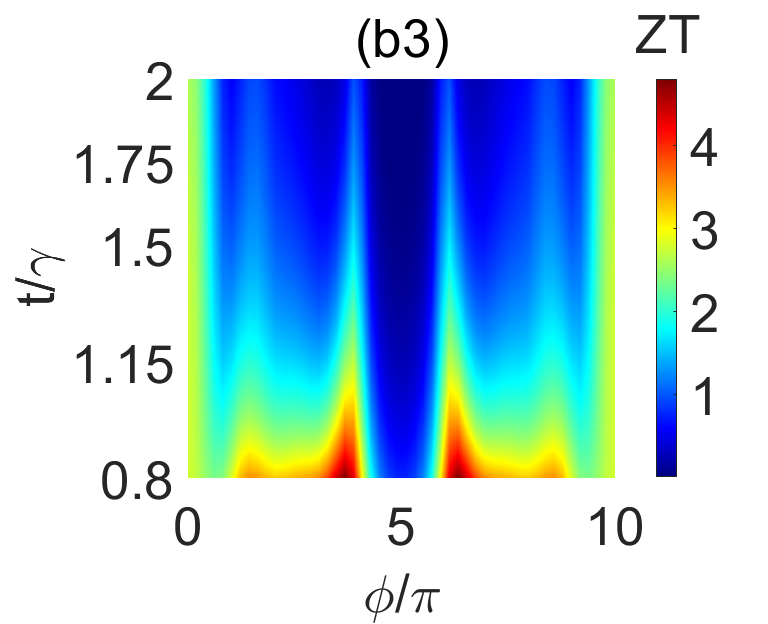}
    \includegraphics[scale=0.21]{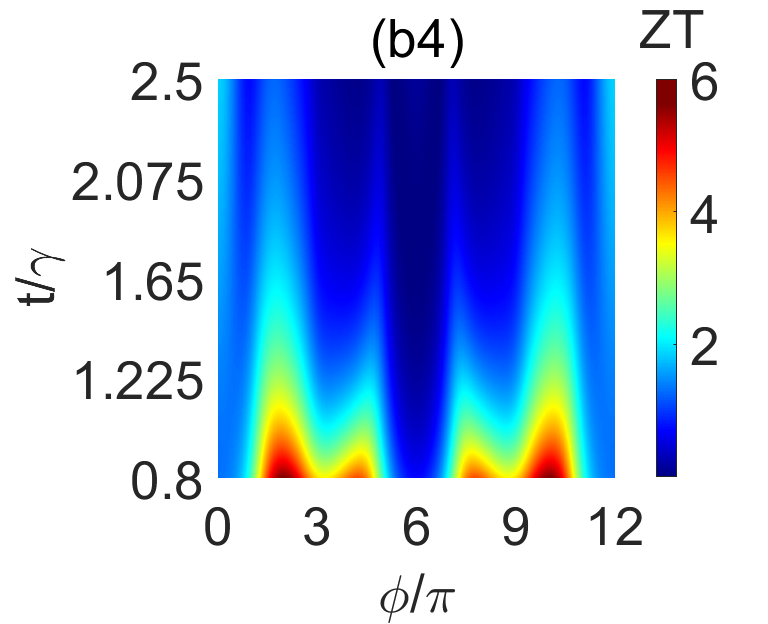}
   
    \includegraphics[scale=0.21]{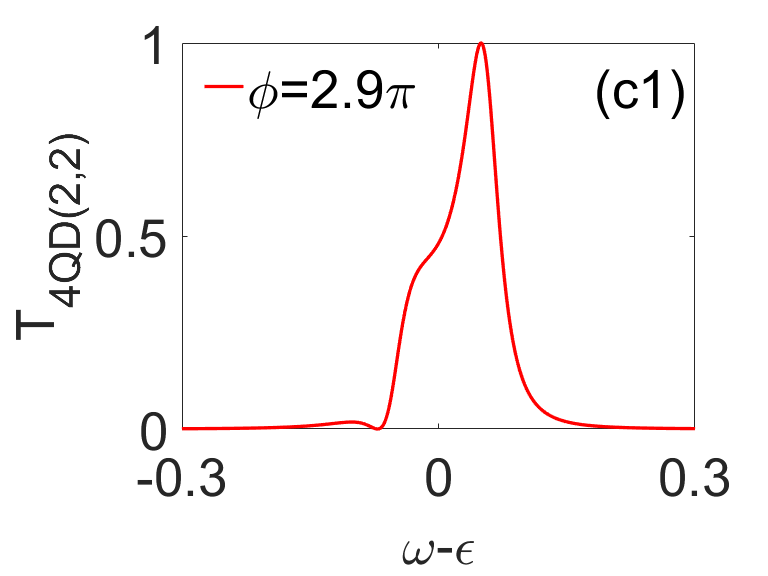}
    \includegraphics[scale=0.21]{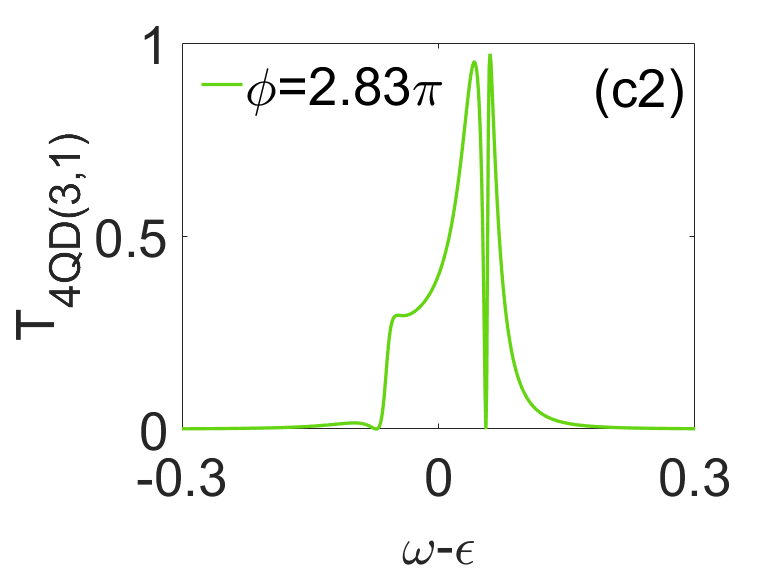}
    \includegraphics[scale=0.21]{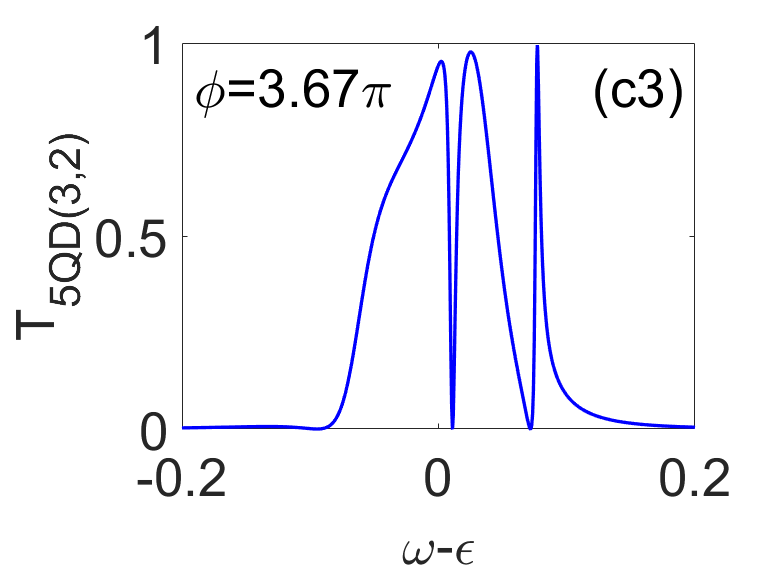}
    \includegraphics[scale=0.21]{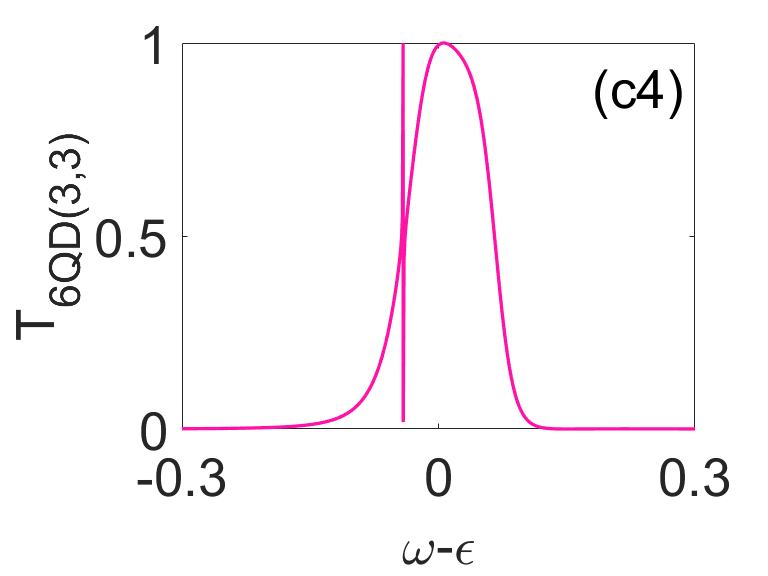}
    \caption{Row (a1)-(a4) denotes the model configurations used for the following studies; Second panel (b1)-(b4) shows the figure of merit ZT  as a function of $\phi/\pi$ and $t/\gamma$. Third panel (c1)-(c4) represents transmission as a function of energy, where ZT is maximum at $\phi=2.9\pi$ for (c1), $\phi=2.83\pi$ for (c2), $\phi=3.67\pi$ for (c3), and $\phi=1.94\pi$ for (c4)  and $t=0.8\gamma$.Parameters used are $\gamma=0.05,\epsilon=6\gamma,\mu=4\gamma,T=6\gamma$ }
    \label{Fig7}
\end{figure*}

\begin{table*}[]
    \centering
    \begin{tabular}{|c|c|c|c|c|c|c|c|}
    \hline
    \multirow{2}{*}{Configuration and phase} & \multirow{2}{*}{$\Delta$} & \multicolumn{3}{|c|}{$P_{max}$(fW)}& \multicolumn{3}{|c|}{$\eta_{max}/\eta_c$}\\

\cline{3-8}
\multirow{7}{*}{4QD(2,2) , $\phi=5\pi/2$} &  & $t/\gamma<1$ & $t/\gamma=1$ &$t/\gamma>1$ &  $t/\gamma<1$ & $t/\gamma=1$ &$t/\gamma>1$ \\
\hline
 & $0$ & $0.22$ & $3.01$ &$4.74$ &  $86.8$ & $83.16$ &$76$ \\
 \cline{2-8}
 & $0.02$ & $0.26$ & $2.87$ &$4.71$ &  $87.2$ & $82.67$ &$75.6$ \\
 \cline{2-8}
 & $0.04$ & $0.23$ & $2.66$ &$4.48$ &  $84.83$ & $81.23$ &$74.8$ \\
 \cline{2-8}
 & $0.06$ & $0.164$ & $2.31$ &$4.20$ &  $80.87$ & $78.92$ &$73.5$ \\
 \cline{2-8}
 & $0.08$ & $0.11$ & $1.94$ &$3.83$ &  $76.23$ & $75.84$ &$71.68$ \\
 \cline{2-8}
 & $0.1$ & $0.07$ & $1.55$ &$3.46$ &  $71.31$ & $72.11$ &$69.41$ \\
\hline 
\hline 
\multirow{7}{*}{4QD(3,1) , $\phi=5\pi/2$} 
 & $0$ & $0.9$ & $1.77$ &$2.08$ &  $93.79$ & $81.47$ &$64$ \\
 \cline{2-8}
 & $0.02$ & $0.47$ & $1.9$ &$2.14$ &  $93.21$ & $80.65$ &$63.66$ \\
 \cline{2-8}
 & $0.04$ & $0.51$ & $2.06$ &$2.18$ &  $90.5$ & $79.21$ &$62.84$ \\
 \cline{2-8}
 & $0.06$ & $0.36$ & $2.11$ &$2.2$ &  $87.03$ & $77.75$ &$61.82$ \\
 \cline{2-8}
 & $0.08$ & $0.24$ & $2.08$ &$2.18$ &  $83.33$ & $76.2$ &$60.72$ \\
 \cline{2-8}
 & $0.1$ & $0.17$ & $1.97$ &$2.11$ &  $79.56$ & $74.5$ &$59.58$ \\
 \hline 
\hline 
\multirow{7}{*}{5QD(3,2) , $\phi=5\pi/2$} 
 & $0$ & $0.27$ & $3.04$ &$3.97$ &  $96.02$ & $81$ &$63.1$ \\
 \cline{2-8}
 & $0.02$ & $0.31$ & $3.02$ &$3.92$ &  $93.97$ & $80.46$ &$62.86$ \\
 \cline{2-8}
 & $0.04$ & $0.29$ & $2.96$ &$3.91$ &  $90.31$ & $79.21$ &$62.28$ \\
 \cline{2-8}
 & $0.06$ & $0.22$ & $2.80$ &$3.78$ &  $86.33$ & $77.35$ &$61.32$ \\
 \cline{2-8}
 & $0.08$ & $0.16$ & $2.43$ &$3.57$ &  $82.26$ & $75.1$ &$60.04$ \\
 \cline{2-8}
 & $0.1$ & $0.12$ & $2.08$ &$3.28$ &  $78.17$ & $72.59$ &$58.49$ \\
\hline 
\hline 
\multirow{7}{*}{6QD(3,3) , $\phi=5\pi/2$} 
 & $0$ & $0.19$ & $2.88$ &$5$ &  $95.3$ & $83$ &$68.5$ \\
 \cline{2-8}
 & $0.02$ & $0.21$ & $2.86$ &$4.93$ &  $94.89$ & $82.7$ &$68.44$ \\
 \cline{2-8}
 & $0.04$ & $0.19$ & $2.8$ &$4.89$ &  $90.63$ & $81.73$ &$68.07$ \\
 \cline{2-8}
 & $0.06$ & $0.16$ & $2.67$ &$4.74$ &  $86.72$ & $80.25$ &$67.47$ \\
 \cline{2-8}
 & $0.08$ & $0.13$ & $2.46$ &$4.56$ &  $82.62$ & $78.41$ &$66.66$ \\
 \cline{2-8}
 & $0.1$ & $0.1$ & $2.23$ &$4.31$ &  $78.46$ & $76.31$ &$65.69$ \\
\hline

    \end{tabular}
    \caption{ Maximum output power $P_{max}$ and maximum efficiency $\eta_{max}$ for different $\Delta$ values   for different geometries at different regimes.}
    \label{tab:my_label}
\end{table*}

\section{Expansion of figure of merit for Fano resonance }\label{appendix B}
The figure of merit ZT can be expressed in terms of Onsager coefficients
\begin{equation}\label{eq_ZT}
    ZT=\frac{\mathcal{L}_{12}^2}{\mathcal{L}_{11}\mathcal{L}_{22}-\mathcal{L}_{12}\mathcal{L}_{21}}
\end{equation}
Where $\mathcal{L}_{ij}$ are the Onsager coefficients defined as follows 
\begin{equation}\label{eq_L11}
   \mathcal{L}_{11}=-T\int_{-\infty}^\infty d\omega T_{SD} f^\prime(\omega) 
\end{equation}
\begin{align}\label{eq_L12}
    \mathcal{L}_{12}=-T\int_{-\infty}^\infty d\omega {T}_{SD}(\omega-\mu)f^\prime(\omega)
\end{align}
\begin{equation}
 \mathcal{L}_{21}  =\mathcal{L}_{12} 
\end{equation}
and $\mathcal{L}_{22}$ is defined as
\begin{align}\label{eq_L22}
    \mathcal{L}_{22}=-T\int_{-\infty}^\infty d\omega (\omega-\mu)^2 {T}_{SD}f^\prime(\omega)
\end{align}
where $T_{SD}$ is the Fano transmission function is given by Eq.(\ref{Fano_transmission}) and the $f^\prime(\omega)=-\left[4T\cosh^2(\frac{\omega-\mu}{2T}) \right]^{-1}$ is the derivative of Fermi distribution function.
Substitute the expression of Fano transmission Eq.(\ref{Fano_transmission}) in the equations (\ref{eq_L11}) ,(\ref{eq_L12}) and (\ref{eq_L22})  we obtained 
\begin{align}
    \mathcal{L}_{11}&\nonumber=\frac{-T}{1+q^2}\left[q^2\int_{-\infty}^\infty d\omega \frac{1}{1+\epsilon^2}f^\prime(\omega)+\int_{-\infty}^\infty d\omega \frac{\epsilon^2}{1+\epsilon^2}f^\prime(\omega)\right.\\ & \left.+q\int_{-\infty}^\infty d\omega \frac{2\epsilon}{1+\epsilon^2}f^\prime(\omega)\right]
\end{align}
\begin{equation}\label{eq_L11_final}
    \mathcal{L}_{11}=\frac{1}{1+q^2}\left(\alpha_1 q^2+\beta_1+\gamma_1 q\right)
\end{equation}
where :
\begin{equation}
    \alpha_1=-T\int_{-\infty}^\infty d\omega \frac{1}{1+\epsilon^2}f^\prime(\omega)
\end{equation}
\begin{equation}
    \beta_1=-T\int_{-\infty}^\infty d\omega \frac{\epsilon^2}{1+\epsilon^2}f^\prime(\omega)
\end{equation}
and 
\begin{equation}
    \gamma_1=-T\int_{-\infty}^\infty d\omega \frac{2\epsilon}{1+\epsilon^2}f^\prime(\omega)
\end{equation}
\begin{align}
    \mathcal{L}_{12}&\nonumber=-T\frac{q^2}{1+q^2}\int_{-\infty}^\infty d\omega (\omega-\mu)\frac{1}{1+\epsilon^2}f^\prime(\omega) \\& \nonumber-\frac{T}{1+q^2}\int_{-\infty}^\infty d\omega \frac{\epsilon^2}{1+\epsilon^2}(\omega-\mu)f^\prime(\omega)\\&-T\frac{q}{1+q^2}\int_{-\infty}^\infty d \omega \frac{2\epsilon}{1+\epsilon^2}(\omega-\mu)f^\prime(\omega)
\end{align}
\begin{equation}\label{eq_L12_final}
    \mathcal{L}_{12}=\frac{1}{1+q^2}(\alpha_2 q^2 +\beta_2 +\gamma_2 q)
\end{equation}
where $\alpha_2=(\omega-\mu)\alpha_1 ,\beta_2=(\omega-\mu)\beta_1$ and $\gamma_2=(\omega-\mu)\gamma_1$

\begin{align}
\mathcal{L}_{22}&\nonumber=  -T\frac{q^2}{1+q^2}\int_{-\infty}^\infty d\omega (\omega-\mu)^2 \frac{1}{1+\epsilon^2}f^\prime(\omega)\\& \nonumber-\frac{T}{1+q^2}\int_{-\infty}^\infty d\omega (\omega-\mu)^2\frac{\epsilon^2}{1+\epsilon^2}f^\prime(\omega)\\&-T\frac{q}{1+q^2}\int_{-\infty}^\infty d\omega (\omega-\mu)^2 \frac{2\epsilon}{1+\epsilon^2}f^\prime(\omega)  
\end{align}
\begin{equation}\label{eq_L22_final}
    \mathcal{L}_{22}=\frac{1}{1+q^2}(\alpha_3 q^2+\beta_3+\gamma_3 q)
\end{equation}
where $\alpha_3=(\omega-\mu)^2\alpha_1 ,\beta_3=(\omega-\mu)^2\beta_1$ and $\gamma_3=(\omega-\mu)^2\gamma_1$

From Eq.(\ref{eq_ZT}), the figure of merit can be written as
\begin{equation}\label{eq_ZT1}
    ZT=\frac{1}{\frac{\mathcal{L}_{11}\mathcal{L}_{22}}{\mathcal{L}_{12}^2}-1}
\end{equation}
From equations (\ref{eq_L11_final}) ,(\ref{eq_L12_final}) and (\ref{eq_L22_final})
\begin{equation}\label{eq_ratio_L11L22/L12^2}
    \frac{\mathcal{L}_{11}\mathcal{L}_{22}}{\mathcal{L}_{12}^2}=\frac{\left(\alpha_1 q^2 +\beta_1 +\gamma_1 q\right)\left(\alpha_3 q^2 +\beta_3 +\gamma_3 q \right)}{\left(\alpha_2 q^2 +\beta_2 +\gamma_2 q \right)^2}
\end{equation}
using Eq. (\ref{eq_ratio_L11L22/L12^2}), we can write $ZT$ in terms of q 
\begin{equation}
    ZT=\frac{1}{f(q)-1}
\end{equation}
where
\begin{equation}
    f(q)=\frac{\left(\alpha_1 q^2 +\beta_1 +\gamma_1 q\right)\left(\alpha_3 q^2 +\beta_3 +\gamma_3 q \right)}{\left(\alpha_2 q^2 +\beta_2 +\gamma_2 q \right)^2}
\end{equation}
\section{Thermoelectric Power Factor(PF) and Thermopower(S)}\label{appendixc}
\begin{figure}[b]
    \centering
    \includegraphics[scale=0.2]{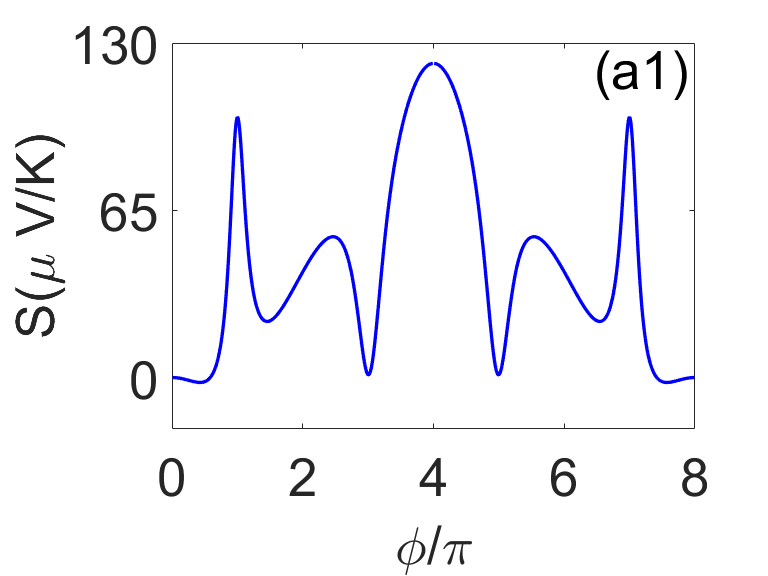}
    \includegraphics[scale=0.2]{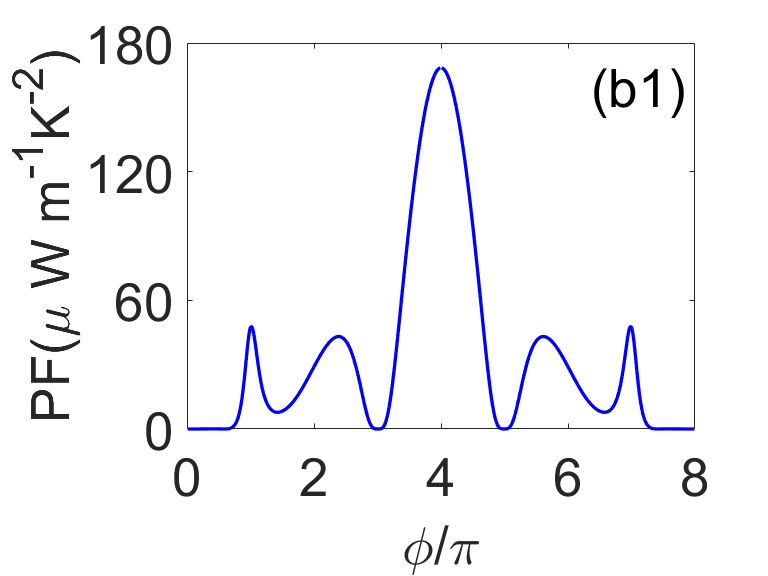}

    \caption{(a1) Seebeck coefficient(S) and (b1) thermoelectric Power factor (PF) as a function of magnetic flux $\phi$ for the $4QD(2,2)$ configuration for $\frac{t}{\gamma}=2$, $\gamma=42.77\mu eV,\mu=8\gamma,\epsilon=6\gamma, T=2\gamma$  }
    \label{power_factor}
\end{figure}
In this appendix, we illustrate the quantitative comparison with molecular-junction experiments \cite{finch2009giant,markussen2010relation,bergfield2009thermoelectric}. We compute the Seebeck coefficient $S$ and thermoelectric power factor $PF$ from the Onsager integrals (see Appendix \ref{appendix B}). We consider a typical configuration of $4QD(2,2)$ for this purpose. The Seebeck coefficient for a two-terminal open circuit condition is defined as $S=\frac{\mathcal{L}_{12}}{T\mathcal{L}_{11}}$. On the other hand, the thermoelectric power factor(PF) is defined as $ S^2 G$, where $S$ is the Seebeck coefficient and $G$ is the electrical conductance. Therefore, the power factor in terms of Onsager coefficients is defined as
\begin{equation}
    PF =S^2 G= \frac{\mathcal{L}_{12}^2}{T^3\mathcal{L}_{11}}
\end{equation}

To place our modeled performance on the same quantitative footing as recent molecular-junction experiments \cite{finch2009giant,markussen2010relation,bergfield2009thermoelectric}, we are using the following conversion formula to convert natural units to physical SI units:
 \begin{equation}
     S(\mu V/K)=(\frac{K_B}{e})S_{natural}=86.17\times S_{natural}
 \end{equation}
 and
 \begin{equation}
     PF_{SI}(\mu W/K^{-2}m^{-1})=\{\frac{k_B}{e}\}^2\times\frac{e^2}{h}\times PF_{nat}
 \end{equation}
 For Normalization, we consider a specific device geometry of a square strip of area $A=1 nm^2$ and length $L=1 nm$; then the power factor in SI units is given below
 \begin{align}
     PF(\mu W/K^{-2} m^{-1})&\nonumber=2.87\times10^{-7}\times10^{9}\times PF_{nat}\\
                            &=287\times PF_{nat}
\end{align}

\bibliography{references}

\end{document}